\documentclass{aa}
\usepackage{graphicx}
\usepackage[varg]{txfonts}
\makeatletter
\renewcommand*\aa@pageof{, page \thepage{} of \pageref*{LastPage}}
\usepackage[colorlinks=true, allcolors=blue]{hyperref}
\title{Forming rocky exoplanets around K-dwarf stars}
   \author{P. Hatalova\inst{1}\fnmsep\inst{2}\thanks{petra.hatalova@geo.uio.no},
          R. Brasser\inst{3}\fnmsep\inst{4},
          E. Mamonova\inst{1}\fnmsep\inst{2}
        \and
          S. C. Werner\inst{1}\fnmsep\inst{2}
          }
   \institute{Centre for Earth Evolution and Dynamics (CEED), University of Oslo, 0315 Oslo, Norway
        \and
             Centre for Planetary Habitability (PHAB), University of Oslo, 0315 Oslo, Norway
        \and
             Origins Research Institute, Research Centre for Astronomy and Earth Sciences, Konkoly Thege Miklos St 15-17, H-1121 Budapest, Hungary
        \and
             CSFK, MTA Centre of Excellence, Konkoly Thege Miklos St 15-17, H-1121 Budapest, Hungary    
             }
\begin{document}

  \abstract
   {New space telescopes, such as the upcoming PLATO mission, aim to detect and study thousands of exoplanets, especially terrestrial planets around main-sequence stars. This motivates us to study how these planets formed. How multiple close-in super-Earths form around stars with masses lower than that of the Sun is still an open issue. Several recent modeling studies have focused on planet formation around M-dwarf stars, but so far no studies have focused specifically on K dwarfs, which are of particular interest in the search for extraterrestrial life.}
   {We aim to reproduce the currently known population of close-in super-Earths observed around K-dwarf stars and their system characteristics. Additionally, we investigate whether the planetary systems that we form allow us to decide which initial conditions are the most favorable.}
   {We performed 48 high-resolution N-body simulations of planet formation via planetesimal accretion using the existing GENGA software running on GPUs. In the simulations we varied the initial protoplanetary disk mass and the solid and gas surface density profiles. Each simulation began with 12\,000 bodies with radii of between 200 and 2000 km around two different stars, with masses of 0.6 and $0.8~M_{\odot}$. Most simulations ran for 20~Myr, with several simulations extended to 40 or 100~Myr.}
   {The mass distributions for the planets with masses between 2 and 12~$M_\oplus$ show a strong preference for planets with masses $M_p<6$~$M_\oplus$ and a lesser preference for planets with larger masses, whereas the mass distribution for the observed sample increases almost linearly. However, we managed to reproduce the main characteristics and architectures of the known planetary systems and produce mostly long-term angular-momentum-deficit-stable, nonresonant systems, but we require an initial disk mass of $15~M_\oplus$ or higher and a gas surface density value at 1~AU of 1500 g cm\textsuperscript{-2} or higher. Our simulations also produce many low-mass planets with $M<2$~$M_\oplus$, which are not yet found in the observed population, probably due to the observational biases. Earth-mass planets form quickly (usually within a few million years), mostly before the gas disk dispersal. The final systems contain only a small number of planets with masses $M_p>10$~$M_\oplus$, which could possibly accrete substantial amounts of gas, and these formed after the gas had mostly dissipated.}
   {We mostly manage to reproduce observed K-dwarf exoplanetary systems from our GPU simulations.}

\maketitle

\section{Introduction} \label{introduction}
As of February 2023, more than 5200 exoplanets have been confirmed from stellar observations\footnote{\url{http://www.exoplanet.eu/}}, and several thousand planetary candidates discovered by missions such as \textit{Kepler}, K2, and TESS are awaiting confirmation. The majority of these planets are so-called super-Earths and mini- or sub-Neptunes, which fall between Earth and Neptune in radius and/or mass; such planets do not have an equivalent in our Solar System. Almost all known exoplanets orbit main-sequence stars of spectral categories F, G, K, or M, and the current programs searching for exoplanets tend to concentrate on such stars. Generally, Earth-mass planets are difficult to detect because of their low mass compared to the star, and Earth-mass planets on Earth-like orbits are even harder to detect because it requires long pointing periods from telescopes \citep[e.g.,][]{petigura2013plateau} and extremely high-precision measurements. However, this situation is slowly changing. New space telescopes, such as the upcoming ESA PLATO mission, aim to detect and study thousands of exoplanets. The ESA-stated primary goal of the PLATO mission is the “detection and characterization of terrestrial exoplanets around bright solar-type stars, with emphasis on planets orbiting in the habitable zone.” F, G, and K dwarfs are considered to be “solar-like stars.” This planned mission, among others, will provide exciting new opportunities for discovering exoplanets with Earth-like orbits and sizes; we must prepare our science accordingly.

While the new space telescopes allow us to detect and study exoplanets, the ground-based telescope Atacama Large Millimeter/sub-millimeter Array (ALMA), which is currently the largest ground-based astronomical project on Earth, contributes to our understanding of the formation of planets by imaging the gas and dust in their protoplanetary disks \citep{brogan20152014}. The initial conditions for planet formation depend on the properties of the protoplanetary disk the planets are forming in. These properties vary with the spectral type of the central star, amongst other things, which indicates that planet formation likely also varies across the stellar spectral types \citep{andrews2013mass}. This motivated us to study planet formation around different types of main-sequence stars. In this paper, we aim to explore the planet formation process and the resulting most common trends of the exoplanet population orbiting K-dwarf stars. Specifically, we model terrestrial planet formation in the inner disk of this type of star. Additionally, we tested the validity of assumptions commonly made for conditions in the early Solar System. 

Several modeling studies have focused on planet formation around M dwarfs \citep[e.g.,][]{laughlin2004core,lissauer2007planets,raymond2007decreased,miguel2020diverse,zawadzki2021rapid}, since they are the most common stars in our Galaxy \citep[e.g.,][]{henry2006solar,winters2014solar}. For this study we focus on K-dwarf stars, which are main-sequence stars with masses and radii between those of red M dwarfs and yellow G-type stars, such as our Sun. They have masses between 0.5 and 0.8 times the mass of the Sun and surface temperatures between 3900 and 5200 K \citep{pecaut2013intrinsic}\footnote{\url{www.pas.rochester.edu/~emamajek/EEM_dwarf_UBVIJHK_colors_Teff.txt}}. K dwarfs are not as common as M dwarfs (which comprise 75\% of the main-sequence stellar population in our Galaxy); they comprise about 15\% of the main-sequence population, meaning they are more abundant than G and F stars, which comprise 7\% and 2\%, respectively \citep{safonova2021quantifying}. Their lifetimes are longer than those of F and G stars; therefore, they are of particular interest in the search for extraterrestrial life. Using N-body simulations, we examined how the properties of planetary systems around K dwarfs are established from planetesimal accretion, that is, during the late stages of assembly when planets accrete from planetesimals and planetary embryos.

The currently known population of planets around this type of star is characterized by compact multi-planet systems of mostly small, dense planets with short periods and generally low orbital inclinations \citep[e.g.,][]{fang2012architecture}. This characterization could be partially due to observational biases, since these planets have been mostly found via the transit method. Additionally, it can be explained by the fact that the rate of solid accretion is proportional to the surface density of planetesimals; therefore, the planetary formation and evolution mostly happens inside and around the snow line (i.e., within a few AU) because at this location there could be a sudden increase in solids \citep[e.g.,][]{stevenson1988rapid,ciesla2006evolution,schoonenberg2017planetesimal}. K dwarfs are cooler than G-dwarf stars, so their snow line is closer to the star. Therefore, we tested the hypotheses that the most massive planetary embryos and protoplanets form there and that the final systems tend to be compact with short orbital periods. This is the reason our simulations are focused on the inner part of the disk around the star.

\begin{figure}
  \resizebox{\hsize}{!}{\includegraphics{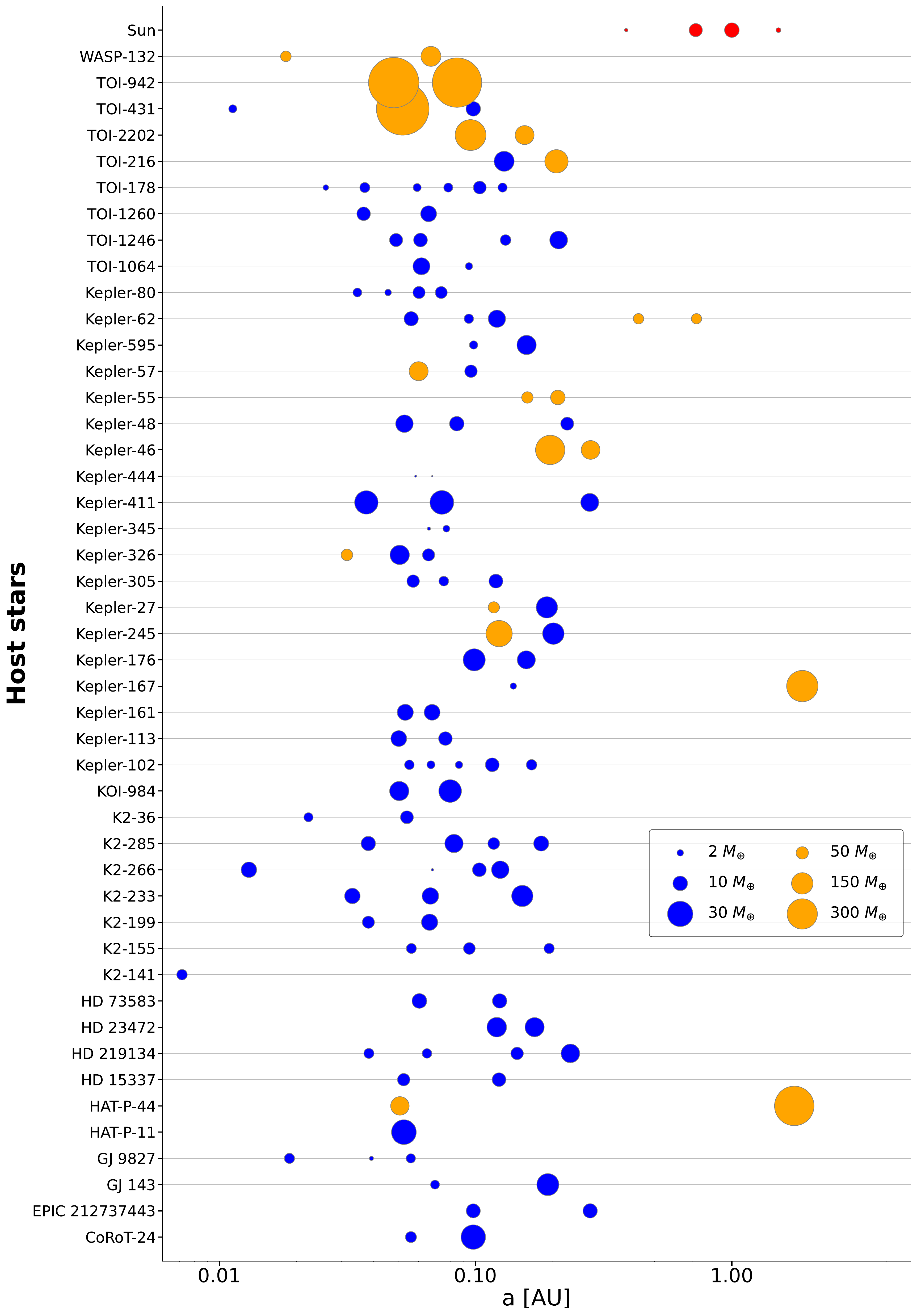}}
  \caption{Currently known multi-planet systems around K-dwarf stars.\ Only planets with known masses are presented. The sample was retrieved from \url{https://exoplanetarchive.ipac.caltech.edu/} on December 4, 2022. Giant planets with masses $>30~M_{\oplus}$ are shown in orange, and all planets with lower masses in blue. Terrestrial Solar System bodies are included for comparison (in red). The point size indicates the masses in $M_{\oplus}$ of the observed planets. Note the different size scale of the different colored bodies (blue and orange). Additionally, the markers representing the Solar System bodies are enhanced by a factor of 10 with regard to the planets indicated in blue for better visualization.}
  \label{KDwarfSystems}
\end{figure}

The 46 observed multi-planet systems around K dwarfs are shown in Fig. \ref{KDwarfSystems} in comparison to the terrestrial Solar System planets. The sample was retrieved from the NASA Exoplanet Archive\footnote{\url{https://exoplanetarchive.ipac.caltech.edu/}}. The systems contain at least two planets each (single planets orbiting K dwarfs are not included in the sample), and there are 139 planets in total (only planets with known masses are presented). Out of these 46 systems, 15 contain at least one giant planet (shown in orange in the figure). Here, we use the definition by \cite{clanton2014synthesizing}, who suggest that ``giant planets'' need to have masses of $\sim 30~M_{\oplus}$ or more. In our sample, about 30\% of the systems contain one or more giants, but since giant planets are typically easier to detect, we assume that this percentage may actually be lower for typical systems around K dwarfs. This assumption is supported by studies that show that giant planets with periods shorter than a few years detected around Sun-like stars have an occurrence rate of only around 10\% \citep{cumming2008keck,winn2015occurrence}. Other studies estimate that giant planets specifically orbiting late K-dwarf stars with masses of $\sim 0.5 - 0.75~M_{\odot}$ have an occurrence rate of under 1\% \citep{gaidos2013understanding,bryant2023occurrence}. Even though not all studies use exactly the same classification for giant planets, it is clear that these planets are not frequent around this type of star. 

It is likely that at least some of the systems in the retrieved sample are not complete, as suggested by studies of completeness of Kepler systems. For example, \cite{zink2019accounting} estimate the average number of planets to be $\sim$ 6 per G or K dwarf within the radius and period parameter space of Kepler detections, whereas the systems in our sample contains on average only three planets (3.0 $\pm$ 1.3). Giant planets in such systems would probably have already been discovered, so there is a higher probability that they are mostly made up of less massive distant planets that have not yet been detected. Earth-mass planets are currently only detectable when they are on very short orbits around lower-mass stars. In our sample, more than 60\% of the planets are ``super-Earths'' or ``sub-Neptunes'' with semimajor axes mostly between 0.04 and 0.2~AU. Almost all the discovered planets are much more massive than Earth and orbit their stars more closely than Mercury. In fact, the vast majority of the known systems are remarkably different from the Solar System \citep[e.g.,][]{petigura2013prevalence,winn2015occurrence,bryan2019excess}. The type of planet with the highest occurrence rate in K-dwarf systems is absent from the Solar System, but it is currently the most abundant class of exoplanet. Planets with radii between 1 and $4~R_{\oplus}$ and orbital periods shorter than 100 days are present around at least 30\%–50\% of main-sequence stars of spectral types G and K \citep{mayor2009harps,mayor2011harps,howard2010occurrence,howard2012planet}.

A typical known K-dwarf planetary system does not contain giant planets, but it comprises on average three planets, mostly super-Earths or sub-Neptunes with very short semimajor axes. This characteristic is in large part due to observational biases, and we expect the existence of thus far undetected, less massive planets to mainly orbit farther from the host stars. In the next sections, our simulated systems are compared to this retrieved sample in order to evaluate our results. Our study is focused on the formation of terrestrial-type planets (Earths and super-Earths). As such, we do not analyze the distribution of planets with masses above $10~M_{\oplus}$, the upper mass limit for super-Earths \citep{stevens2013posteriori}, in the observed systems when comparing them with our simulated systems.

The interaction between the growing planetary embryos and the gas in the protoplanetary disk induces a torque that causes these embryos to migrate toward the star. The onset of migration occurs when the planets reach approximately the mass of Mars \citep[i.e., about $0.1~M_{\oplus}$;][]{ward1986density,tanaka2002three,paardekooper2011torque}. The observed planets on very short orbits described earlier suggest that orbital migration (Type I migration in the case of smaller planets) was important during the evolution of these planetary systems, as it is not likely that these planets formed at their current orbits \citep[e.g.,][]{izidoro2014terrestrial}. After the gas disk dissipates, the migration stops and rocky planets may continue growing via collisions between embryos \citep{agnor1999character} because their gravitational interactions might lead to an excitation of their eccentricities, making collisions possible. The gas disk damps their eccentricities and inclinations, so when the gas is gone, the evolution of embryos gradually becomes chaotic.

Our understanding of terrestrial planet formation has improved over the last few decades, but the formation of multiple close-in super-Earths around stars with lower masses than the Sun is still an open issue. Specifically, the mechanism that prevents these planets from migrating too close to their host star is not yet satisfactorily known due to the uncertain effects of torques near the disk's inner edge \citep[e.g.,][]{brasser2018trapping}. The gas disk does not extend all the way to the star, probably due to the star’s magnetosphere. The magnetic field disrupts the disk out to a radius at which the magnetic energy density and the kinetic energy density of the gas are equal \citep[e.g.,][]{long2005locking}. At this radius the Type I migration slowly stops, so this gas-free cavity works as a migration trap where there is no migration, nor eccentricity or inclination damping. For a typical T Tauri star, the magnetospheric boundary is located at around 0.05–0.1~AU \citep{romanova2006magnetospheric,romanova20193d}. Migrating protoplanets usually end up captured in mean motion resonance \citep[MMR;][]{terquem2007migration}. In multiple N-body simulation studies, several planets migrate inward together as a resonant chain of low-mass protoplanets. When the innermost of them reaches the inner disk, their migration slows until they eventually stall near the disk’s inner edge 
\citep{ogihara2009n,cossou2014hot,coleman2014formation,coleman2016formation}. Simulations also show that either the further evolution of these planets will keep them in resonances and dynamically stable for billions of years \citep{esteves2020origins} or they will become dynamically unstable, typically at the end of or shortly after the gas disk dispersal. Late dynamical instabilities break resonances and often lead to orbital crossings and giant impacts \citep{izidoro2017breaking,izidoro2021formation}. This scenario is commonly referred to as ``breaking the chains'' \citep[e.g.,][]{terquem2007migration,ogihara2009n,mcneil2010formation,cossou2014hot,izidoro2017breaking,izidoro2021formation}. The final configurations of systems that underwent late dynamical instabilities are expected to be nonresonant. Most currently known super-Earths are not found in resonant systems \citep{lissauer2011architecture,fabrycky2014architecture}. On the other hand, systems such as TRAPPIST-1 and Kepler-223 have several planets in resonant chains, and, according to multiple studies \citep{cossou2014hot,izidoro2017breaking,izidoro2021formation,ogihara2018formation}, they represent a small fraction of systems that did not become unstable after the dispersal of the gas disk. Resonant systems are naturally produced by migration; therefore, this is an important effect to consider when it comes to terrestrial planet formation.

In the next sections, we first explain the methods used in this study and introduce our model. We describe the protoplanetary disk model and density profile together with the initial conditions for each of our setups. Then we present and discuss the results of our simulations and the dynamical evolution, final architecture, and characteristics of the systems we have built, as well as their significance and implications. We also compare the simulated systems to observations. Finally, in the last section, we present the conclusions of our study.

\section{Methods} \label{methods}
For the N-body simulations we employed the GENGA software \citep{grimm2014genga,grimm2022genga}. GENGA is a GPU implementation of a hybrid symplectic N-body integrator, developed based on another N-body code MERCURY \citep{chambers1999hybrid}, for simulating planet and planetesimal dynamics in the final stages of planet formation, and the evolution of planetary systems. It integrates planetary orbits over long timescales with excellent energy conservation. Gravitational interactions between planetary bodies are computed as perturbations of the Keplerian orbits \citep{wisdom1991symplectic}. The GENGA code has been successfully used for simulations of the terrestrial planets formation in the Solar System \citep[e.g.,][]{clement2020embryo,woo2021growing,woo2022terrestrial}. Planet formation simulations usually start from the runaway growth phase when planetesimals collide and form bigger objects called planetary embryos \citep{kokubo1995orbital}. Then they continue with oligarchic growth \citep{kokubo1998oligarchic}, when a small number of largest embryos begin to dominate the dynamics of the disk until they reach their isolation mass by accreting all planetesimals in their feeding zones, and eventually protoplanets and planets form \citep[e.g.,][]{kokubo2000formation,chambers2006semi}.

We model the formation of terrestrial planets around a $0.6~M_{\odot}$ and a $0.8~M_{\odot}$ star from planetesimals and planetary embryos to evolved planets using planetesimal accretion. We do not actually expect any substantial differences in the simulated outcomes for the different stellar masses, but we have chosen these values to cover most of the K-dwarf stars mass range, which will be potentially useful for our future studies. We have not adapted the structure or characteristics of the disk to account for the lower or higher stellar mass as we assume these to be just second-order differences. In the end, we do not draw any conclusions for the outcomes (and their comparison to the observed systems) based on the different stellar masses as the uncertainties on exoplanet radii, masses, and orbital parameters do not allow for it.

In this study we do not attempt to reproduce any specific observed planetary systems. Rather, we consider the generic outcome of planet formation for this type of star and the results of our simulations are then compared to the known systems (as described in Sect. \ref{introduction}). All together we ran 48 high-resolution N-body simulations that varied the initial protoplanetary disk mass, and solid and gas surface density profiles. We investigated whether the planetary systems that were formed allow us to decide which initial conditions and other parameters are responsible for their final characteristics. Simulations start with planetesimals and planetary embryos in their predefined initial locations. Then the bodies begin interacting gravitationally around the star, which results in perturbations of their orbits, their close encounters and collisions, and often ejections of some of the bodies. We followed their collisional growth for up to 20 million years, until a system of planets or protoplanets (and some leftover planetesimals) is formed, where the planets generally do not collide with each other anymore.

Current computational hardware including recently improved GPU hardware is increasingly powerful and allows us to run the necessary calculations for planet formation processes faster than ever before. Nevertheless, the high-resolution long-term simulations still take a relatively long time to run; therefore, we limited the simulations to 20 million years. This time period is long enough for our purposes as previous studies already show that this type of planet forms quickly. For example, \cite{ogihara2015reassessment} examine the formation of close-in super-Earths around G dwarfs using N-body simulations and find that Type I migration can cause planets to form in only a few~megayears. Our N-body simulations include the gas disk effects (Type I migration as well) on the formation of this kind of planet (close-in super-Earths), but around a bit less massive stars. We therefore expect that planets around K-dwarf stars may form within a similar time frame.

\subsection{Initial conditions and parameters} \label{initial}
All the simulations begin with 12\,000 bodies grouped in radius between 200-2000 km depending on the initial mass of the disk of solids, which we vary between $5~M_{\oplus}$ and $40~M_{\oplus}$. Even though this disk mass may seem excessive, recent studies have been performed with similar initial masses. For example, \cite{morbidelli2022contemporary} manage to produce the total mass of planetesimals exceeding $40~M_{\oplus}$ at the silicate sublimation line in the forming Solar System (at $\sim$ 1~AU, but closer to the star for K dwarfs) by reducing the viscosity parameter $\alpha$ and suggested that this large amount of mass could explain the formation of the frequently observed super-Earths. In our simulations all planetesimals are fully self-gravitating. The higher the initial disk mass, the larger (and therefore more massive) bodies we start with to preserve the number of 12\,000 particles. The planetesimal number was chosen based on a series of simulations testing the speed of the code. Since more bodies result in more calculations and longer computational time scaling roughly as $N^2$ \citep{grimm2022genga}, using 12\,000 particles achieves a high-enough resolution, but still within a reasonable time limit of around 2 months and available computation time. The fact that simulations with higher initial disk masses start with larger bodies should not affect the simulation outcome. According to \cite{miguel2020diverse} the initial size of planetesimals does not affect the final population, at least for bodies between 100 and 500 km, and our test-runs showed the same for larger sizes. With larger initial bodies the planetary growth happens faster \citep[e.g.,][]{woo2021growing}, but the formation timescale is not the main focus of this study. At the beginning of a simulation, the masses of the planetesimals and embryos are calculated from the initial radii and density; the latter is chosen to be 3 g cm\textsuperscript{-3}. This value is commonly assumed for rocky planetesimals formed in the inner part of a protoplanetary disk. As planetesimals collide and grow, the radius of the new planetary body is set by conserving the mass and mixing the densities of the merged bodies. Since the density is the same for all the particles, the final planets will all have the same densities if any compressional effects are ignored; hence, these simulations do not allow us to properly assess the radii of the final bodies. This is a limitation that will be improved in the future studies.

\begin{figure}
  \resizebox{\hsize}{!}{\includegraphics{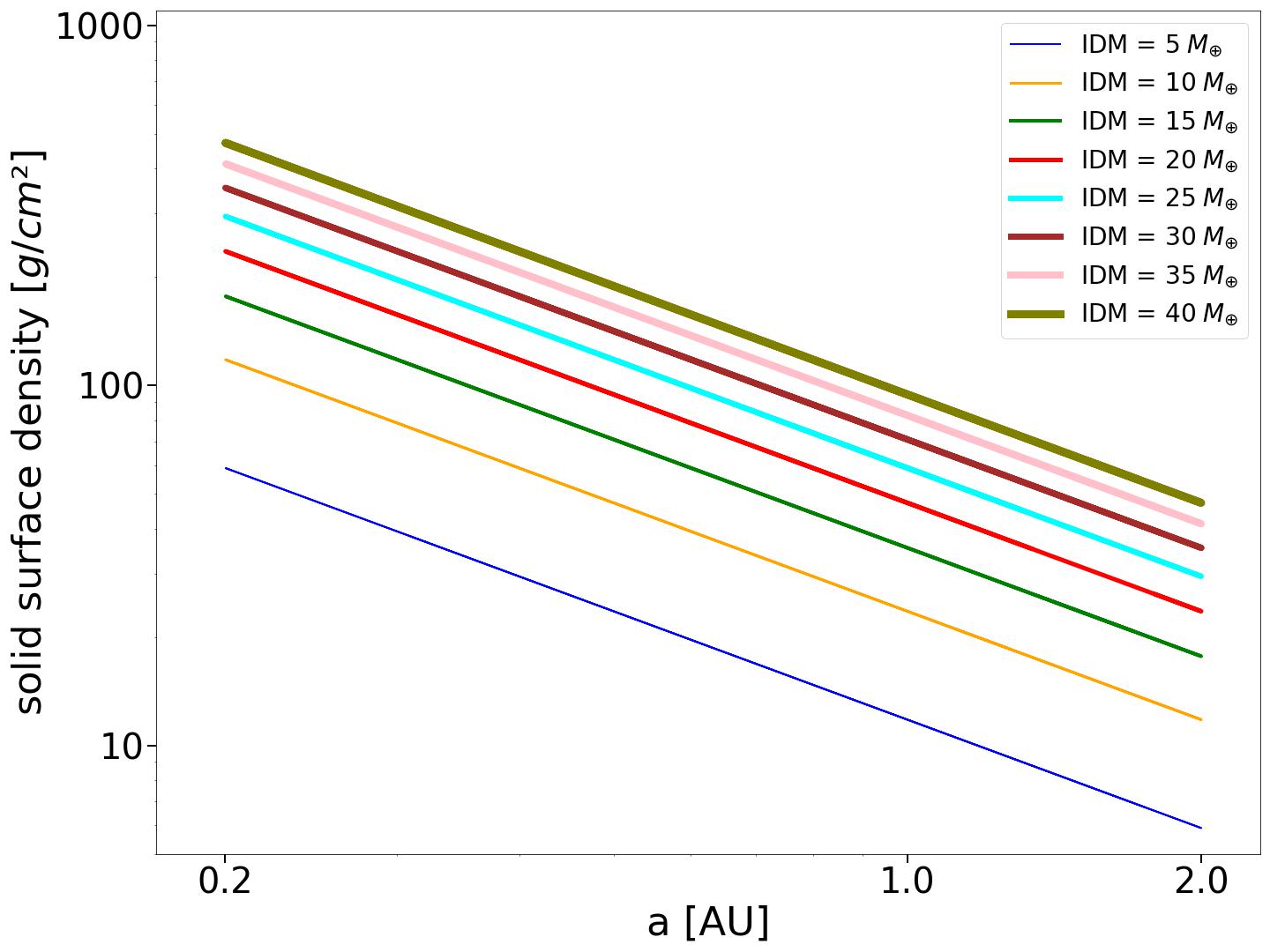}}
  \caption{Initial solid surface density distribution for simulations with the starting planetesimal locations between 0.2 and 2~AU and different initial disk masses (IDM). The increasing linewidth with the values on the y-axis represents the growing fraction of larger bodies in the disk with increasing initial disk mass.}
  \label{solidSurfaceDensity}
\end{figure}

The planetesimals are distributed between 0.2 and 2~AU with a surface density ($\Sigma_{P}$) profile $\Sigma_{P}\propto r^{-1}$ for the initial disk. Models for the viscous evolution of the gas in the disk suggest a shallow planetesimal surface density slope of $\Sigma_{P}\propto r^{-0.9}$ \citep{shakura1973black}, whereas the minimum mass solar nebula (MMSN) hypothesis provides a steeper profile of $\Sigma_{P}\propto r^{-1.5}$ \citep{weidenschilling1977distribution,hayashi1981structure}. Other similar studies use these surface density profiles or even a steeper slope of $\Sigma_{P}\propto r^{-2.1}$ from  \cite{lenz2019planetesimal}. \cite{zawadzki2021rapid} run their simulations with $\Sigma_{P}\propto r^{-0.6}$ and $\Sigma_{P}\propto r^{-1.5}$, and show that the solid surface density of the final planetary systems appear to be independent from the initial distribution of embryos. Therefore, we decided to use a mean value of the last two slopes. Figure \ref{solidSurfaceDensity} shows the initial solid surface density distribution for simulations with the starting planetesimal locations between 0.2 and 2~AU, and different initial disk masses. The values of surface density at 1~AU are in the range from 11.8 (for the initial disk mass $5~M_{\oplus}$) to 94.4 g cm\textsuperscript{-2} (for the initial disk mass of $40~M_{\oplus}$). The higher-end values are significantly higher than in other studies, for example \cite{mulders2020earths} do not exceed 50.0 g cm\textsuperscript{-2} in their N-body simulations around a solar mass central star. We extended the range to these hypothetical values to study whether massive planets can form in such conditions.

\begin{figure}
  \resizebox{\hsize}{!}{\includegraphics{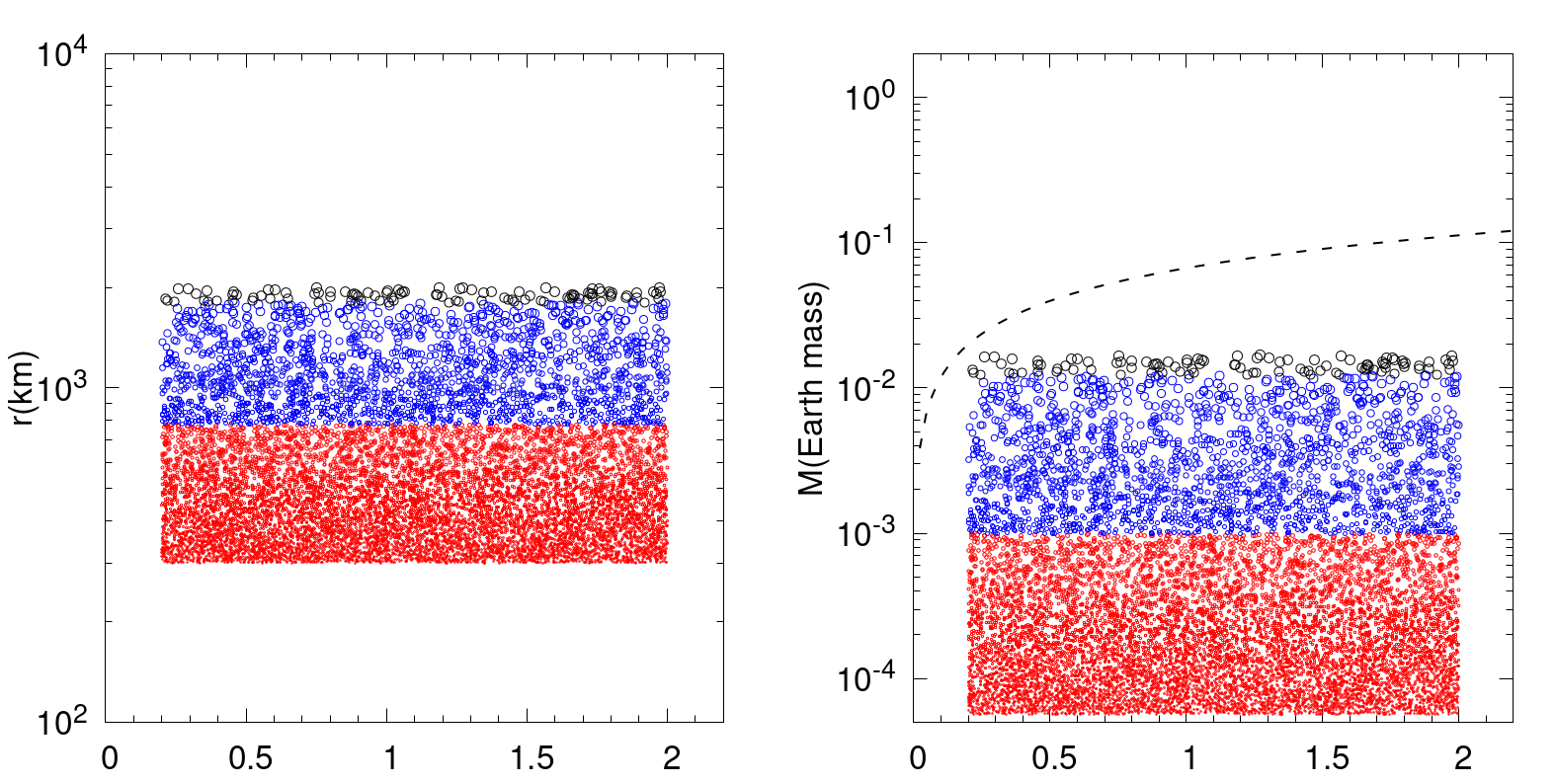}}
  \caption{Initial distribution, masses, and radii of 12\,000 bodies for a simulation with an initial solid disk mass of $10~M_{\oplus}$. The x-axis represents the distance from the central star in AU, and the y-axis the radii (left panel) and masses (right panel). Red, blue, and black indicate different mass ranges of the objects, i.e., mass $M<10^{-3}~M_{\oplus}$ (considered as planetesimals), $10^{-3}M_{\oplus}<M<10^{-2}~M_{\oplus}$ (proto-embryos), and $M>10^{-2}~M_{\oplus}$ (embryos), respectively. There is no difference in the treatment of these objects during the N-body simulations. The isolation mass, which is the mass of embryos that accreted all the bodies within their feeding zone, is represented by the dashed line.}
  \label{fig:initial}
\end{figure}

During the simulations, bodies are removed when they come closer to the central mass than the inner truncation radius (they are assumed to collide with the star), or move farther out than the outer truncation radius of the disk, or when they collide and the less massive body merges into the more massive one. We assume perfect accretion during collisions, ignoring fragmentation, since it seems that this effect does not radically alter the outcome of the simulations \citep{kokubo2010formation,chambers2013late,quintana2016frequency}. The formation of the planetesimals themselves is suggested to happen very early in the lifetime of a protoplanetary disk, already after a few thousand years at 2–3~AU \citep{lenz2019planetesimal}. In the case of large planetesimals or planetary embryos, age determination of iron meteorites, representing the metal cores of differentiated planetesimals, suggests that these were formed in the inner Solar System within the first megayear \citep{halliday2006meteorites,kruijer2014protracted,schiller2015early}. This time period is still short, so for our purposes we set the start time of the simulations to 0 and assume that planetesimals and embryos have not moved much while they were forming. In this manner their initial distribution of orbital elements reflects that of a young disk. The planetesimals had initial values of the inclination and orbital eccentricity randomly ranging between 0 to 0.1 degree and 0 to 0.01, respectively, but these values increase quickly after the simulations start due to particle interactions. An example of the initial planetesimal/embryo distribution with their masses and radii is shown in Fig. \ref{fig:initial}, where the different colors indicate mass ranges of the different objects (i.e., planetesimals, protoplanetary embryos, and embryos). Planetary embryos are often defined as objects of at least the mass of the Earth’s Moon, $M_{Moon}=0.012~M_{\oplus}$ \citep[e.g.,][]{woo2021growing,woo2022terrestrial,voelkel2021linking}, we also use this definition. There is no difference in treatment of these objects during the N-body simulations.

Our simulations represent a disk model with the particles initially distributed between 0.2 and 2~AU and the inner truncation radius of 0.05~AU; in other words, the inner edge of the simulations is set to quarter of the semimajor axis of the innermost body at the start of the simulations. All other parameters are then varied in the model. The time step necessary for the N-body integration runs depends on the orbital period of the possible innermost object. It is recommended to use 1/20 of the orbital period at the cutoff distance for sufficient accuracy \citep{wisdom1991symplectic}. However, a shorter time step results in a longer computation time and such N-body simulations already require many weeks, even months of computation time. Therefore, we used a longer time step than recommended. Results that are presented in the following sections do not seem affected by this compromise. This issue is discussed more in Sect. \ref{limitations}, where we present results of a run with the recommended time step. Table \ref{tab:coparameters} summarizes the main parameters and initial conditions of our model. The inner and outer truncation radius are cutoff distances for the simulations; bodies with a separation to the central mass smaller or larger than these values are taken out of the simulation. Initial semimajor axes $a_{min}$ and $a_{max}$ limit the planetesimal spatial distribution at the beginning of a simulation. We varied the initial mass of the solid disk and gas surface density profile, presented in Table \ref{tab:parameters}, and simulated 20~Myr of planetary evolution around two different central masses. We ran all combinations of these parameters.

\begin{table}
\caption{Main parameters and initial conditions of the model.}   
\label{tab:coparameters}      
\centering                                     
\begin{tabular}{c c c}   
\hline\hline                       
Time step & 0.7305 days \\
Inner truncation radius & 0.05 AU \\
Outer truncation radius & 3 AU \\
Initial semimajor axes - $a_{min}$ & 0.2 AU \\
Initial semimajor axes - $a_{max}$ & 2 AU \\
Gas disk decay time & 1~Myr \\
\hline                                             
\end{tabular}
\end{table}

\begin{table}
\caption{Parameters varied in the simulations.}   
\label{tab:parameters}      
\centering                                     
\begin{tabular}{c c c}   
\hline\hline                       
Central mass & Initial disk mass & Gas surface density \\ \newline
[$M_{\odot}$] & [$M_{\oplus}$] & [g cm\textsuperscript{-2}] \\
\hline                                
0.6 & 5, 10, ..., 40 & 1000, 1250 \\
0.8 & 5, 10, ..., 40 & 1000, 1250, 1500, 1750 \\
\hline                                             
\end{tabular}
\end{table}

The initial conditions for planet formation depend on the properties of the protoplanetary disk. For the estimates of the initial mass of the protoplanetary disk we followed \cite{manara2018protoplanetary}. In their study, they focus only on the solid part of the disks and present recalculated disk dust masses (the original disk masses are from \cite{pascucci2016steeper} and \cite{ansdell2016alma}) around various spectral types of stars, including K dwarfs with different masses, located in two young star-forming regions. The dispersion of the disk mass estimates is large, and for K-dwarf stars this ranges from less than $5~M_{\oplus}$ to more than $70~M_{\oplus}$, for one of the star-forming regions and to more than $110~M_{\oplus}$ for the other one. But for the majority of the disks the dust masses are less than $10~M_{\oplus}$. Of course, planetary systems are concentrated within a few AU of the central star, whereas most of the disk mass is farther out; therefore, we should consider only a small portion of this estimate for our simulations. On the other hand, these estimated disk dust masses might be highly underestimated due to multiple reasons, as several studies have pointed out. For example, they are derived from disk measurements that are only sensitive to certain grain sizes, so it is possible that
a significant fraction of the dust is undetected \citep[e.g.,][]{williams2012astronomical}. Also, the estimates depend on the values of dust opacity and disk temperature, which are still debatable \citep[e.g.,][]{andrews2013mass,pascucci2016steeper}. Some dust mass might be confined to optically very thick inner regions of disks \citep[e.g.,][]{tripathi2017millimeter}; however, the measurements are based on the emission from the outer regions ($R > 10$~AU) of disks \citep[e.g.,][]{bergin2017determination}. After taking all these uncertainties into account, we chose initial solid mass values in a range of 5 to 40~$M_{\oplus}$ for the simulations.

\subsection{Gas surface density profile}
The gas disk implementation in GENGA follows that of \cite{morishima2010planetesimals}. GENGA supports all important gas disk effects, such as aerodynamic gas drag, disk-planet interaction including the gas disk tidal damping, Type I migration, and the global disk potential. All these effects are included in the simulations. The gas surface density dissipates exponentially in time ($t$) and uniformly in space ($r$) following
\begin{equation} \label{eq:1}
\Sigma_{\rm gas}(r,t)=\Sigma_{\rm gas,0} \left( \dfrac{r}{1 \text{ AU}} \right)^{-1} \text{exp} \left( -\dfrac{t}{\tau_{\rm decay}} \right).
\end{equation}

In Eq. \ref{eq:1}, $\Sigma_{\rm gas,0}$ is a gas surface density ($\Sigma_{\rm gas}$) at 1~AU and $t=0$. The MMSN, which is used to define the initial conditions of the protoplanetary disk required to make the planets around the Sun, has $\Sigma_{gas,0}$ = 1700 g cm\textsuperscript{-2} based on \cite{hayashi1981structure}, whereas \cite{morishima2010planetesimals} assume $\Sigma_{gas,0}$ = 2000 g cm\textsuperscript{-2}. We decided to test four different values: 1000, 1250, 1500, and 1750 g cm\textsuperscript{-2}, which are mostly lower than assumed MMSN values and reflect the lower masses of our stars compared to the Sun. The quantity $\tau_{\rm decay}$ represents the gas disk decay timescale, which we fixed at 1~Myr. \cite{woo2021growing,woo2022terrestrial} examine decay timescales of 1~Myr and 2~Myr for simulations of the Solar System formation, and show that decay timescale $\le$ 1~Myr can better explain the specific characteristics of our planetary system. They also show that the mass loss to the Sun due to Type I migration is lower in the case of the shorter decay time. Since both our stars are smaller than the Sun, and we are trying to form more massive terrestrial planets than exist in the Solar System, we chose the shorter decay time of 1~Myr. However, it is important to mention that real disks may not dissipate smoothly at all orbital radii, instead inside-out dissipation can alter the orbital architectures of planetary systems \citep{liu2017dynamical} and even trigger instability \citep{liu2022early}. After the gaseous part of the protoplanetary disk dissipates, gas drag and gas dynamical friction disappear as well. This causes the accretion rate to slow down and eccentricity and inclination damping of planetary embryos will cease too \citep[e.g.,][]{bitsch2010orbital}. In the version of GENGA employed here, the gas disk inner edge is hard-coded to 0.1~AU. As mentioned in Sect. \ref{introduction}, the magnetospheric boundary is located at around 0.05–0.1~AU. The gas disk inner edge represents the boundary in our simulations and we chose to keep this value of 0.1~AU for our runs.

This gas surface density profile is used in our model, the inner and outer truncation radius are at 0.05~AU and 3~AU, respectively. That means that, since the gas disk inner edge is set to 0.1~AU, there is a cavity without gas between the inner truncation radius and the inner edge of the gas disk. Many known planetary systems around K-dwarf stars contain planets with orbital distances shorter than 0.1~AU (see Fig. \ref{KDwarfSystems}); therefore, we decided to set the inner truncation radius to 0.05~AU. Figure \ref{KDwarfSystems} shows that some planets are observed even closer to their host stars than 0.05~AU; however, due to limited computation time we did not use an even lower value of the inner radius.

\section{Results}
In this section, we present our simulated planetary systems around the K-dwarf stars after 20~Myr of planet formation. In total, we performed 48 runs with different combinations of parameters with the target of finding parameters that reproduce the known K-dwarf system characteristics. We varied the initial protoplanetary disk mass, and solid and gas surface density profiles. But first we start by discussing the dynamical evolution of the simulations.

\subsection{Dynamical evolution of the simulated systems} \label{evolution}
\begin{figure*}
\centering
  \includegraphics[width=17cm]{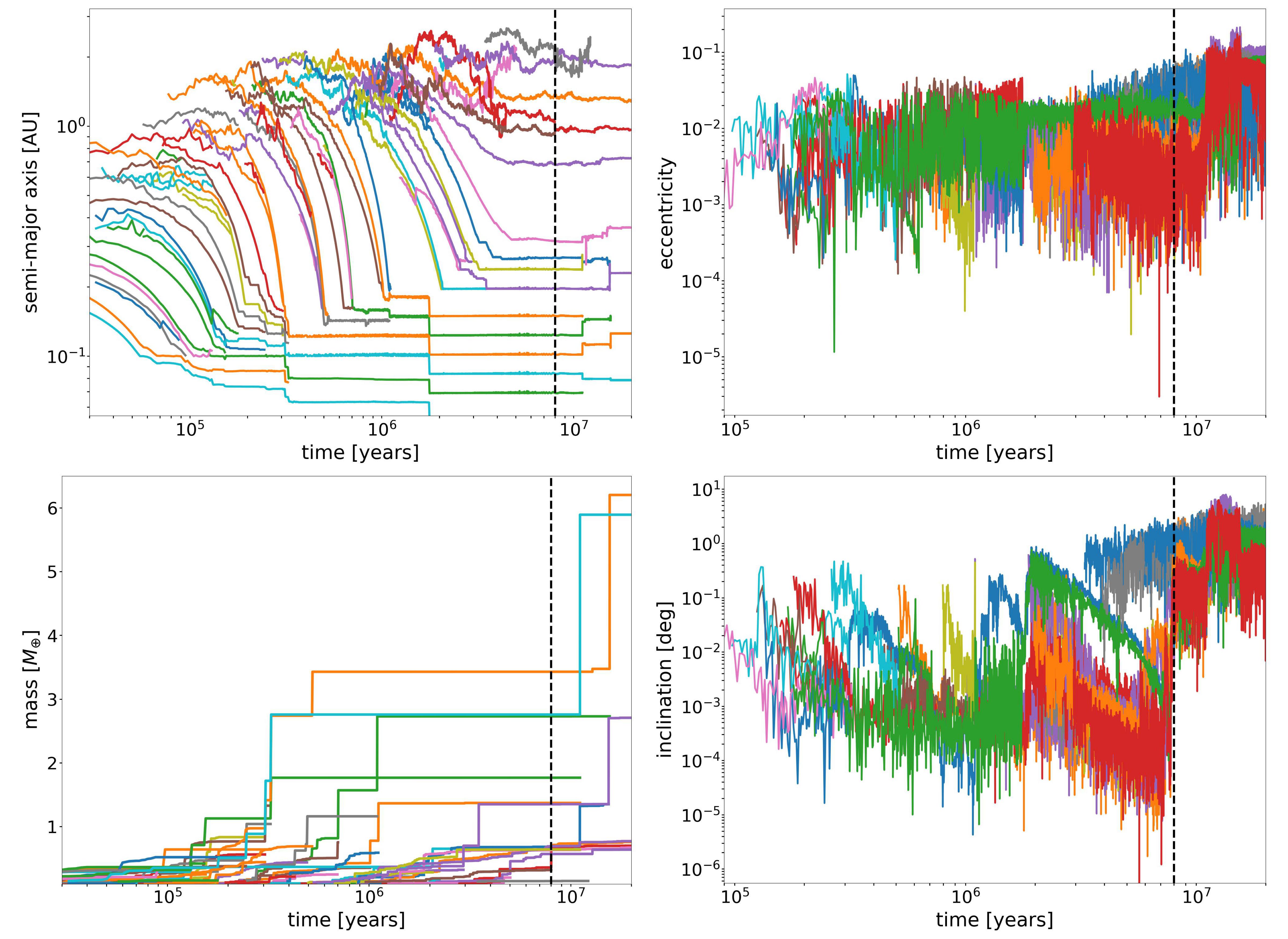}
  \caption{Dynamical evolution of planets in one of the typical systems that underwent several late dynamical instabilities after the gas dispersal. The plots show the temporal evolution (20~Myr) of planetary bodies, specifically their semimajor axes, masses, eccentricities, and orbital inclinations. Each line color represents an individual object; however, the color is random in each plot. The dashed black line shows the estimated time of the gas dissipation at $\sim$ 8~Myr. For better visualization, the panels on the left (semimajor axes and masses) show the evolution of all bodies with masses > $0.1~M_{\oplus}$, and the panels on the right (eccentricities and orbital inclinations) show only bodies with masses > $0.5~M_{\oplus}$.}
  \label{fig:dynEvolution79}
\end{figure*}

The dynamical evolution (20~Myr) of a typical simulation is presented in Fig. \ref{fig:dynEvolution79}. The panels on the left show the evolution of semimajor axes and masses of all bodies with a mass $>0.1~M_{\oplus}$, and the panels on the right present eccentricities and orbital inclinations, and show only bodies with masses $>0.5~M_{\oplus}$ to better visualize these orbital characteristics. The top left plot shows a typical evolution of such a simulation based on the migration model as described in Sect. \ref{introduction}. During the gas disk phase, planetary embryos grow and migrate inward, occasionally end up falling onto the central star as we see just before 2~Myr. The planets quickly settle into chains of MMRs. According to \cite{izidoro2017breaking}, the chains are typically established within 1.5~Myr; this is the case in our simulations as well. As the gas in our disk decays exponentially with the decay time of 1~Myr, it will never completely dissipate, but we see that at $\sim$~5~Myr the amount of gas in the disk must be relatively low as migration mostly stops. Additionally, eccentricity and inclination damping disappears together with the gas and this is clearly demonstrated in both eccentricity and inclination plots at around 7 or 8~Myr when both properties, but inclination particularly, increase. Therefore, we estimated the time of the (most) gas dispersal at $\sim$~8~Myr. After this time, the plots display several dynamical instabilities occurring at around 11~Myr and later. They lead to orbit crossings, collisions and merging of some of the planets, as displayed in both left plots. Almost all our systems undergo such instability phases, but usually evolve to a stable configuration fairly quickly (in the order of $10^{5}$ years). Eventually, the planetary system becomes more dynamically excited and less compact; however it still contains eight planets, six of them interior to 1~AU. In this simulation, the three most massive planets have several $M_{\oplus}$ and the rest are under $1~M_{\oplus}$. Both the first and the third (from the central star outward) subsequent planet pairs are in 2:1 resonance. The final orbital eccentricities are typically around 0.05 (median = 0.05 $\pm$ 0.02) and inclinations around 1.42 deg (median = 1.42 $\pm$ 0.74 deg). These values are within the current eccentricities and inclinations of the three largest terrestrial planets of the Solar System.

\begin{figure*}
\centering
  \includegraphics[width=17cm]{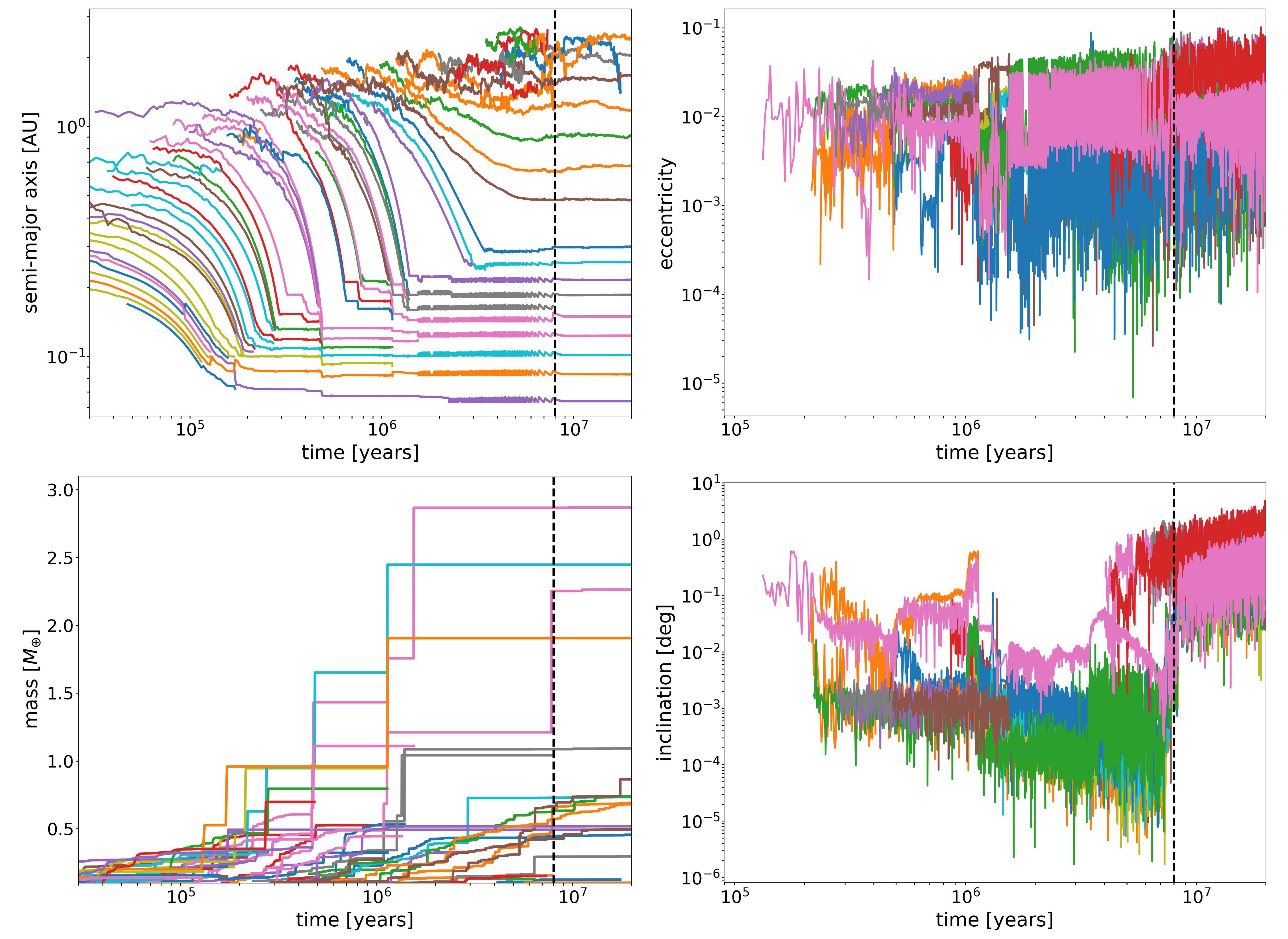}
  \caption{Dynamical evolution of planets in a less common system that did not undergo any late dynamical instabilities after the gas dispersal. The plots show the temporal evolution (20~Myr) of planetary bodies, specifically their semimajor axes, masses, eccentricities, and orbital inclinations. Each line color represents an individual object; however, the color is random in each plot. The dashed black line shows the estimated time of the gas dissipation at $\sim$ 8~Myr. For better visualization, the panels on the left (semimajor axes and masses) show the evolution of all bodies with masses > $0.1~M_{\oplus}$, and the panels on the right (eccentricities and orbital inclinations) show only bodies with masses > $0.5~M_{\oplus}$.}
  \label{fig:dynEvolution61}
\end{figure*}

While Fig. \ref{fig:dynEvolution79} displays the dynamical evolution of a typical simulation that underwent several dynamical instabilities after the gas disk dissipation, Fig. \ref{fig:dynEvolution61} shows a less common evolution of a system that did not do through any instabilities. The panels again show 20~Myr of evolution in semimajor axes, masses, eccentricities and inclinations, both before and after the gas dispersal at $\sim$~8~Myr. The plots show some collisions after the gas disk phase as well, but they are very rare and involve only small leftover bodies farther from the central star. No dynamical instabilities that would affect the architecture of the system are visible, but we still see some planet growth in the bottom left plot, which displays bodies with masses > $0.1~M_{\oplus}$. After 20~Myr, the final planetary system consists of 11 planets (> $0.5~M_{\oplus}$), with the most massive planets being half the mass of the most massive planets in the previous system (Fig. \ref{fig:dynEvolution79}). This system is also more compact than the previous example. The inner planets are in a 4:3-4:3-4:3---5:4 resonant chain, that is, the three innermost subsequent planet pairs are all in 4:3 period ratio ($P_{out}/P_{in}$) and the fifth pair is in 5:4 resonance. Multiple other pairs are very close to being in resonance. In this study, we allowed for a 1\% deviation of a resonant loci away from the exact commensurability based on \cite{brasser2022long} (Fig. 2), independently of the eccentricities. The final orbital eccentricities are typically around 0.02 (median = 0.02 $\pm$ 0.03) and the inclinations around 0.84 deg (median = 0.84 $\pm$ 1.15 deg). These values are a bit lower than the typical values for the system that underwent several dynamical instabilities, which is expected as the instabilities excite the planets. But the values are still very similar, which can be explained by the fact that the previous system had enough time to stabilize again after the last instability phase. The dynamical evolution examples presented in Figs. \ref{fig:dynEvolution79} and \ref{fig:dynEvolution61} are representative of all our simulations. Although we present the most contrasting examples, we see that their evolutionary paths show many similarities, hinting at a generic pathway.

\begin{figure*}
\centering
  \includegraphics[width=17cm]{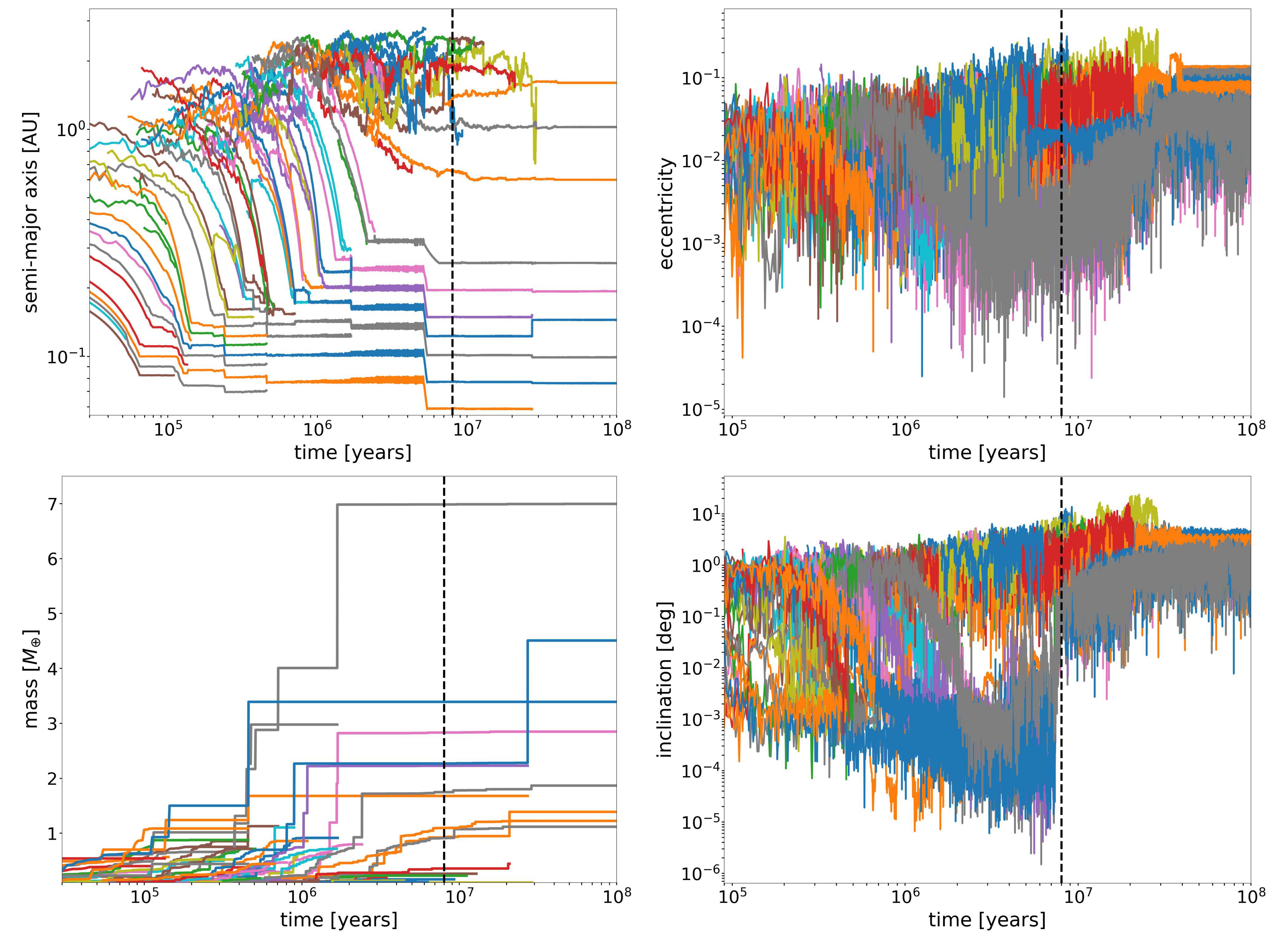}
  \caption{Dynamical evolution of planets in an extended simulation (100~Myr) that underwent multiple dynamical instabilities much later after the gas dispersal, at approximately 21~and 28~Myr. The plots show the temporal evolution of planetary bodies, specifically their semimajor axes, masses, eccentricities, and orbital inclinations. Each line color represents an individual object; however, the color is random in each plot. The dashed black line shows the estimated time of the gas dissipation at $\sim$ 8~Myr. For better visualization, the panels on the left (semimajor axes and masses) show the evolution of all bodies with masses > $0.1~M_{\oplus}$, and the panels on the right (eccentricities and orbital inclinations) show only bodies with masses > $0.5~M_{\oplus}$.}
  \label{fig:dynEvolution92}
\end{figure*}

To demonstrate that a dynamical instability can also occur much later in the simulation time, even in a simulation that did not undergo an instability after the gas dispersal during the standard time of our simulations (20~Myr), we present one example of a few simulations that were extended to 100~Myr (see Fig. \ref{fig:dynEvolution92}). The plots show several collisions resulting in planetary growth at approximately 21~Myr and then later at around 28~Myr. The final outcome of the simulation is a less compact, dynamically stable system (for 70~Myr) with 5 super-Earths and 3 Earth-sized planets with orbital eccentricities reaching values of $\sim$ 10\% and inclinations up to a few degrees, so within the current eccentricities and inclinations of the three largest terrestrial planets of the Solar System.

\begin{figure}
  \resizebox{\hsize}{!}{\includegraphics{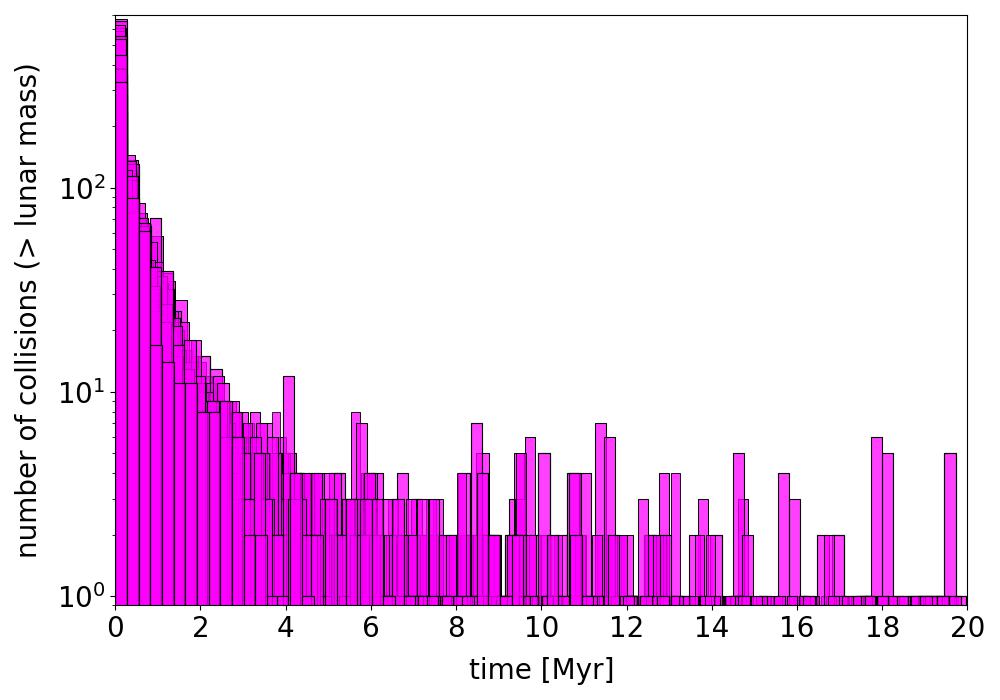}}
  \caption{Histogram showing a number of collisions during the 20~Myr of evolution in all our simulations. Only collisions where both bodies are above the lunar mass ($M_{Moon}=0.012~M_{\oplus}$) are presented.}
  \label{fig:histoAll}
\end{figure}

To determine which systems underwent a late instability phase with giant impacts and which did not, \cite{izidoro2017breaking,izidoro2021formation} use the distributions of the last collision epochs in their simulated systems. They subsequently divide the systems into "stable", with no collisions occurring after the gas dispersal, and "unstable" with some collisions happening afterward. We cannot use this criterion for our simulations as basically all our systems experienced collisions after the gas disk phase, and often long after. Most such collision events happened at much lower collision frequency than a dynamical instability that results in disruption of resonant chains, and thereby significantly changing the architecture of a system. Also, we cannot state when exactly the gas disk phase ends, whereas in their simulations it ends abruptly. In addition, their simulations start with only embryos, no planetesimals that exert dynamical friction on protoplanets and can damp their eccentricities and inclinations \citep{wetherill1993formation,ida1993scattering,o2006terrestrial}. Therefore, it is difficult to compare the evolution of our versus their simulated systems. According the collision criterion of \cite{izidoro2017breaking,izidoro2021formation}, all our systems would be classified as unstable, even though some of them remained in resonance and clearly appear to be long-term stable. The histogram in Fig. \ref{fig:histoAll} shows a number of embryo-embryo collisions in all our simulations during their 20~Myr of evolution. The number of collisions decreases mostly smoothly from the start of the simulations until $\sim$~8~Myr. Then the number stays at a continuous low level of collisions and the collision rate is roughly constant, with occasional potential dynamical instabilities until the end of the 20~Myr period. We say "potential" because it is questionable whether the spikes in the number of collisions are actually statistically significant. Our systems that get disturbed by the instabilities then stabilize relatively quickly and readjust, and those with a lower number of planets sometimes become resonant again (some gas might still be present in the system). Additionally, we extended one-third of all our runs to 40~Myr and one-fifth to 100~Myr. At 40~Myr, we see that in around one-third of the extended simulations one or several instabilities occurred after 20~Myr of evolution (see again Fig. \ref{fig:dynEvolution92}); while between 40~Myr and 100~Myr, we see no more instabilities. \cite{izidoro2017breaking} show that at least 75\% (and probably 90–95\%) of their simulated systems must be unstable to match the period ratio distribution of observations. By investigating each simulation and its dynamical evolution individually, we recognized at least $\sim$~85\% of our systems that underwent at least one late dynamical instability, which resulted in the resonant chains becoming unstable. However, due to the reasons mentioned previously, the number is a bit uncertain.

\subsection{Effects of disk parameters} \label{effects}
\begin{figure}
  \resizebox{\hsize}{!}{\includegraphics{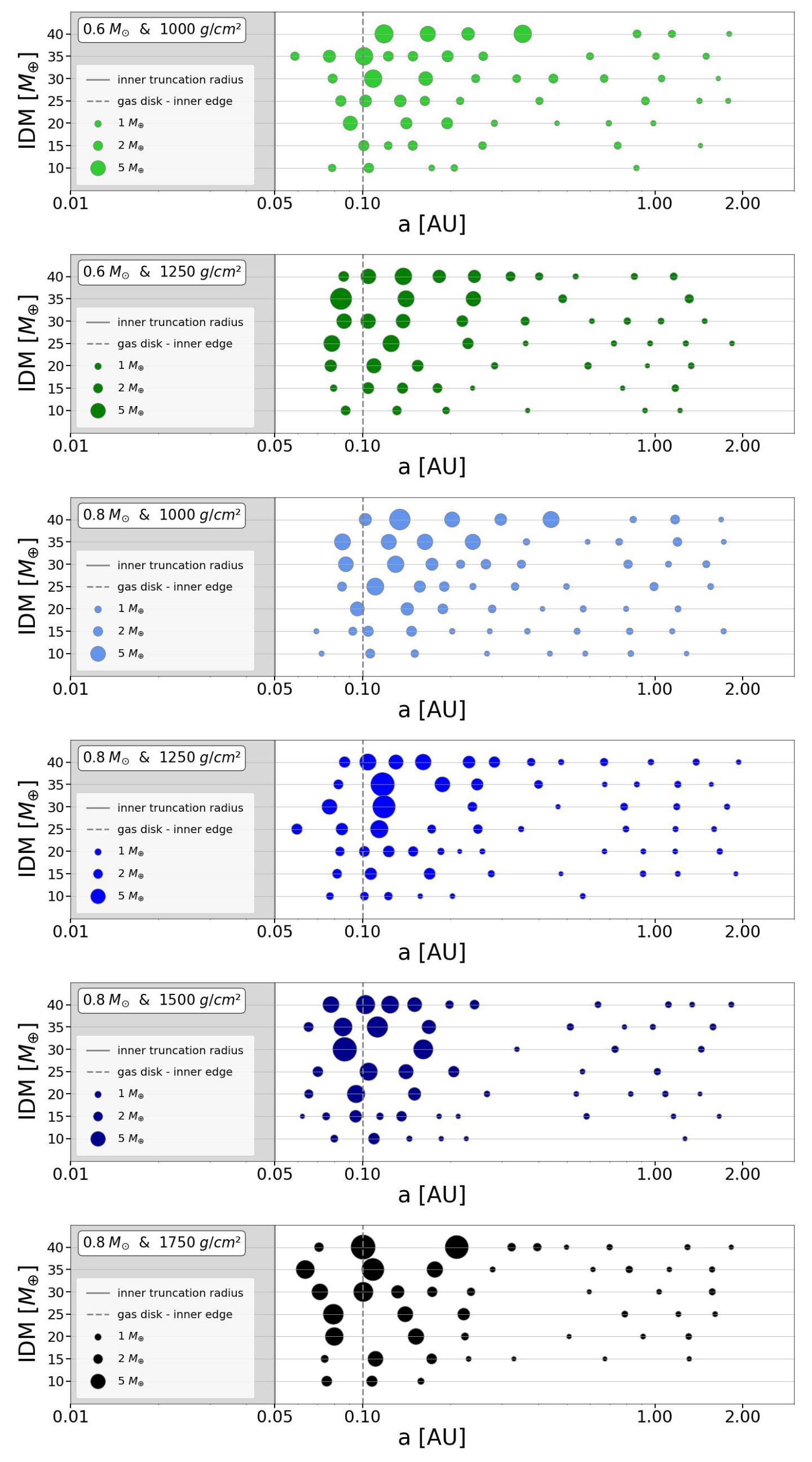}}
  \caption{Architecture of our simulated planetary systems around a $0.6~M_{\odot}$ and a $0.8~M_{\odot}$ K-dwarf star, for two or four different gas surface density values and different initial disk masses (IDM) after 20~Myr of N-body integration using GENGA. Outcomes for the initial disk mass $5~M_{\oplus}$ are not displayed. The point size indicates the masses in $M_{\oplus}$ of the formed planets. The gray zone is the region inside the inner truncation radius, and the dashed gray line shows the inner edge of the gas disk.}
  \label{sims}
\end{figure}

\begin{figure}
  \resizebox{\hsize}{!}{\includegraphics{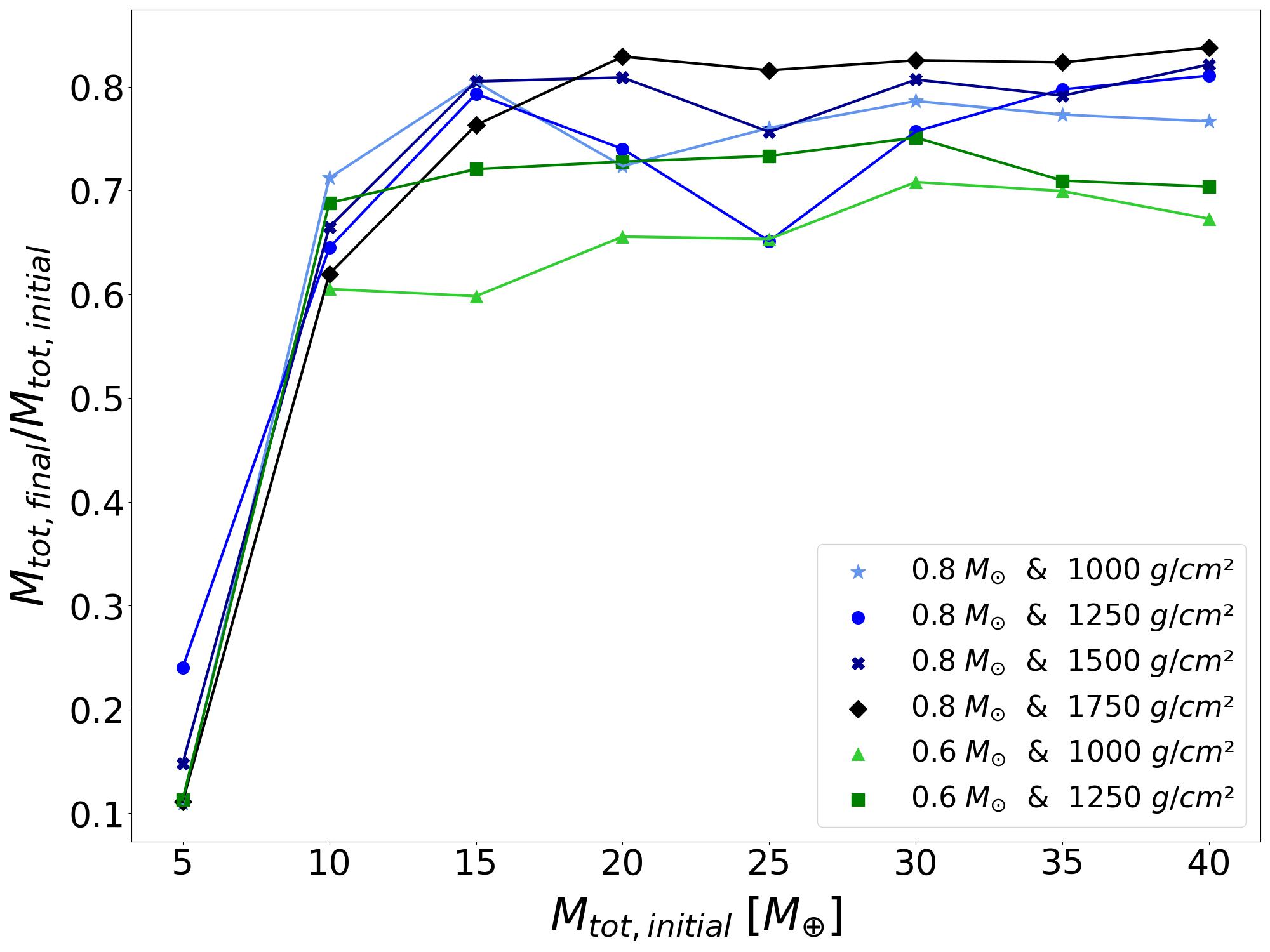}}
  \caption{$M_{\rm tot, final}/M_{\rm tot, initial}$ ratio versus the initial solid disk mass, $M_{\rm tot, initial}$. The total mass of a system, $M_{\rm tot, final}$, is calculated by summing up the masses of all individual planets in each system. Values for $0.6~M_{\odot}$ and $0.8~M_{\odot}$ star masses and various gas surface densities are presented.}
  \label{totalMass}
\end{figure}

In this subsection, we briefly present and compare the various generated systems, and discuss the effects of the disk parameters and initial conditions on the final systems before exploring their architecture more in the next subsection. The simulated planetary systems are displayed in Fig. \ref{sims}. The architectures of the systems around a $0.6~M_{\odot}$ and $0.8~M_{\odot}$ central star are plotted for different masses of the initial disk, and two (for $0.6~M_{\odot}$ star) and four (for $0.8~M_{\odot}$ star) gas surface density values. In this paper, a ``planet'' is a body with $M_P\geq0.5~M_{\oplus}$. We used this cutoff to facilitate comparison with observations: at present smaller planets are very rarely detected, and our retrieved observational sample contains only a few of them. In Fig. \ref{sims}, the symbol size is proportional to the mass of the formed planets. Many of our planets are similar in mass to the close-in super-Earths observed around these stars. Specifically, the simulations with higher initial disk masses reproduce the observed sample quite well (this will be discussed in detail in Sect. \ref{discussion}), whereas the lowest disk mass ($5~M_{\oplus}$) forms only one or two very low-mass planets (or none at all) barely reaching the ``planet cutoff''; hence, they are not displayed in Fig. \ref{sims}, but will be discussed a bit later. The increasing initial disk mass clearly results in generally more massive planets in a system, particularly in the case of planets very close to the host star. In Fig. \ref{totalMass}, the ratios of final masses of the systems $M_{\rm tot, final}$ to initial disk masses $M_{\rm tot, initial}$ are plotted against $M_{\rm tot, initial}$ to show the planet formation efficiency for each simulation. We can see that the simulations with the lowest initial disk mass are not very successful in forming planets. On the other hand, the simulations with higher initial disk masses are quite efficient in producing planets. The ratio $M_{\rm tot, final}$/$M_{\rm tot, initial}$ ranges from approximately 0.6 to 0.84 (when not considering the initial disk mass of $5~M_{\oplus}$). In other words, 60 to 84\% of the initial mass will end up in the planets. Figure \ref{totalMass} also shows that higher central mass and higher gas density usually produce higher total mass of the simulated systems. This general trend is mostly followed by all the parameter combinations except for $0.8~M_{\odot}$ \&\;1250\;g cm\textsuperscript{-2} (darker blue line with circles), which displays somewhat erratic behavior. Particularly, when initial disk mass = 25~$M_{\oplus}$ the total final mass is considerably lower than expected according to the trend observed in the figure. This can be explained by the fundamentally chaotic nature of the formation process. We discuss this more in Sect. \ref{reproducibility}, where we look closer at the evolutionary track of this simulated system.

\begin{figure}
  \resizebox{\hsize}{!}{\includegraphics{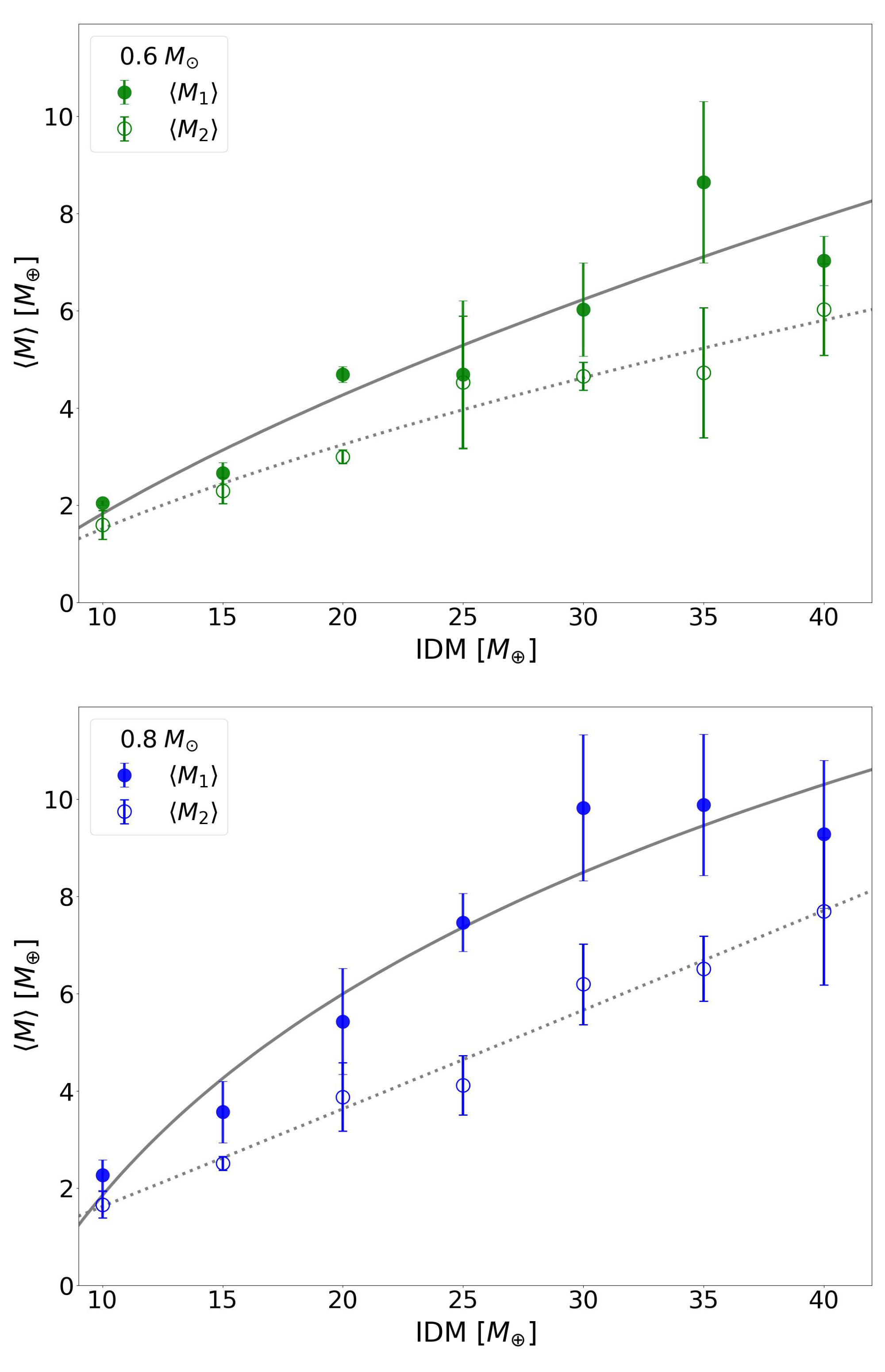}}
  \caption{Average masses of the largest $\langle M_1 \rangle$ (filled circles) and second-largest $\langle M_2 \rangle$ (open circles) planets against the corresponding initial disk masses (IDM) for the two central star masses $0.6~M_{\odot}$ (in green, the top plot) and $0.8~M_{\odot}$ (in blue, the bottom plot), and the empirical fits for $\langle M_1 \rangle$ (solid line) and $\langle M_2 \rangle$ (dotted line). Since the systems starting with $5~M_{\oplus}$ typically contain only one planet, they are not included in the plots.}
  \label{scaling}
\end{figure}

In addition, we examined the scaling between disk mass and planet mass in our systems by plotting the average masses of the largest $\langle M_1 \rangle$ and second-largest $\langle M_2 \rangle$ planets against the corresponding initial disk masses and their empirical fits (see Fig. \ref{scaling}). This was studied by \cite{kokubo2006formation} in a migration-free setting. They showed that the planet mass scales roughly linearly with the available mass in the disk (i.e., $M_{\rm pl}\propto M_{\rm disk}^{0.97-1.1}$ for a fixed stellar mass). \cite{raymond2007decreased} independently found the same result for a star with $0.4~M_{\odot}$. In our simulations, $\langle M_1 \rangle$ and $\langle M_2 \rangle$ also typically increase with the disk mass, but predominantly not linearly. Using the least-squares fit method we obtained $M_{\rm pl}\propto M_{\rm disk}^{0.6}$ for $0.6~M_{\odot}$ star and $M_{\rm pl}\propto M_{\rm disk}^{0.1-1.0}$ for $0.8~M_{\odot}$ star, so the growth mostly follows a power-law function rather than linear. However, the values are quite scattered and for the $0.8~M_{\odot}$ star, and the fit is significantly different for the largest versus second-largest planets. The comparison with non-migrating simulations shows that migration results in larger variations in planet masses (of the largest planets) and generally a slower increase in planet mass with the initial disk mass. The mass of the most massive planets in the systems will be discussed a bit further in this subsection.

From a theoretical perspective, most models proposed to explain the formation of exoplanets are based on processes that are quite inefficient. Some of the initial solid mass is often lost to the star due to rapid Type I migration. Migration is happening also in our simulations, but the gas-free cavity between the inner cutoff of the simulation range and the inner edge of the gas disk creates a planet trap, which prevents a significant mass loss to the star. The bodies keep piling up around the inner edge of the gas disk until they form planets similar to the close-in super-Earths with orbital distances mostly within $\sim$ 0.3~AU and masses of several $M_{\oplus}$ \citep[e.g.,][]{ogihara2015reassessment}. Figure \ref{sims} shows that many planets migrated farther inward than the inner edge of the gas disk. As mentioned in Sect. \ref{introduction} when discussing the breaking-the-chains scenario, planets often become locked into resonant chains while migrating. The innermost planet enters the cavity and stops being affected by the disk torques, that is to say, it slowly stops migrating. But if the planet is in resonance with one or several planets still inside the gas disk experiencing the inward Type I migration, the torques induced on those planets might push the inner planet farther inward through the resonance lock
\citep[e.g.,][]{terquem2007migration,brasser2018trapping}. This way one or several planets might end up located very close to the central star (as observed) without colliding with the star. At the same time, planets that formed later or farther away might be prevented from migrating too close, either by joining the resonant chain or if migration simply stops due to the dissipation of the gas disk, so not all planets in a system end up too close to the star (as displayed in Fig. \ref{sims}). 

\begin{figure}
  \resizebox{\hsize}{!}{\includegraphics{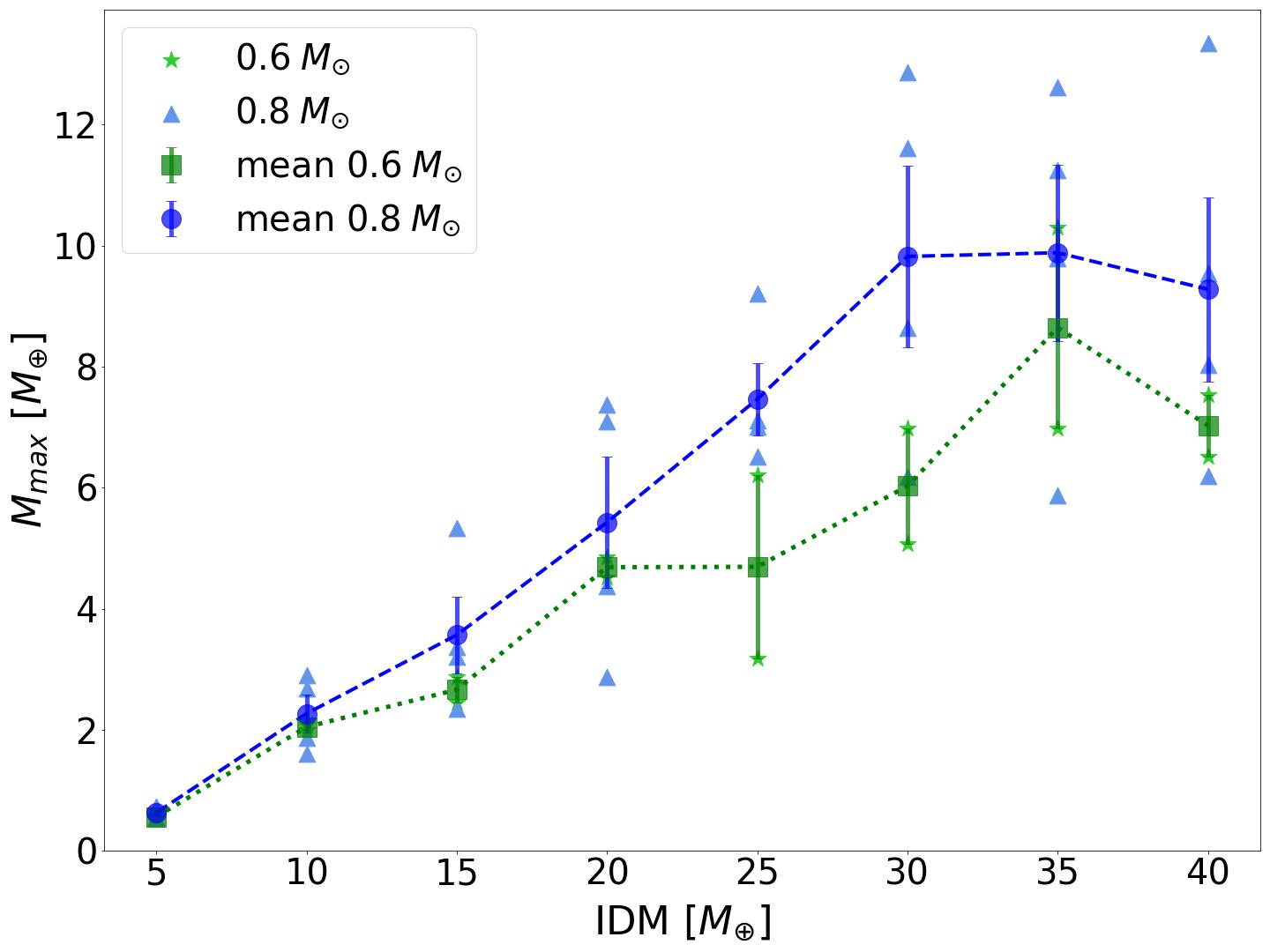}}
  \caption{Mass of the most massive planet, $M_{\rm max}$ [$M_{\oplus}$], in a system for each initial disk mass value (IDM) and both star masses, 0.6 and 0.8 $M_{\odot}$, shown in green and blue, respectively. Mean values and their standard errors are included (green squares with the dotted line and blue circles with the dashed line for 0.6 and 0.8 $M_{\odot}$, respectively).}
  \label{maxMass}
\end{figure}

Tables \ref{table:1000}, \ref{table:1250}, and \ref{table:1500} present some of the final characteristics of the simulated systems grouped by the value of the gas surface density and central mass. Average planet mass $\langle M_p \rangle$ in an individual system mostly grows systematically with increasing initial disk mass. The same applies to the mass of the most massive planet $M_{\rm max}$ in each system. In the case of the gas surface density values higher than 1000 g cm\textsuperscript{-2}, the increase is not that systematic. Both the average mass and the mass of the most massive planet reach a maximum when the disk mass is 30-35~$M_\oplus$, and then often drop. Figure \ref{maxMass} displays this tendency for $M_{\rm max}$. Interestingly, in most cases it is actually not the highest initial disk mass that produces the highest $M_{\rm max}$. It seems that there is an upper limit for the mass of the most massive planet in a system and that adding more disk mass will not result in a much more massive planet. Generally, both higher central mass and gas density result in more massive planets. At the same time, with more massive planets in a system the dispersion in mass is also higher as all systems contain small planets as well. In addition, it seems that more gas in the disk and higher central mass results in more irregularity for this typically increasing trend. The highest $\langle M_p \rangle$ is 4.97 $\pm$ 3.58~$M_{\oplus}$ and the highest $M_{\rm max}$ is 13.34~$M_{\oplus}$, so we managed to form some more massive planets than generally considered super-Earths. The average planet in the observed K-dwarf sample (when considering only the super-Earths) has $\langle M_p \rangle$ = $5.9 \pm 2.5~M_{\oplus}$ (planet mass is calculated as an average value from the whole sample, not individually per system). In order to be able to compare the calculated values for the simulated systems presented in Tables \ref{table:1000}, \ref{table:1250} and \ref{table:1500} with the average observed mass, we have to consider observational biases and take into account the fact that $\langle M_p \rangle$ values in the table also include small planets located farther from the star than the currently used methods are able to detect (as discussed in Sect. \ref{introduction}). This is done in Sect. \ref{discussion}.

\begin{table}
\caption{Simulations with 1000 g cm\textsuperscript{-2}.\ Listed are the average planet mass $\langle M_p \rangle$ with its standard deviation, the mass of the most massive planet, $M_{\rm max}$, and number of planets, $N$, in each system for the different initial disk masses (IDM) and two central star masses.}
\label{table:1000}      
\centering                                      
\begin{tabular}{c|c c c|c c c}          
\hline\hline                        
 & \multicolumn{3}{|c|}{Central mass} & \multicolumn{3}{c}{Central mass} \\\newline
 & \multicolumn{3}{|c|}{0.6 [$M_{\odot}$]} & \multicolumn{3}{c}{0.8 [$M_{\odot}$]} \\
\hline  
IDM & $\langle M_p \rangle$ & $M_{\rm max}$ & $N$ & $\langle M_p \rangle$ & $M_{\rm max}$ & $N$ \\\newline
[$M_{\oplus}$] & [$M_{\oplus}$] & [$M_{\oplus}$] & & [$M_{\oplus}$] & [$M_{\oplus}$] & \\
\hline 
5 & & & & 0.55 & 0.55 & 1\\    
10 & 1.21 $\pm$ 0.48 & 2.09 & 5 & 0.89 $\pm$ 0.44 & 1.88 & 8\\ 
15 & 1.50 $\pm$ 0.62 & 2.45 & 6 & 1.10 $\pm$ 0.62 & 2.35 & 11\\   
20 & 1.87 $\pm$ 1.42 & 4.53 & 7 & 1.81 $\pm$ 1.35 & 4.38 & 8\\   
25 & 1.81 $\pm$ 0.91 & 3.19 & 9 & 2.11 $\pm$ 1.68 & 6.52 & 9\\
30 & 2.36 $\pm$ 1.92 & 6.99 & 9 & 2.62 $\pm$ 1.73 & 6.18 & 9\\
35 & 2.45 $\pm$ 1.68 & 6.99 & 10 & 3.01 $\pm$ 2.23 & 5.87 & 9\\
40 & 3.84 $\pm$ 2.62 & 7.53 & 7 & 3.83 $\pm$ 2.78 & 9.55 & 8\\
\hline                                             
\end{tabular}
\end{table}

\begin{table}
\caption{Simulations with 1250 g cm\textsuperscript{-2}.}              
\label{table:1250}      
\centering                                      
\begin{tabular}{c|c c c|c c c}           
\hline\hline                        
 & \multicolumn{3}{|c|}{Central mass} & \multicolumn{3}{c}{Central mass} \\\newline
 & \multicolumn{3}{|c|}{0.6 [$M_{\odot}$]} & \multicolumn{3}{c}{0.8 [$M_{\odot}$]} \\
\hline  
IDM & $\langle M_p \rangle$ & $M_{\rm max}$ & $N$ & $\langle M_p \rangle$ & $M_{\rm max}$ & $N$ \\\newline
[$M_{\oplus}$] & [$M_{\oplus}$] & [$M_{\oplus}$] & & [$M_{\oplus}$] & [$M_{\oplus}$] & \\
\hline 
5 & 0.57 & 0.57 & 1 & 0.60 $\pm$ 0.08 & 0.68 & 2\\    
10 & 1.15 $\pm$ 0.60 & 2.00 & 6 & 1.07 $\pm$ 0.40 & 1.61 & 6\\ 
15 & 1.54 $\pm$ 0.88 & 2.88 & 7 & 1.49 $\pm$ 1.02 & 3.21 & 8\\   
20 & 2.08 $\pm$ 1.48 & 4.85 & 7 & 1.35 $\pm$ 0.81 & 2.87 & 11\\   
25 & 2.29 $\pm$ 2.26 & 6.20 & 8 & 2.04 $\pm$ 1.92 & 7.73 & 8\\
30 & 2.50 $\pm$ 1.77 & 5.07 & 9 & 3.24 $\pm$ 3.72 & 11.61 & 7\\
35 & 4.97 $\pm$ 3.20 & 10.30 & 5 & 3.10 $\pm$ 3.65 & 12.61 & 9\\
40 & 2.81 $\pm$ 1.82 & 6.52 & 10 & 2.70 $\pm$ 1.91 & 6.19 & 12\\
\hline                                             
\end{tabular}
\end{table}

\begin{table}
\caption{Simulations with 1500 g cm\textsuperscript{-2} (left) and 1750 g cm\textsuperscript{-2} (right).}              
\label{table:1500}      
\centering                                      
\begin{tabular}{c|c c c|c c c}           
\hline\hline                        
 & \multicolumn{3}{|c|}{Central mass} & \multicolumn{3}{c}{Central mass} \\\newline
 & \multicolumn{3}{|c|}{0.8 [$M_{\odot}$]} & \multicolumn{3}{c}{0.8 [$M_{\odot}$]} \\
\hline  
IDM & $\langle M_p \rangle$ & $M_{\rm max}$ & $N$ & $\langle M_p \rangle$ & $M_{\rm max}$ & $N$ \\\newline
[$M_{\oplus}$] & [$M_{\oplus}$] & [$M_{\oplus}$] & & [$M_{\oplus}$] & [$M_{\oplus}$] & \\
\hline 
5 & 0.74 & 0.74 & 1 & 0.56 & 0.56 & 1 \\    
10 & 1.11 $\pm$ 0.85 & 2.90 & 6 & 2.07 $\pm$ 0.74 & 2.68 & 3 \\ 
15 & 1.21 $\pm$ 0.90 & 3.37 & 10 & 1.64 $\pm$ 1.65 & 5.34 & 7 \\   
20 & 2.02 $\pm$ 2.16 & 7.10 & 8 & 2.76 $\pm$ 2.75 & 7.37 & 6 \\   
25 & 3.15 $\pm$ 2.19 & 7.02 & 6 & 3.40 $\pm$ 3.10 & 9.21 & 6 \\
30 & 4.84 $\pm$ 4.99 & 12.86 & 5 & 3.10 $\pm$ 2.71 & 8.64 & 8 \\
35 & 3.46 $\pm$ 3.29 & 9.79 & 8 & 3.60 $\pm$ 3.86 & 11.25 & 8 \\
40 & 3.29 $\pm$ 2.71 & 8.03 & 10 & 3.72 $\pm$ 4.86 & 13.34 & 9 \\
\hline
\end{tabular}
\end{table}

\subsection{Architecture of the simulated systems} \label{architecture}
Knowledge of architecture of planetary systems is important for constraining the theories of planet formation and evolution. Systems of planets with nearly coplanar and circular orbits, such as the Solar System, are consistent with the standard model of planet formation via planetesimal accretion in a gaseous protoplanetary disk. While planetary systems with differently inclined, noncircular orbits suggest past events that increased eccentricities and inclinations in the system \citep[e.g.,][]{juric2008dynamical,ford2008origins,chatterjee2008dynamical}, such as resonant encounters between planets, or planet–planet scattering. Figure \ref{sims} presents the architecture of the 42 simulated systems after 20~Myr of planetary evolution. We ran 48 different combinations of parameters, but we omit the runs with the initial disk mass of $5~M_{\oplus}$. In the end, our simulated population contains 335 planets in 42 systems. We examine their architecture but do not describe each individual system, instead focusing on the general trends. At the end of this subsection, we look specifically at the final eccentricities and inclinations of the simulated planets. All simulations started with planetesimals and planetary embryos distributed between 0.2 and 2~AU. Planets in the final systems are located between $0.06-0.09$~AU to 2~AU (just a reminder, during the planet formation if bodies cross the inner radius of 0.05~AU or outer radius of 3~AU, they are taken out of the simulation). Basically, in almost all systems the innermost planet, in rare cases two planets, is found inside the cavity.

\begin{figure}
  \resizebox{\hsize}{!}{\includegraphics{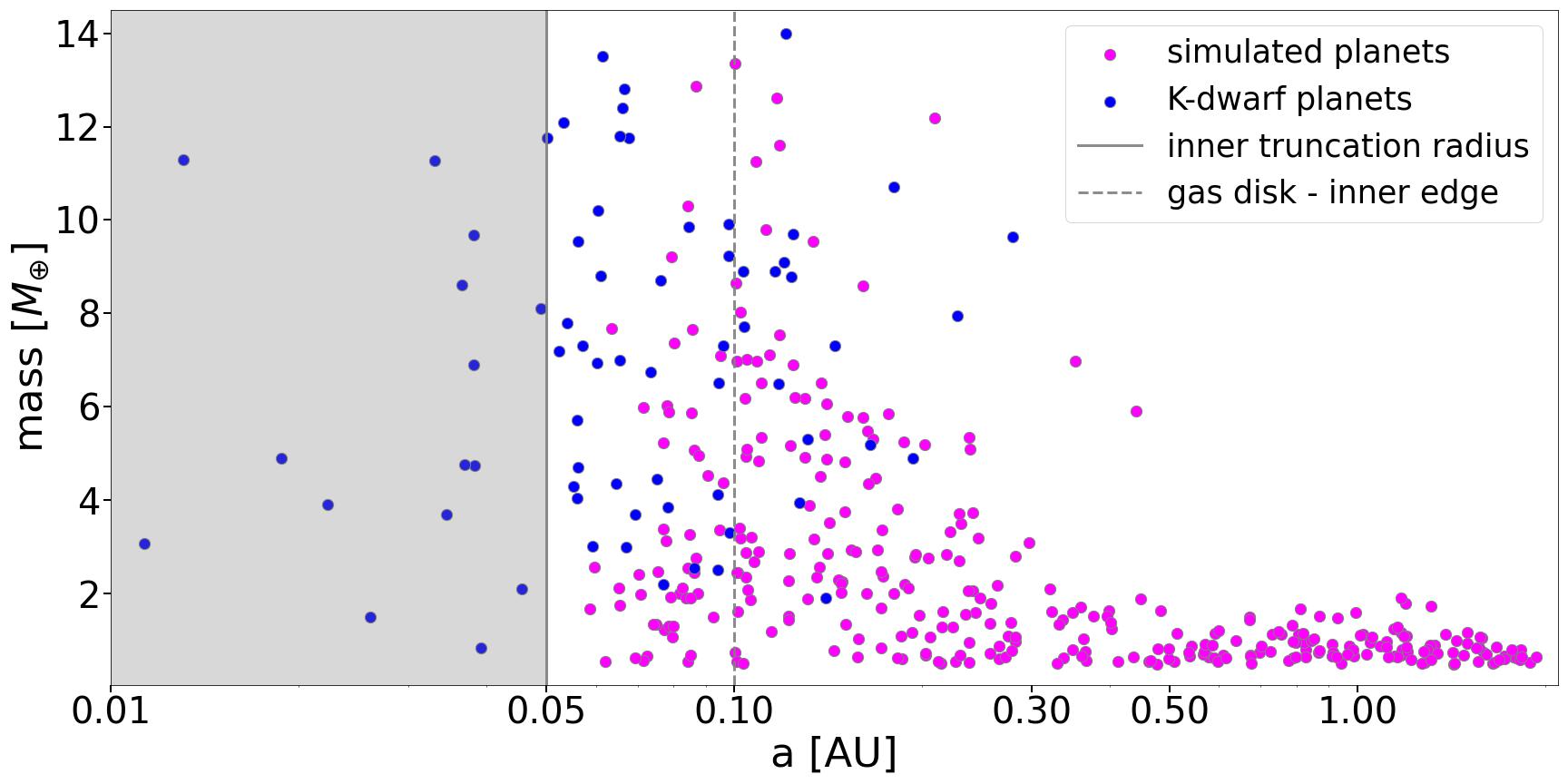}}
  \caption{Mass-distance relationship for the planetary systems. Magenta and blue circles respectively represent our simulated sample and the observed sample around K dwarfs. The gray zone is the region inside the inner truncation radius, and the dashed gray line shows the inner edge of the gas disk.}
  \label{mass_distance}
\end{figure}

The planets can be divided into two loose categories: more massive planets of several $M_{\oplus}$ located close to the star almost exclusively within 0.3~AU, and lower-mass planets of at most $2~M_{\oplus}$, but often much lower mass, which can mostly be found farther from the star (see Fig. \ref{mass_distance}). We find many lower-mass planets close to the star as well, but no really massive planets with masses above $2~M_{\oplus}$ beyond 0.5~AU. \cite{stevens2013posteriori} proposed $2~M_{\oplus}$ as the lower limit for super-Earths. This is not a universally accepted definition but if we adopt it, then we basically did not manage to form any super-Earths beyond 0.5~AU. In many cases, there is actually something like a spatial division between the two categories of planets with a region containing no planets (see again Fig. \ref{sims}). The ``zone of more massive planets'' typically extends farther out for larger initial disk masses. Generally, in this zone a system contains either a few very massive planets that are far from each other or several planets with lower masses packed more tightly together. This seems to be independent of the gas surface density, central mass or initial disk mass. Relative spacing between the planets is then approximately the same, although the lowest values of disk masses show generally a bit larger distances between their less massive planets (the dynamical spacing between planets is examined in Sect. \ref{stability}). 

\begin{figure}
  \resizebox{\hsize}{!}{\includegraphics{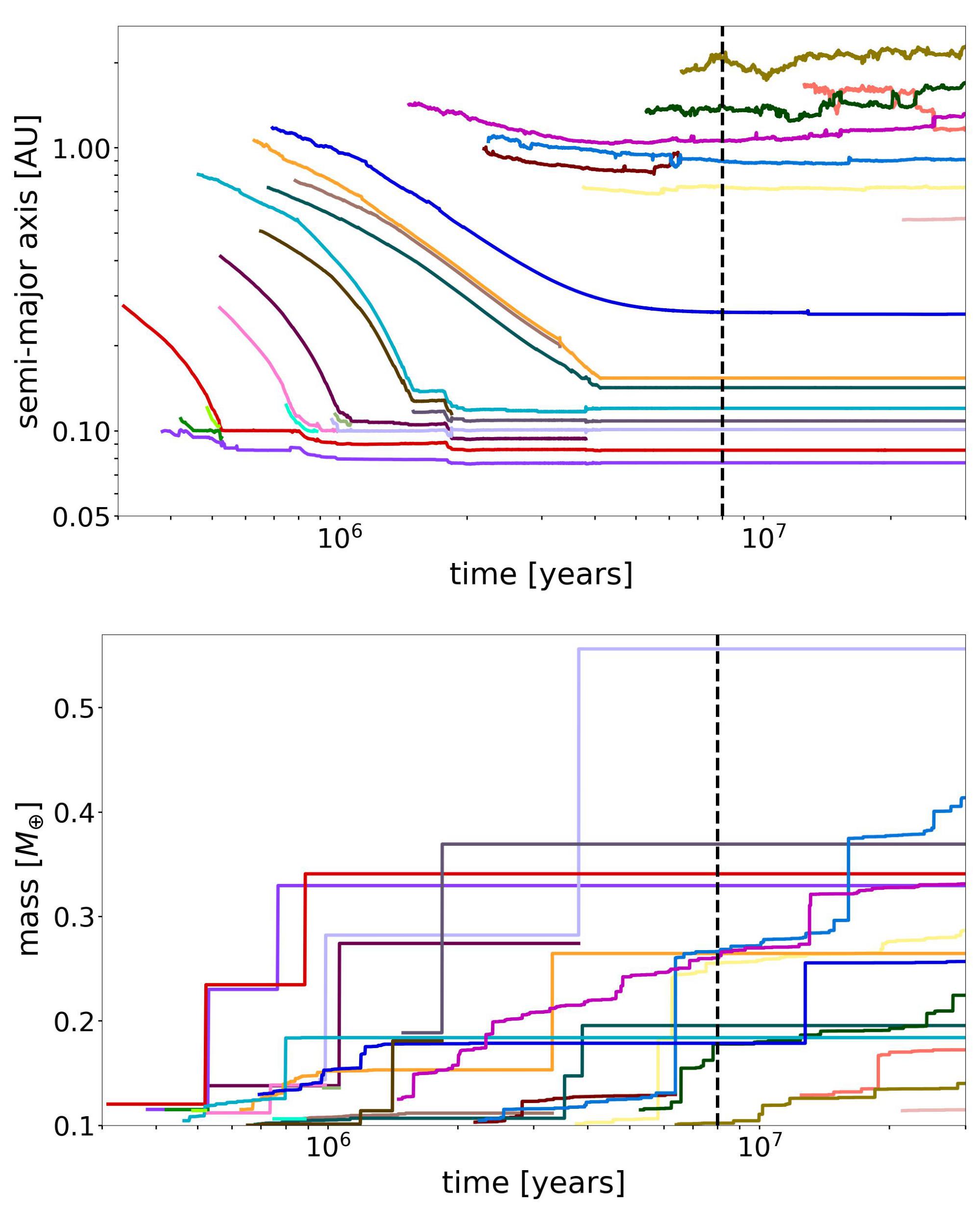}}
  \caption{Dynamical evolution of one of the single planet systems. The plots show the temporal evolution (30~Myr) of planetary bodies, specifically their semimajor axes and masses. Only bodies with masses > $0.1~M_{\oplus}$ are displayed. Each line color represents an individual object; the same object is indicated by the same color in both plots. The dashed black line shows the estimated time of the gas dissipation at $\sim$ 8~Myr. Only one object (in light purple) has a mass above $0.5~M_{\oplus}$ and is classified as a planet in this study.}
  \label{dynEvolution_116}
\end{figure}

In the simulated population, we have four systems with a single planet. This is an interesting result as single planet systems are quite common around K-dwarf stars, and not only around them. The large number of Kepler systems with single transiting planets versus multiple transiting planets is known the as Kepler dichotomy \citep{johansen2012can}. Of course, these systems might just be incomplete, even though some studies claim that the dichotomy is real \citep[e.g.,][]{fang2012architecture,johansen2012can,moriarty2016kepler}. On the other hand, other studies demonstrate that super-Earth systems are naturally multiple and the Kepler dichotomy is just an observational effect, a consequence of the mutual inclinations becoming excited due to the dynamical instabilities the systems experience as they evolve \citep[e.g.,][]{izidoro2017breaking,izidoro2021formation}. These studies predict an insignificant number of real single-planet systems in the Kepler sample, which agrees with the outcomes of our simulations. However, our single-planet systems are formed by the lowest initial disk masses ($5~M_{\oplus}$) and have very low masses ($0.7~M_{\oplus}$ at most), which suggests that their initial disks may not have contained enough material to form multiple planets. The dynamical evolution of one these simulations is presented in Fig. \ref{dynEvolution_116}, and it is representative of all the simulations that end with one planet. As after 20~Myr of evolution there were still many planetesimals available in the simulation, we extended the simulation time to 30~Myr, but not longer due to the extremely long computation time necessary for running the simulations starting with lowest disk masses (i.e., a larger fraction of smaller bodies). In this system, the evolution follows the general trend with embryos migrating inward, but there is not much growing happening. Most of the closest (to the central star) and heaviest embryos do not experience any collisions after 4~Myr. Embryos orbiting farther from the star keep growing until the end of the simulation time, though not at a significant rate. Basically, the embryos grew too slow and then it became too late, because the planet growth phase mostly stopped. Nevertheless, the material is still available in the disk. The simulation started with a disk mass of $5~M_{\oplus}$, and during the 30~Myr of planet formation, only $\sim 0.4~M_{\oplus}$ is ejected from the system. The rest of the initial material remains in the system as leftover embryos and planetesimals, mostly contained in bodies of the mass of Mars and above. At the end of the simulation, several of the bodies are piled up close to the inner edge, but only one of them actually classifies as a planet in our study, with a mass of $\sim 0.6~M_{\oplus}$. A longer simulation time would probably not improve the chances of additional planet growing much; however, a higher surface density or a longer gas disk decay time could increase the number of planets in the system as it would increase the planetesimal accretion or make the planet growth phase longer. Of course, if we defined a planet differently, then we would get a different number of planets in the system (and in all the other "single-planet" systems). It seems that not enough initial building material to form larger bodies causes that all the formed planetary bodies tend to be small and similarly sized, not even reaching the mass of Earth. Our results suggest that the large number of detected single-planet systems may simply be an observational effect as these relatively low-mass planets should have even lower-mass companions.

We also explore the number of planets in the individual systems. The majority of our systems have a relatively high number of planets (up to 12 in one case) because they have a high number of small planets farther from the star. Extending the simulation time would slightly reduce their number in some cases as collisions would still occasionally occur, as discussed in Sect. \ref{evolution}. Since the main purpose of this study is to reproduce the observed planet population around K dwarfs, we mostly focus on the region close to the star. The typical number of planets in a system (when not considering systems with the initial disk mass of $5~M_{\oplus}$) is between 7 and 9. For a $0.6~M_{\odot}$ star the average number of planets is 7.57 $\pm$ 1.81 and 7.43 $\pm$ 1.72 for 1000 and 1250 g cm\textsuperscript{-2}, respectively. For a $0.8~M_{\odot}$ star the average number of planets is 8.86 $\pm$ 1.07, 8.71 $\pm$ 2.14, 7.57 $\pm$ 1.99 and 6.71 $\pm$ 1.98 for 1000, 1250, 1500 and 1750 g cm\textsuperscript{-2}, respectively. It appears that the higher gas surface density the lower the final number of planets, which is the same trend as found in a non-migration case of similar N-body simulations \citep{kokubo2006formation,raymond2007decreased}, and in the case of the higher star mass for the same gas surface density value more planets form (or survive). Generally, systems with massive planets harbor fewer planets than systems with less massive planets, and these fewer massive planets are usually spaced farther apart.

 \begin{figure*}
\centering
   \includegraphics[width=17cm]{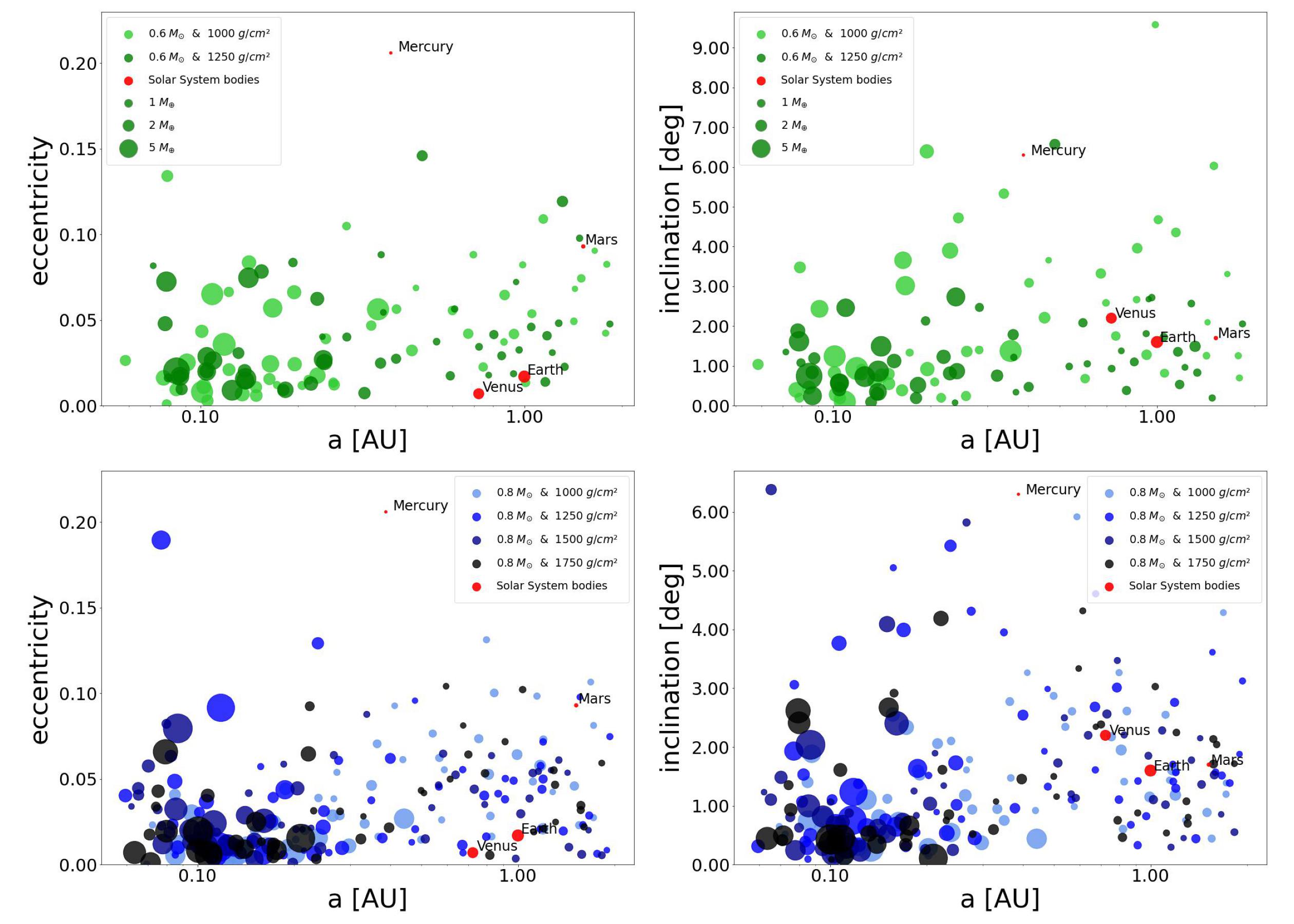}
     \caption{Eccentricities (on the left) and inclinations (on the right) of the simulated planets around a $0.6~M_{\odot}$ star (in green; the top row) and a $0.8~M_{\odot}$ star (in blue; the bottom row) versus their semimajor axes. Different shades of green and blue denote the different initial values of the gas surface densities at 1~AU, i.e., 1000, 1250, 1500, and 1750 g cm\textsuperscript{-2}; the higher initial density, the darker color of the circles. The point size indicates the masses in $M_{\oplus}$ of the formed planets. Eccentricities and inclinations of terrestrial Solar System planets are displayed for comparison (in red); their sizes are enlarged by a factor of 2 for better visualization.}
     \label{inc_ecc}
\end{figure*}

Figure \ref{inc_ecc} displays the final eccentricities and inclinations of planets around $0.6~M_{\odot}$ star (in green) and $0.8~M_{\odot}$ star (in blue) plotted against their semimajor axes. Current eccentricities and inclinations (with respect to the invariable plane) of the Solar System terrestrial planets are displayed for comparison. Essentially, all the planets in our sample are within the range of the Solar System planets for both orbital elements (except for a few exceptions). As Mercury's orbit has the highest eccentricity and inclination of all the eight planets, we can use it as the upper limit. The $0.8~M_{\odot}$ star plots (the bottom row) display several very massive planets with quite high inclinations and, particularly, eccentricities (compared to the Earth and Venus), which suggest more violent evolution of their planetary systems with possibly recent events of dynamical instabilities. We do not want to investigate these parameters in detail, only show that they have reasonable values based on the studies of similar known exoplanets and the Solar System values. Median value for eccentricities in the simulated population is 0.03 $\pm$ 0.03 with a maximum of 0.19, and for the observed sample around K dwarfs it is $0.08\substack{+0.08\\ -0.01}$\ (1$\sigma$ ranges), but the observed sample contains only a small number of planets with calculated eccentricity. Nevertheless, studies show that most Kepler planets tend to have relatively low eccentricities \citep[e.g.,][]{fabrycky2014architecture,hadden2014densities,mills2019california,van2019orbital}. In addition, for observed planets the inclination is defined as the angle of the plane of the orbit relative to the plane perpendicular to the line-of-sight from Earth to the object, and this makes the comparison with the simulated population a bit difficult. Regardless, as discussed in Sect. \ref{introduction}, the currently known population of planets around similar stars generally have low orbital inclinations, mostly less than 3 deg relative to a common reference plane \citep{fang2012architecture}. Median value for inclinations in the simulated population is 1.11 $\pm$ 1.38 deg with a maximum of 6.57 (if we do not consider the single outlier inclination in the top right plot with a value of 9.58 deg), which agrees with the study. When comparing values of these parameters specifically for different central masses and gas surface densities, we see that both eccentricities and  tend to be a bit lower for the runs with initially higher gas surface density and also for higher central mass, which at least in the case of gas surface density can be explained by the fact that more gas in the disk should result in more eccentricity and inclination damping (as discussed in Sect. \ref{introduction}). Inclinations show this trend as well, but not as systematically. Either way, orbits of the simulated planets are similar to the coplanar and circular orbits of the Solar System planets. This is consistent with planets forming in a protoplanetary disk, followed by evolution mostly without significant or lasting perturbations from other bodies. Figure \ref{inc_ecc} also shows how much closer to the star the majority of the simulated planets orbit, compared to, for example, Mercury; only lower-mass planets are located farther out.
\subsection{Reproducibility of the outcomes} \label{reproducibility}
Now we briefly explore the reproducibility of our simulated outcomes. Even though we might expect that simulations with similar initial conditions produce planetary systems with similar characteristics, we have to take into account the chaotic nature of this stage of planetary evolution. We examine an evolutionary track of the simulated system with the starting parameters of $25~M_{\oplus}$,  $0.8~M_{\odot}$, and\;1250~g cm\textsuperscript{-2}, whose total final mass is much lower than expected according to the general trend observed in Fig. \ref{totalMass}. This overall increasing trend of $M_{\rm tot, final}$ with the initial disk mass is followed relatively well by all the parameter combinations except for $0.8~M_{\odot}$ \&\;1250~g cm\textsuperscript{-2} (darker blue line), which displays quite irregular behavior. Particularly, in the case of initial disk mass = 25~$M_{\oplus}$ the total final mass is considerably lower than expected; therefore, we focused on this simulation. We ran another three simulations with exactly the same parameters (also planetesimals and planetary embryos have the same sizes and are distributed in the same way) in order to reproduce the first outcome. 

\begin{figure}
  \resizebox{\hsize}{!}{\includegraphics{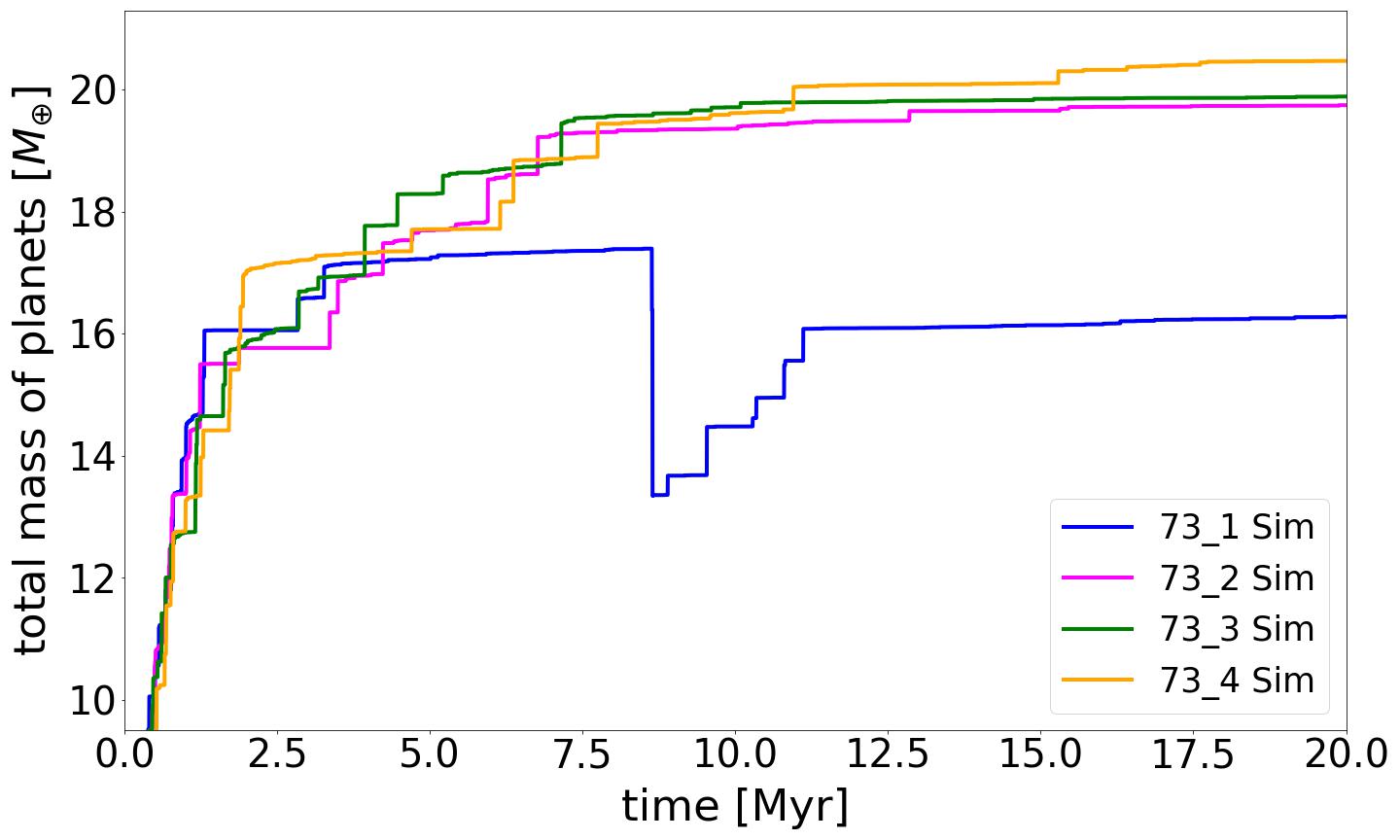}}
  \caption{Temporal evolution (20~Myr) of the total planet mass (sum of all planetary bodies with $M_P\geq0.5~M_{\oplus}$ in a system) in the simulated system called 73 Sim (initial disk mass = $25~M_{\oplus}$, central mass = $0.8~M_{\odot}$, gas surface density = 1250 g cm\textsuperscript{-2}). Shown are the evolutionary tracks of four simulations with exactly the same parameters. The total mass included in Fig. \ref{totalMass} is the final state of 73\_1 Sim, represented by the blue line.}
  \label{fig:evolution}
\end{figure}

\begin{figure}
  \resizebox{\hsize}{!}{\includegraphics{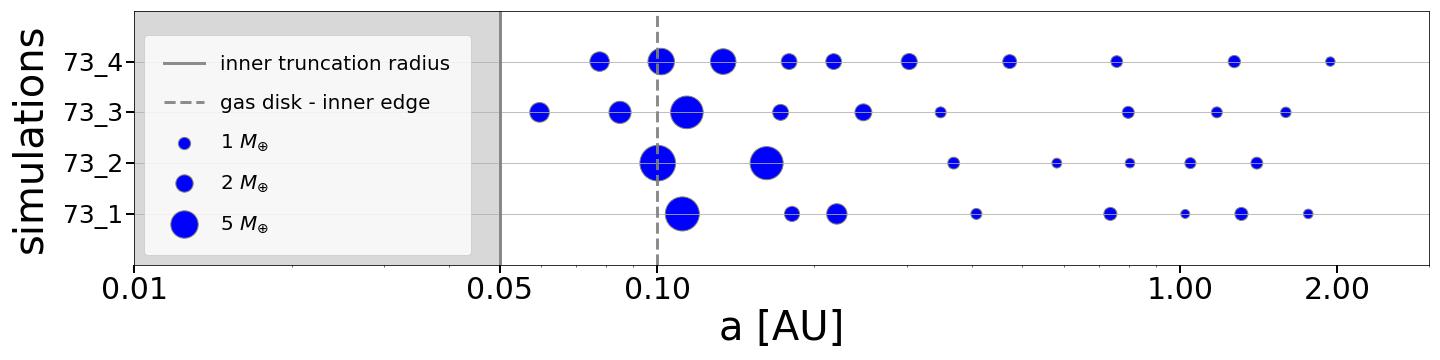}}
  \caption{Outcomes of four 73 Sim simulations with initial disk mass = $25~M_{\oplus}$, central mass = $0.8~M_{\odot}$, and gas surface density = 1250 g cm\textsuperscript{-2}. The original simulation is 73\_1 Sim. The size of the circles indicates the mass of the formed planets.}
  \label{4sims}
\end{figure}

\begin{figure*}
\centering
  \includegraphics[width=17cm]{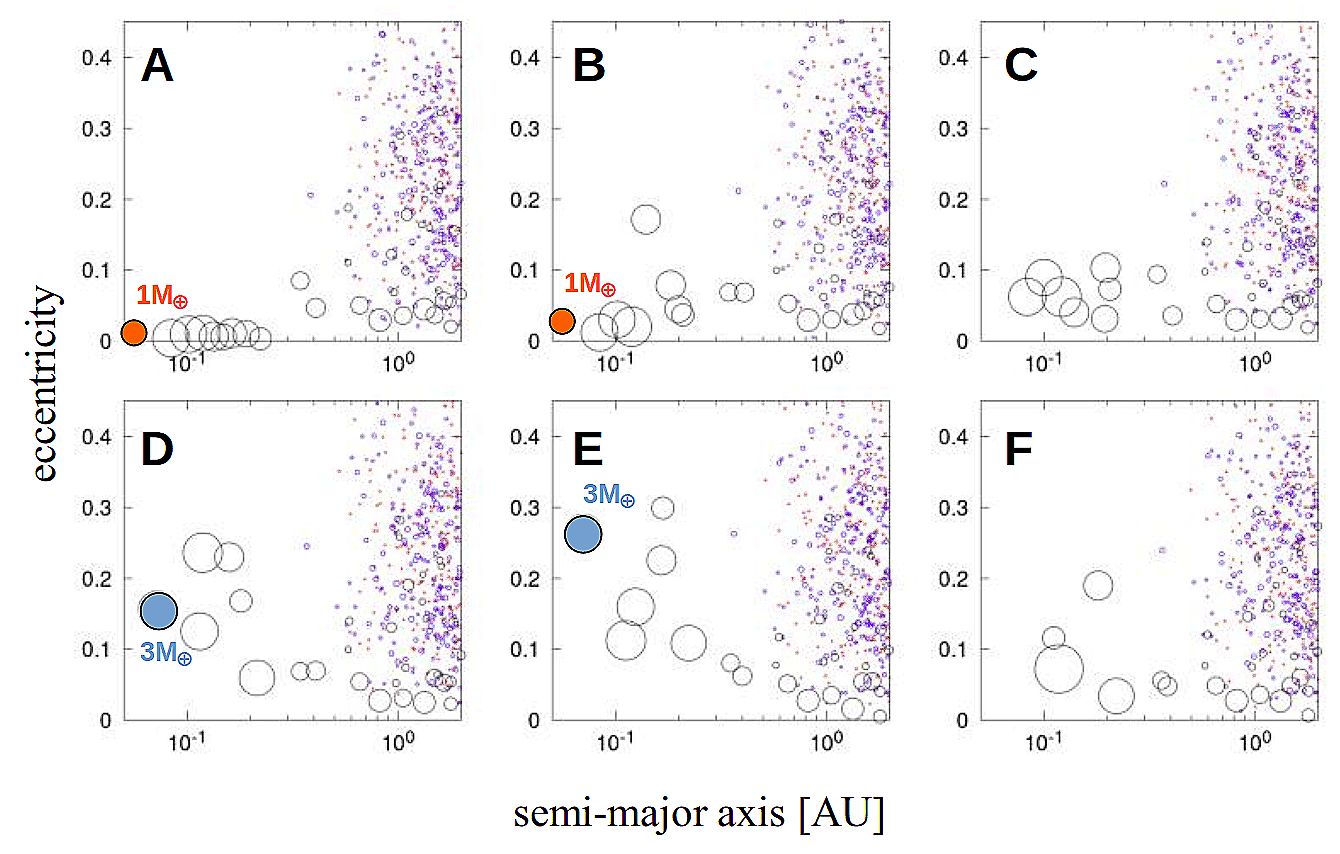}
    \caption{Snapshots of the 73\_1 Sim simulation showing eccentricity versus the semimajor axis of the planetary bodies. Panel (A) shows simulation at $\sim$ 8.6~Myr of evolution, and each subsequent panel displays the simulation a few thousand years later (B-F). Snapshot A presents a resonant chain of planets located very close to the star, which gets disrupted by another planet that gets too close and excites the group (B). This dynamical instability results in a $\sim1~M_{\oplus}$ planet at first (C), and later on a $\sim3~M_{\oplus}$ planet (F) falling onto the central star.}
  \label{fig:snaps}
\end{figure*}

Figure \ref{fig:evolution} displays the temporal total planet mass evolution of these four runs in total. We plot their individual evolutionary tracks against time; the original simulation is 73\_1 Sim, represented by the blue line. Immediately we see that this track behaves very differently compared to the additional three runs. The sudden decrease in the total mass at approximately 8.6~Myr of evolution is followed by a gradual increase, but the final mass of the system is still much lower than for the other simulations, where the final masses are very similar. Each of their masses would actually neatly followed the increasing trend in Fig.~\ref{totalMass}. After investigating GENGA output files, it is clear that this 73\_1~Sim feature in Fig. \ref{fig:evolution} is caused by a sudden orbital instability that happens when several relatively massive planets, piled-up very tightly close to the inner edge of the disk and locked in a resonance chain, are suddenly disrupted by another planet that gets too close and excites the group. The resonant chain breaks and two planets (with $\sim1~M_{\oplus}$ and $\sim3~M_{\oplus}$) fall into the star by venturing closer than 0.05~AU from the central mass (see Fig. \ref{fig:snaps}). Planets that form later and migrate toward the inner edge might get caught up in resonance with the planets already anchored there, and pump up their eccentricities. Since the innermost planets have no eccentricity damping, it is expected that their eccentricities grow large and they merge. This is what happens in this simulation: the planets merge, but the excitation causes two of the planets to be lost to the star. Some of the smaller bodies then continue colliding and growing for some time, which is represented by the step-wise increase in the total mass following the sudden drop, but too much mass is already lost. The other three simulations do not experience such an instability (when so much mass is removed from the system) and show regular increase in the total mass all the way. This is in line with the chaotic nature of the accretion process. As 73\_1 Sim behaves differently than the other three simulations, we treated it as a special case and used 73\_3 Sim in the final statistics and comparison of the simulations to the observations. Overall, we see that most significant collisions and merging happen within the first $\sim$ 11~Myr in all four runs, but some small events happen later as well. The various outcomes of the simulations are displayed in Fig. \ref{4sims}. It is obvious that simulations with the same initial conditions and parameters can produce planetary systems with differing characteristics. In two cases, the final systems include fewer massive planets in the region close to the inner edge of the disk; in two other cases, a higher number of smaller planets is located in this region. These two types of systems can still have the same total mass. The erratic behavior makes is difficult to decide which initial conditions and other parameters might be responsible for the final architectures of the formed planetary systems. On the other hand, random variations between the outcomes of successive simulations can help explain the diversity of exoplanet systems, as discussed in detail in \cite{raymond2020solar}.

\subsection{Time needed to grow an Earth-mass planet}
\begin{figure}
  \resizebox{\hsize}{!}{\includegraphics{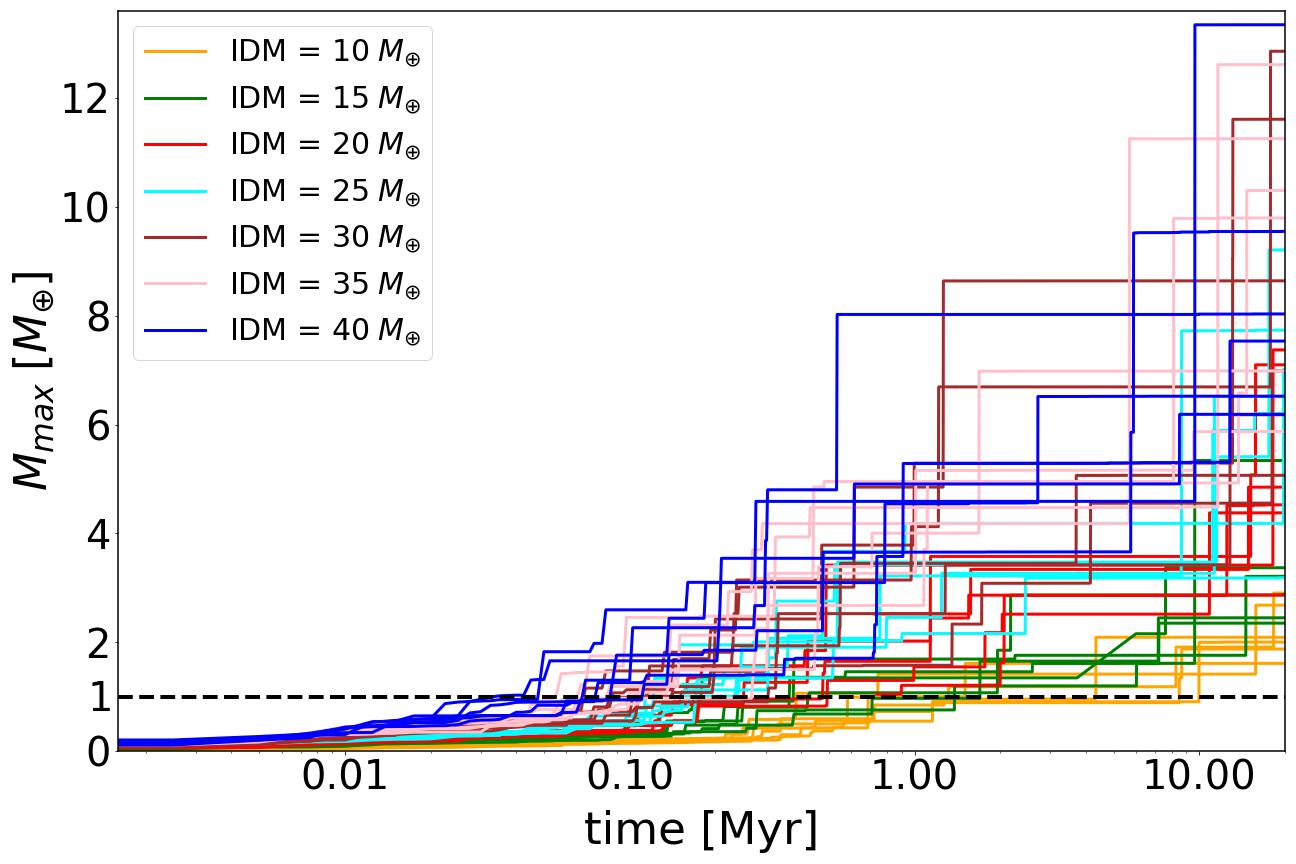}}
  \caption{Temporal evolution (20~Myr) of the mass of the most massive surviving body in the system. It shows the time it takes to grow an Earth-mass planet for each system (where the solid lines representing each simulation meet the dashed line). The simulations are grouped by their initial disk mass (IDM), indicated by the same color. Bodies in the systems starting with the initial disk mass $5~M_{\oplus}$ never reach 1$~M_{\oplus}$ and are therefore not included in the plot.}
  \label{growEarth}
\end{figure}

\begin{figure}
  \resizebox{\hsize}{!}{\includegraphics{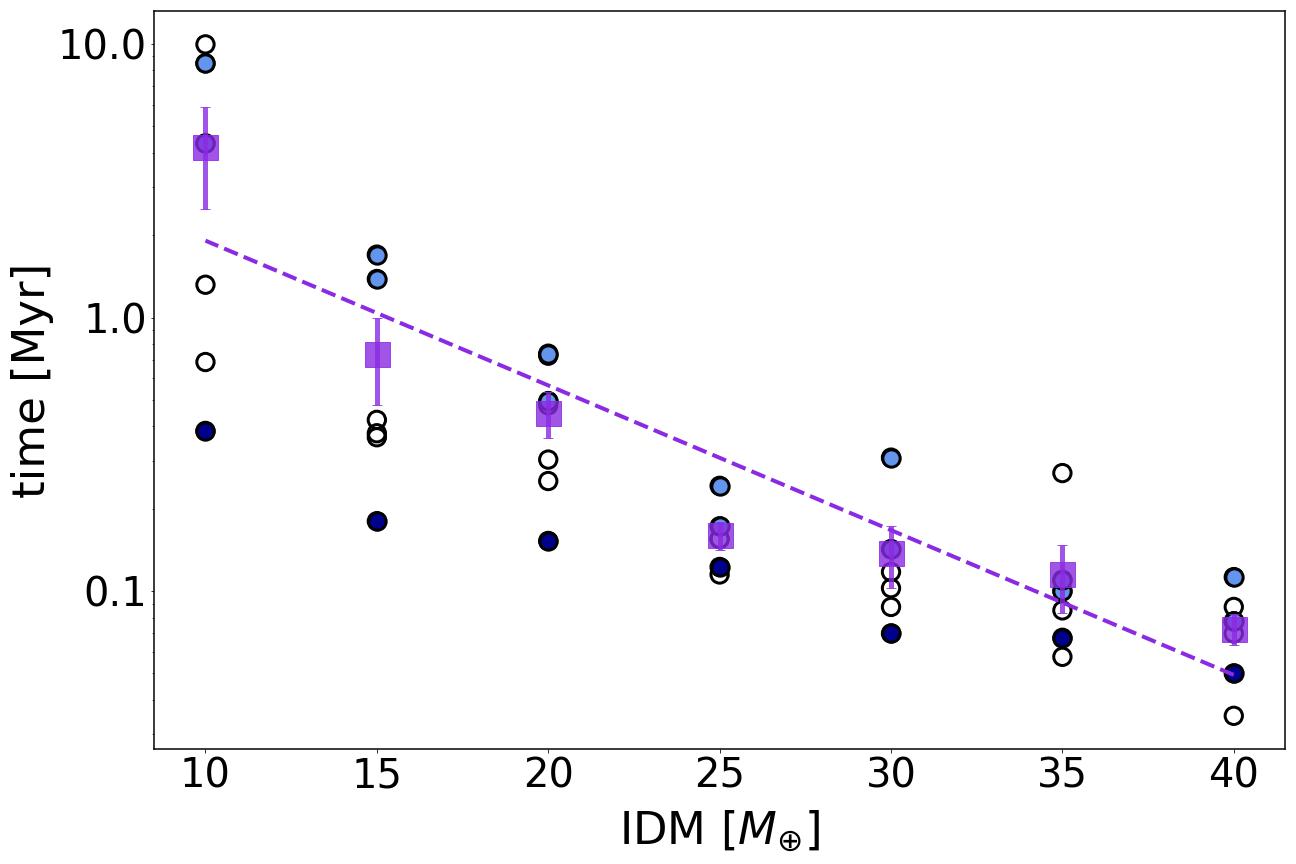}}
  \caption{Time it takes to grow an Earth-mass planet depending on the initial disk mass (IDM) for the individual planetary systems (circles) and for the average values of each mass bin (purple squares with error bars). The dashed purple line indicates the descending trend of needed time with increasing disk mass. The light blue and dark blue circles represent the times from the simulations starting with the gas surface densities = 1000 g cm\textsuperscript{-2} and 1750 g cm\textsuperscript{-2}, respectively.}
  \label{IdmVsTime}
\end{figure}

In this subsection, we examine the time needed to grow an Earth-mass planet in our simulations. Our findings are presented in Fig. \ref{growEarth}, which displays the mass of the most massive body in each system during the 20~Myr of simulation time. The runs are grouped by their initial disk mass and it is clear that higher disk mass generally results in shorter time necessary for growing a planet with the mass of Earth (or higher). Figure \ref{IdmVsTime} then shows the actual time it takes to grow an Earth-mass planet depending on the initial disk mass for the individual planetary systems (circles) and for the average values of each disk mass bin (purple squares with error bars). The light blue  and dark blue circles represent the times from the simulations starting with the gas surface density = 1000 g cm\textsuperscript{-2} and 1750 g cm\textsuperscript{-2}, respectively. This clearly shows that the time is typically shorter for the higher gas surface density. The average time ranges from less than 0.1~Myr for the initial disk mass of $40~M_{\oplus}$ to around 4~Myr for the disk mass of $10~M_{\oplus}$, and displays the generally decreasing trend with the increasing initial disk mass. This is expected as higher surface density of solids as well as larger initial bodies lead to the planetary growth happening faster as discussed already in Sect. \ref{initial}. Embryos grow faster because the number of collisions required to reach a certain mass is lower when the simulations start with larger planetesimals. We start all our simulations with the same number of particle (i.e., the same resolution), but since we increase the initial mass of the disk, we ran the simulations with larger and larger planetesimals. This causes the simulations with higher initial mass of the disk running much faster than simulations with lower disk masses. Even though, there is no commonly accepted formation time of an Earth-mass planet, the shortest times ($\sim$ 0.1~Myr and less) are probably not realistic. The simulations starting with the disk mass $10~M_{\oplus}$ show the largest dispersion of the time needed to grow an Earth-mass planet from $\sim$ 400\,000 years to even close to 10~Myr in two cases. However, the vast majority of the simulations form their Earth-mass planet(s) within the lifetime of the gas disk, and the planetary bodies then often keep colliding and growing. 

In Fig. \ref{growEarth} we also see that most collisions happen within a few million years ($\sim$ 2~Myr), which is in agreement with other similar studies \citep[e.g.,][]{ogihara2015reassessment,zawadzki2021rapid}. In some runs, occasional collisions occur even after 10~Myr. Generally in our simulations, $\sim$ 80 to 95\% of collisions happen during the first 2~Myr, and only $\sim$ 3 to less than 1\% occur after the first 10~Myr (for the runs starting with $10~M_{\oplus}$ and $40~M_{\oplus}$, respectively). So, particularly for the higher initial disk masses, almost all collisions happen during the first 10~Myr of the evolution. According to the core accretion scenario \citep{perri1974hydrodynamic,mizuno1978instability} a protoplanet starts efficiently capturing a massive gas envelope from the protoplanetary disk to become a gas giant when its mass is $\sim 10~M_{\oplus}$ \citep[e.g.,][]{mizuno1978instability,stevenson1982formation,bodenheimer1986calculations,hubickyj2005accretion}. In our simulated sample, there are only about eight planets that actually reach the mass necessary to get into this runaway gas accretion phase, and this happens in all cases after more than $\sim$ 6~Myr of evolution when most of the gas is already gone. This could explain the fact that giant planets are quite rare around this kind of star, as discussed in Sect. \ref{introduction}. Smaller planets of only several $M_{\oplus}$ are not assumed to accrete substantial amounts of gas.

\subsection{Period ratio distribution and logarithmic spacing}
Now, we discuss the simulated systems and further compare them to the observed systems. The primary goal of this study is to reproduce the known exoplanets around K-dwarf stars with their characteristics and by exploring how the initial conditions of the late stage of planet formation affect the properties of the simulated planetary systems. To evaluate the results we used the sample of the observed K-dwarf systems described in Sect. \ref{introduction} for comparison. The sample is actually quite small; therefore, in some cases we extend the sample by known systems with at least two planets with known masses discovered mostly by the Kepler telescope using transit method around M-, K-, and G-dwarf stars. At the present time, transiting exoplanets provide the most informative and complete collection of data on exoplanet characteristics, but only within $\sim$ 0.2~AU of their host stars. This sample (hereafter the MKG sample) contains 384 planets from 143 planetary systems and it was also retrieved from NASA Exoplanet Archive\footnote{\url{https://exoplanetarchive.ipac.caltech.edu/}}. In this section, we compare the results of our simulations directly to the observed samples. However, since all observational data are affected by observational biases and most of the known exoplanets were discovered by Kepler mission, we apply the detection bias of this survey to our simulated data and then compare them to the Kepler observations again in the following section.

\begin{figure}
  \resizebox{\hsize}{!}{\includegraphics{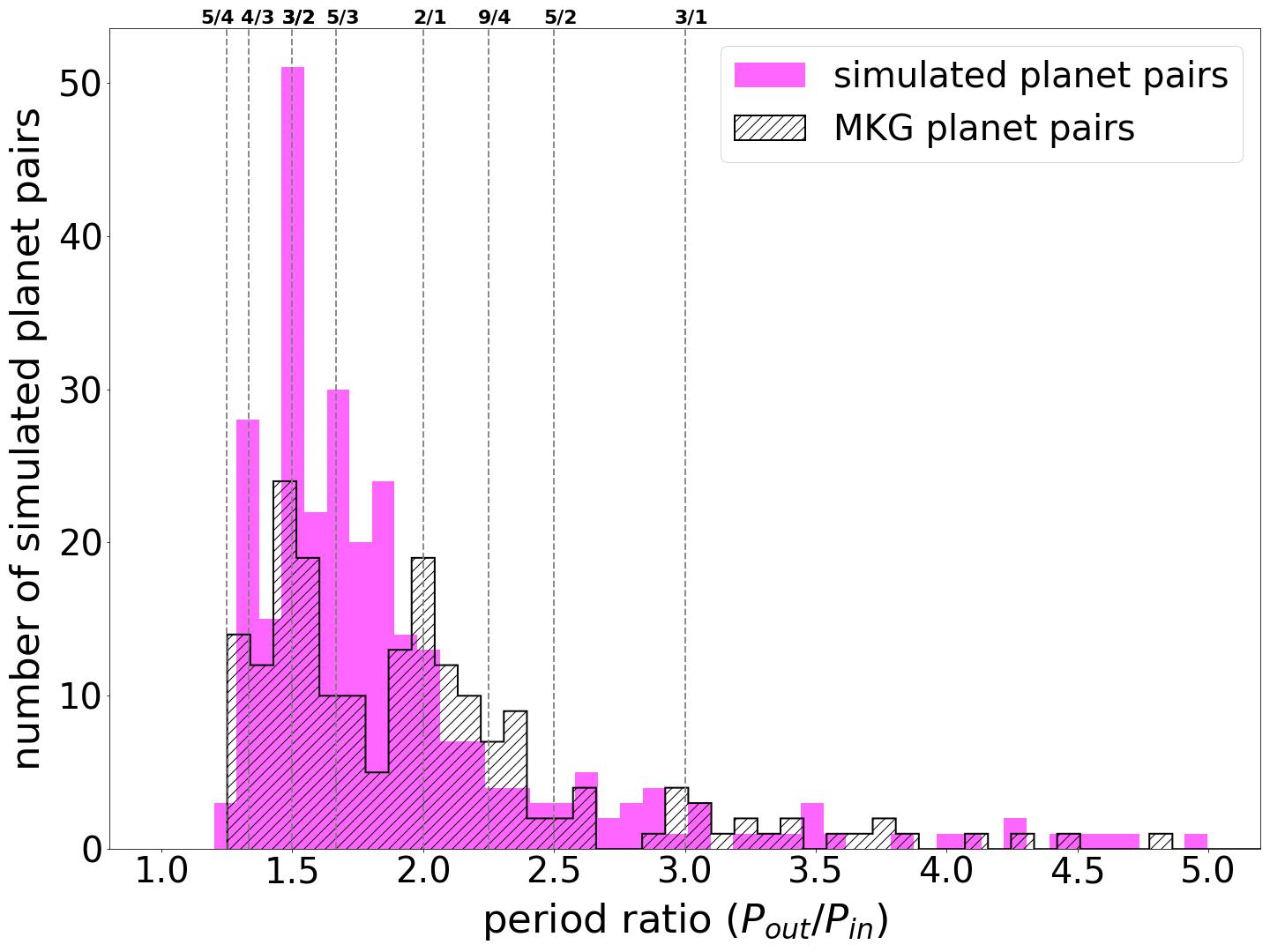}}
  \caption{Period ratios $P_{out}/P_{in}$ of all pairs of subsequent planets for our simulated population of planets (in magenta) compared to the period ratios of the pairs from our MKG sample (see the hatched area of the histogram). Main orbital resonances between the planet pairs are indicated by the dashed gray lines and identified by the fractions at the top of the plot.}
  \label{resonances}
\end{figure}

First we examine period ratios and spacing between planets in the simulated population. These are two of the most important characteristics of planetary systems. We compute period ratios $P_{\rm out}/P_{\rm in}$ of all subsequent pairs of planets in the individual systems for our simulated sample and compare them to the period ratios of the MKG pairs. Figure \ref{resonances} shows that many planetary pairs in both samples are located in or near orbital resonances, indicated by the dashed lines. Orbital migration often moves planets into resonances \citep{terquem2007migration}, which can then either stabilize or destabilize a system if the resonant chain becomes too long \citep{matsumoto2012orbital,goldberg2022criterion}. We see that the MKG sample has a quite broad distribution of period ratios quantitatively similar to our sample that is, however, a little narrower. The regions of the most occupied ratios partially overlap, but in the simulated population we do not see the decrease in occurrences at around 1.8, which is quite significant in the MKG sample. In both samples, the majority of the pairs are in 3:2 resonance, but this peak is much more prominent in our simulated sample compared to the MKG systems. Many exoplanets have been found to be in resonances, specifically, 3:2 resonance has been confirmed to be well populated by multiple studies \citep{lithwick2012resonant,batygin2013analytical,fabrycky2014architecture}. In our sample, the 5:3, 4:3, and much less pronounced 2:1 resonances, and in the observed sample 2:1 and 5:4/4:3 resonances, are also very common. A weaker peak near the 2:1 is present in the Kepler distribution according to \cite{lissauer2011architecture} as well. Other studies of the Kepler sample also confirmed weaker peaks near the 5:3, 4:3 and 5:4 resonances \citep[e.g.,][]{aschwanden2017exoplanet}. Nevertheless, as already discussed in Sect. \ref{introduction}, most currently known super-Earths are not found in resonant systems \citep{lissauer2011architecture,fabrycky2014architecture}. They spend time trapped in resonances, but most are removed by late instabilities according to breaking-the-chains scenario \citep[e.g.,][]{terquem2007migration,ogihara2009n,mcneil2010formation,cossou2014hot,izidoro2017breaking,izidoro2021formation}. In our simulations, almost all systems go through one or several instabilities, but the instabilities often disrupt only part of the chain, or the planets manage to readjust back to a resonant chain after the instability. At 20~Myr, around 75\% of the planet pairs in our systems are not in resonances. Also, around 75\% of the systems do not contain multiple-planet resonant chains, even though some of them still contain individual pairs in resonances. As shown in Fig. \ref{fig:dynEvolution79}, even a system that underwent several instabilities may have some planets in resonances. If we define systems with surviving resonant chains as the stable ones, then we get around 75\% of unstable systems. Previously we showed that 85\% of our systems potentially underwent at least one late instability (Sect. \ref{evolution}). These numbers are consistent with the estimates in \cite{izidoro2017breaking}, where they presented that at least 75\% of their simulated systems must be unstable to match the observations; and probably even 90–95\%. We also showed that extending the simulation time would probably increase the numbers (Sect. \ref{evolution}).   

\begin{figure}
  \resizebox{\hsize}{!}{\includegraphics{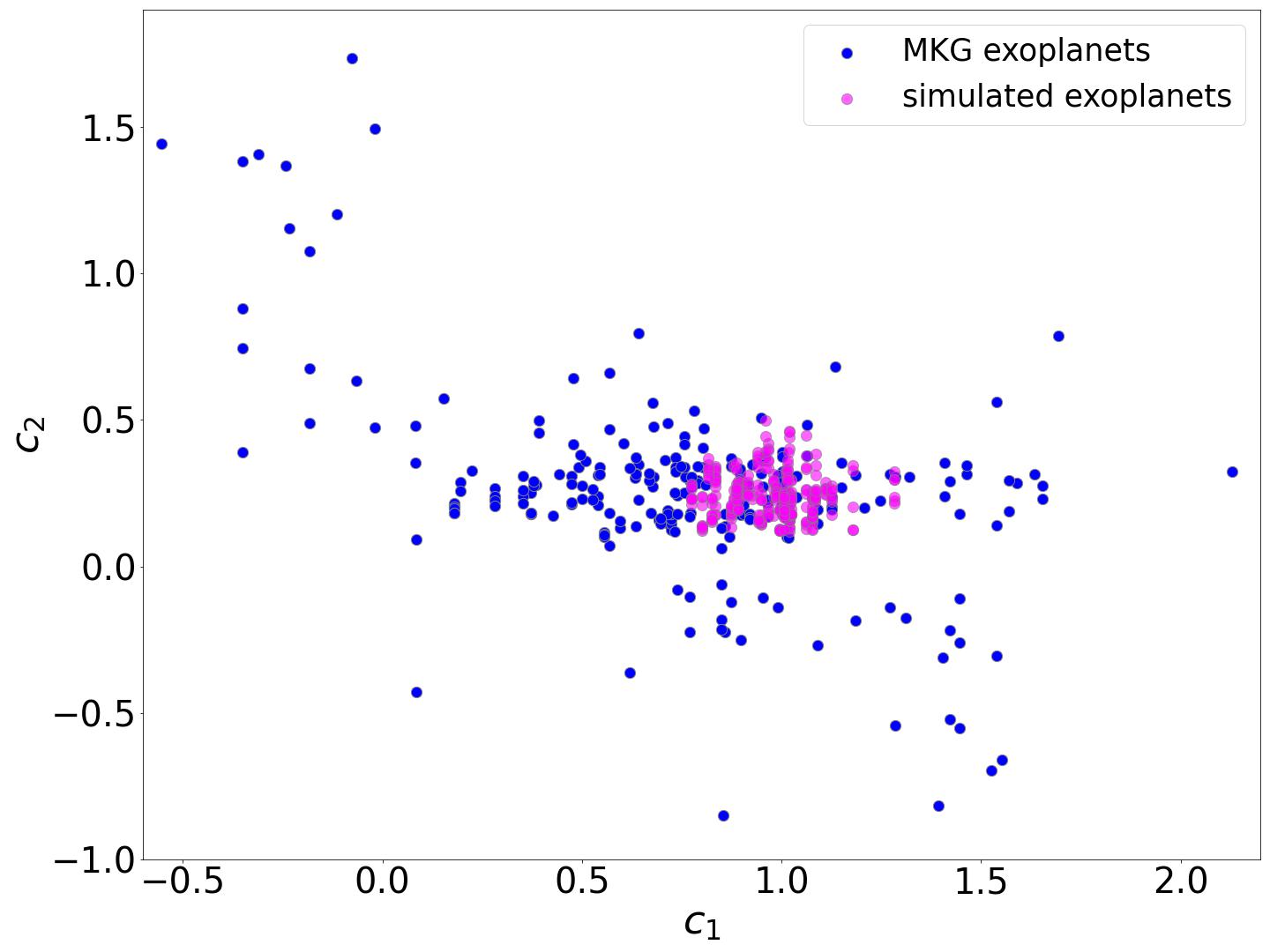}}
  \caption{Logarithmic spacing between pairs of neighboring planets in each planetary system. The constant $c_2$ is plotted against the values of constant $c_1$. Magenta and blue circles represent our simulated sample and the observed sample around MKG-dwarf stars, respectively.}
  \label{c1c2}
\end{figure}

Planetary systems generally follow a simple logarithmic spacing rule \citep[e.g.,][]{bovaird2013exoplanet,mousavi2021exoplanets}. The relation can be written as follows: $\log P_{n}= c_1 + c_2(n-1)$, where $n$ is the planet number from the inside out, and $c_1$ and $c_2$ are (fitting) constants, where $c_1 = \log P_{1}$ (i.e., the log of the period of the innermost planet). We compute logarithmic spacing between pairs of neighboring planets in each planetary system (see Fig. \ref{c1c2}) for both the simulated and the observed systems, and plot the obtained values of $c_2$ versus $c_1$. The figure shows our simulated sample together with the sample around MKG dwarfs. The observed sample around K dwarfs is included in the calculations, but is not displayed as it looks similar to the MKG sample (with fewer planets though). For the simulated systems, $c_1$ ranges from 0.8 to about 1.3 and the typical value is close to 1 (mean = $0.98\pm0.10$) -- that is, $P_{1} \sim$ 9.5 days -- while $c_2$ ranges from 0.1 to 0.5 and is typically around 0.24 (mean = $0.24\pm0.08$) -- that is, $P_{\rm out}/P_{\rm in} = 10^{c_2} \sim$ 1.7. For known systems around K and MKG dwarfs, the values of $c_1$ and $c_2$ show a large range: for the K-dwarf sample, the mean of $c_1$ is $0.68\pm0.42$ and of $c_2$ it is $0.27\pm0.33$, and for the MKG sample, the mean of $c_1$ is $0.75\pm0.46$, and of $c_2$ it is $0.26\pm0.33$. This means that the typical period of the innermost planet is $\sim$ 5 days and $\sim$ 6 days, and the typical $P_{\rm out}/P_{\rm in}$ ratio is $\sim$ 1.9 and $\sim$ 1.8 for the K and MKG samples, respectively, whereas $P_{1} \sim$ 9.5 days and $P_{\rm out}/P_{\rm in} \sim$ 1.7 for the simulated population. We note that the innermost planets in the simulated systems are farther from their host stars than in both observed samples, which results in generally higher values of $c_1$. This can be explained by the fact that the simulations are limited by the inner truncation radius. Many of the known planets are located closer to the star than the inner truncation radius of our simulations. Also, there is a small difference in the period ratios: the orbits of two adjacent planets are generally a bit closer to each other in the simulated systems compared to the known systems ($P_{\rm out}/P_{\rm in} \sim$ 1.7 versus $\sim$ 1.8/1.9). Small planets are usually packed tighter than larger planets and the simulated planets are generally less massive than the planets in the observed samples, so this behavior might be expected. In addition, some of the currently known systems are possibly incomplete and these "missing" planets can affect the typical period ratio of the whole system. Nevertheless, the difference is so small that we did not explore it more. We also clearly see (in Fig. \ref{c1c2}) that the simulated systems are much more similar to each other than the observed ones; the latter are very diverse. This is also expected since the initial conditions and parameters of the simulations are limited to a several specific values; therefore, they might not be able to produce very diverse planetary systems.

\subsection{Long-term stability of the simulated systems} \label{stability}
\begin{figure}
  \resizebox{\hsize}{!}{\includegraphics{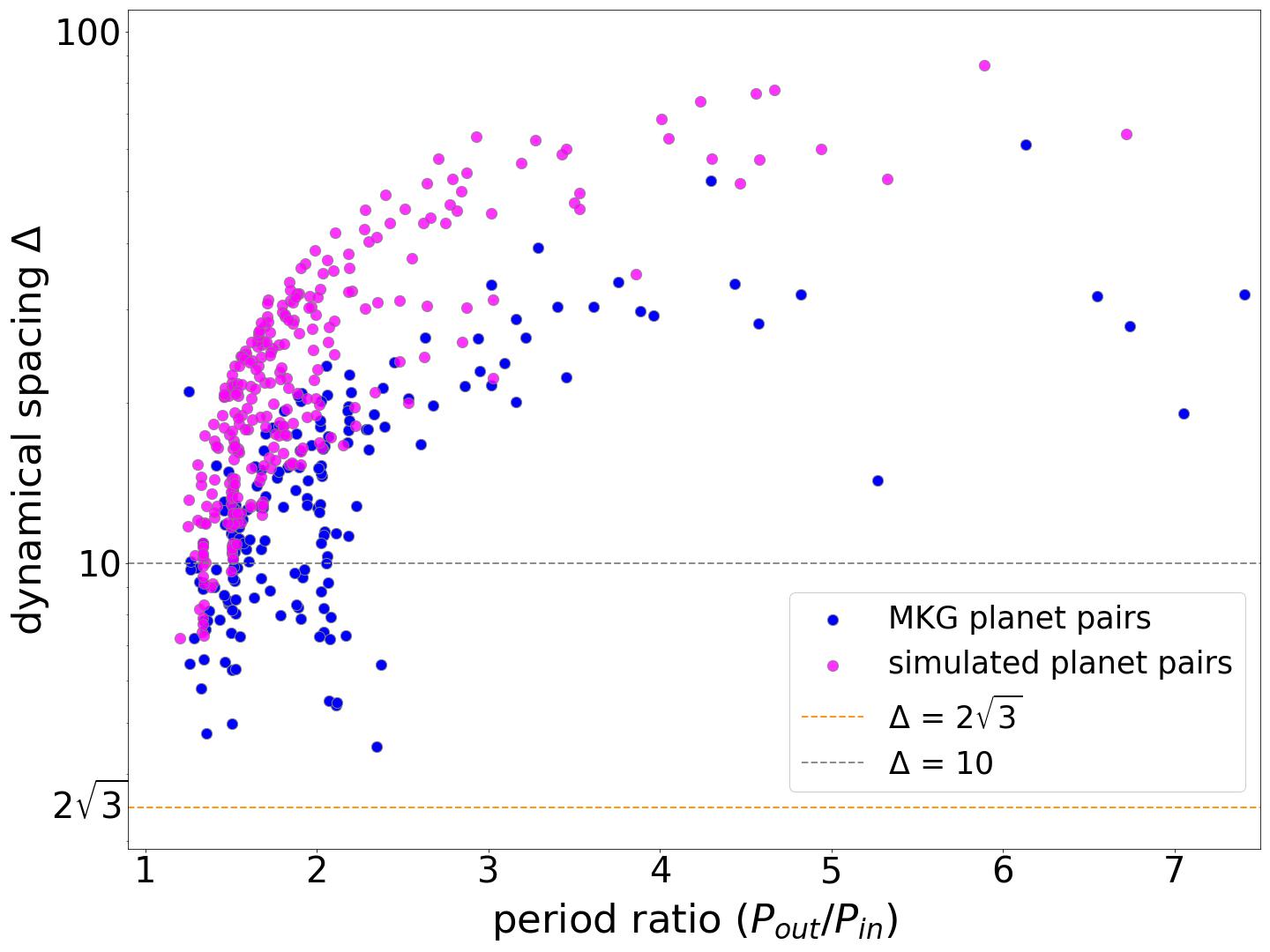}}
  \caption{Dynamical spacing, $\Delta,$ between all pairs of adjacent planets in each system in units of mutual Hill radius plotted against their period ratios, $P_{out}/P_{in}$. The figure shows our simulated sample (in magenta) together with the observed sample around MKG dwarfs (in blue). Two important values of dynamical spacing are represented by the dashed lines: the Hill stability criterion $\Delta$ = $2\sqrt{3}$ \citep{chambers1996stability} in orange and $\Delta = 10$ in gray.}
  \label{delta}
\end{figure}

To assess a long-term orbital stability of our simulated objects in their positions in multi-planet systems without resorting to computationally expensive simulations, usually simple, approximate conditions can be used, such as a minimum separation between neighboring planets in mutual Hill radii or dynamical spacing $\Delta$ \citep{chambers1996stability,pu2015spacing}. Following the Hill stability criterion \citep{chambers1996stability}, which defines the critical value of the mutual separation as $2\sqrt{3}$ Hill radii, we can access the Hill-stability of the systems. Planetary orbits in Hill-stable systems should be unable to cross each other. None of the systems, neither observed nor simulated, contains a pair of planets with a mutual distance below this critical value (see Fig. \ref{delta}), so none are supposedly Hill-unstable. Still, in any system, two adjacent planets are less likely to be long-term (gigayear-scale) stable when their dynamical spacing is smaller than $\Delta = 10$ \citep{pu2015spacing}. Only $\sim$ 5\% (13 out of 288) of all simulated pairs are less likely to be stable compared to almost a quarter of the K-dwarf systems and almost a third of the MKG systems. Yet, if they are long-term stable, these planets are likely to be in orbital resonance with each other, as is the case for many exoplanetary systems. In general, the notion of Hill stability should be interpreted with caution. For example, Jupiter and Saturn have dynamical spacing $\Delta \sim 8$ and still they are very dynamically stable as a pair \citep{laskar2008chaotic}.

Figure \ref{delta} displays the dynamical spacing $\Delta$ computed between all pairs of adjacent planets in each system. Planets in Kepler multi-planet systems tend to have a spacing of around 20 Hill radii, and are usually not closer together than 10 Hill radii \citep[e.g.,][]{lissauer2011architecture,fabrycky2014architecture,weiss2018california}. For the simulated systems, only $\sim$ 5\% of the adjacent pairs have $\Delta < 10$ and the typical $\Delta$ value is $\sim$ 21 (median = 20.9 $\pm$ 15.9), in agreement with previous studies. For the observed K-dwarf sample, $\sim$ 24\% of the adjacent pairs have $\Delta < 10$ and the typical value is $\sim$ 15 (median = 15.2 $\pm$ 9.0), and for the observed MKG-dwarf systems, $\sim$ 31\% of the adjacent pairs have $\Delta < 10$ and the typical value is $\sim$ 13 (median = 12.8 $\pm$ 12.1). This is similar to the findings of \citep{fang2012architecture}. Even so, the simulated systems tend to be wider spaced than the observed ones. The dispersion is very large for all three typical $\Delta$. This is because all three populations include neighboring pairs of very massive planets orbiting very close to each other, as well as low-mass planets located far away from each other. Again, the possibility that at least some of the currently known systems might be incomplete can also play a role here. Additionally, if we focus only on the zone of more massive planets within 0.3~AU of the star (which is where basically all observed planets around K dwarfs are found), then we get a lower typical value of $\Delta$ $\sim$ 15 (median = 15.2 $\pm$ 5.6) for the simulated systems. This value agrees very well with $\Delta$ $\sim$ 15 from the K-dwarf sample. Planetary orbits are usually spaced farther apart as distance from the star increases, our results support this as well. As the observed samples contain mostly only planets very close to the star while the simulated systems contain both very close but also distant planets, this can explain the generally larger dynamical separation of the complete simulated population.

\begin{figure}
  \resizebox{\hsize}{!}{\includegraphics{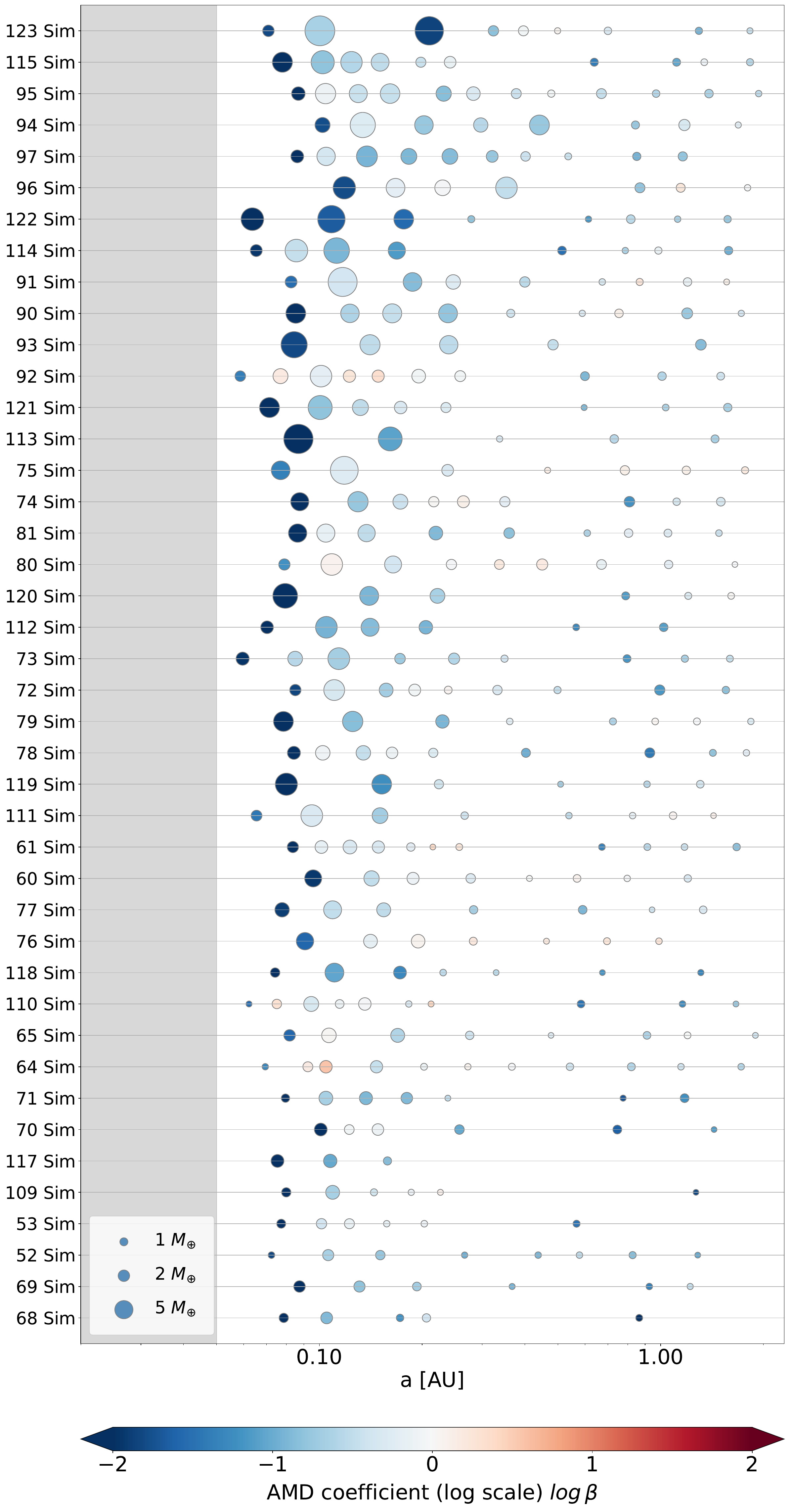}}
  \caption{AMD stability of the simulated systems. Each planet is represented by a circle, the size of which is proportional to the mass of the planet. The color represents the AMD stability coefficient, $\beta$ (in log scale), of the inner pair associated with the planet. The innermost planet is represented by the AMD stability coefficient associated with the star. Numbers in the simulation names (y-axis) are random, but the systems are ordered by their initial disk mass value, which decreases from top to bottom. Systems starting with the disk mass = $5~M_{\oplus}$ contain only one or two very small planets and are therefore not included in the plot (but they are AMD-stable).}
  \label{amd}
\end{figure}

The Hill stability criterion is intended for two-planet cases, not for multi-planet systems, as it does not consider mutual inclinations between orbits. Mutual orbital inclinations are known to influence the evolution of planetary systems and their long-term stability strongly. Therefore, we also use the angular momentum deficit \citep[AMD;][]{laskar1997large,laskar2000spacing,laskar2017amd,petit2017amd} to predict the long-term stability of our simulated systems. The AMD is a measure of the deviation of the system from being perfectly circular and coplanar. This is a more sophisticated approach that accounts for both eccentricity and mutual inclination, and it is applicable to multi-planet systems. A planetary system is AMD-stable if the AMD in the system is not sufficient to allow for collisions between its planets. We used the AMD stability criterion \citep{laskar2017amd} for 42 simulations (systems starting with a disk mass of $5~M_{\oplus}$ contain only one or two very small planets and are therefore not included), and in Fig. \ref{amd} we plot the values of the AMD stability coefficient, $\beta$ (in log scale), which is defined as
\begin{equation} \label{eq:2}
\beta=\dfrac{{C}}{\Lambda^{'} C_c},
\end{equation}
where $C$ is the total AMD of a system, $\Lambda^{'}$ is the circular momentum of the outer planet and $C_c$ is the critical AMD, such that for smaller AMD, collisions are forbidden. A pair is considered as AMD-stable if its AMD stability coefficient $\beta < 1$ or $\log \beta < 0$; thus, collisions are not possible. A system is AMD-stable if every adjacent pair of planets is AMD-stable. In the case of the innermost planet, the pair is formed with the star. The majority of the simulated systems ($\sim$ 55\%) meets the criterion for AMD stability at the end of 20~Myr, which means that these systems quite quickly settled into a long-term orbital arrangement. The systems that are AMD-unstable are often very close to stability with only one or two pairs with $\log \beta$ positive, but very close to 0. The closest pair (the innermost planet and the star) is in all cases very AMD-stable; hence, we do not expect the planet to be in danger of colliding with the star, unless the AMD is transferred inward from a more distant AMD-unstable pair. However, AMD-unstable systems are not necessarily unstable. They might be stabilized by the presence of MMRs, which can prevent pairs from colliding or prevent tidal evolution. In fact, around 80\% of all pairs in the AMD-unstable systems are in or very near MMR. It seems that our simulations have formed mostly long-term stable planetary systems already after 20~Myr. Additionally, we extended the simulation time for some of the AMD unstable systems; one-third of all runs was extended to 40~Myr and one-fifth to 100~Myr. At 40~Myr, we see changes in AMD stability coefficient values in every system. AMD has been transferred between the planet pairs, so that some of them have become more AMD-stable and some less. At 100~Myr, we see almost no changes in AMD of the systems compared to the values at 40~Myr. Generally, we do not see much improvement in the AMD stability of the systems after extending the simulation time.

\subsection{Peas in a pod}
\begin{figure}
  \resizebox{\hsize}{!}{\includegraphics{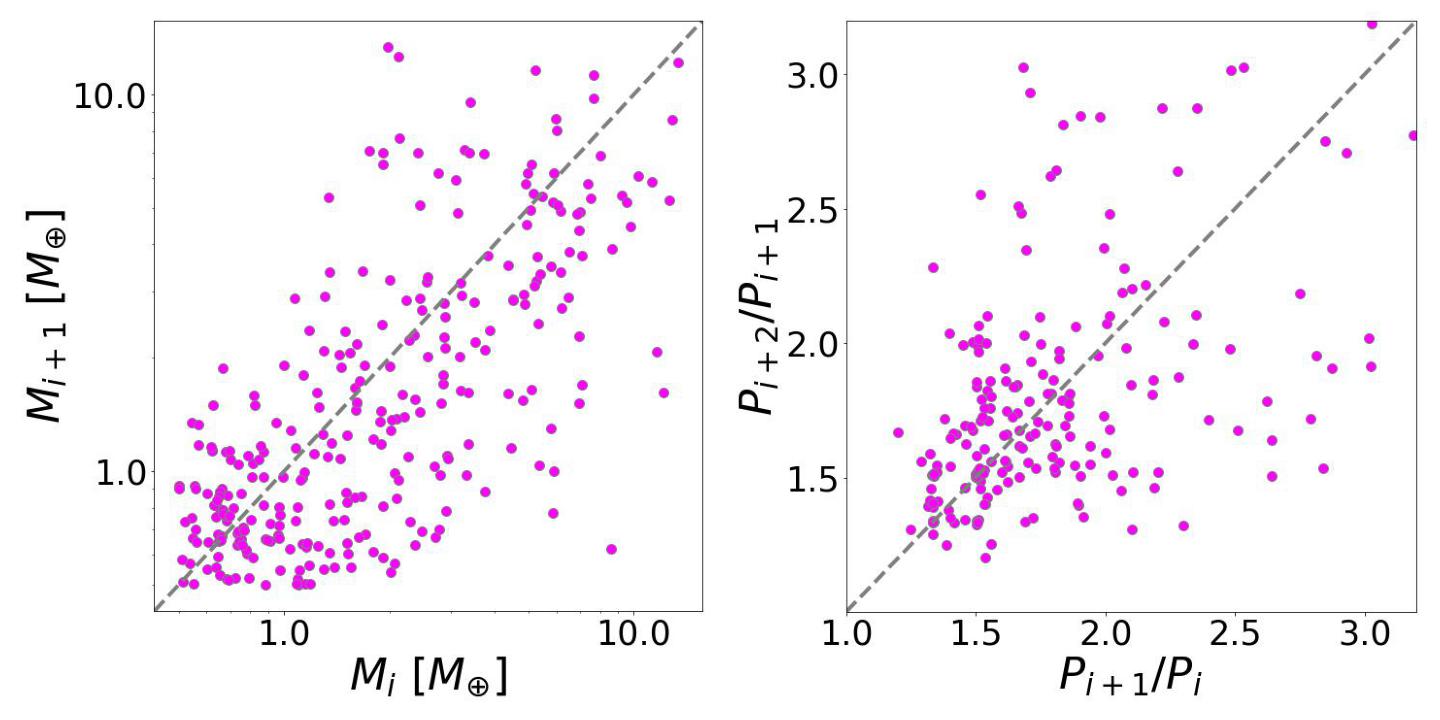}}
  \caption{Mass ratios of the outer/inner neighboring planet of the entire sample (the top plot) and period ratios of the outer/inner planet pair within 0.5~AU of the star (the bottom plot) in the simulated population. The dashed gray line is the 1:1 line.}
  \label{peas}
\end{figure}

Exoplanets within a system tend to be of similar sizes and masses, often ordered in size and/or mass and evenly spaced. In addition, smaller planets are frequently packed in tight configurations, while large planets often have wider orbital spacing (relative spacing between the planets then being approximately the same). This ``peas in a pod'' pattern, proposed for the first time by \cite[e.g.,][]{weiss2018california,millholland2017kepler}, is not universally accepted but recent observations indicate the presence of several of these correlations in the architecture of exoplanetary systems \citep{millholland2021split,otegi2022similarity}. Some correlations have also been reproduced in planet formation simulations \citep{mishra2021new,mulders2020earths}. Since our simulations do not allow us to assess radii of the final bodies, we examine masses of adjacent planets in the simulated systems and their spacing to evaluate whether our data follow at least some of these similarity trends. Figure \ref{peas} shows the mass ratios of all outer/inner neighboring planets (the top plot) and the period ratios of the outer/inner planet pairs (the bottom plot). According to the peas in a pod pattern, $M_{i+1}/M_{i} \approx 1$ and $(P_{i+2}/P_{i+1})/(P_{i+1}/P_{i}) \approx 1$ are expected for these ratios. For mass ratios, we analyzed all 288 planet pairs in all simulated systems that contain more than one planet, and using the Pearson correlation test, we calculated that there is a moderate positive correlation for masses of adjacent planets with the R-value of 0.58 and P-value of 4.27$\times10^{-27}$ for all planets in the data set. For period ratios, we have found no possible correlation for the entire sample with R-value of 0.04 and P-value of 0.53. This is likely due to the spatial division between the two categories of planets, the close-in more massive planets and the less massive distant planets with a region often containing no planets in the middle (as discussed in Sect. \ref{architecture}). However, if we consider only the more massive planets within 0.5~AU of the star (i.e., 127 planet pairs), then we again find a moderate positive correlation for periods of adjacent planet pairs with R-value of 0.54 and P-value of 4.48$\times10^{-11}$. Thus, we have shown that the architectures of our simulated systems often follow the peas in a pod pattern, at least when it comes to planetary mass and period ratios (only within 0.5~AU) of adjacent planets/planet pairs. \cite{mamonova2023patterns} analyzed an observed sample containing planets around M- and K-dwarf stars, and found a strong correlation of the R-value of 0.896 and P-value of 5.6$\times10^{-15}$ in mass ratios but no correlation in period ratios. This agrees with our findings as we find no correlation in period ratios, if we do not restrict the population.  

\section{Discussion} \label{discussion}
In this section, we present the final comparison of the simulated and observed populations and analyze whether our effort to reproduce the known exoplanet sample around K-dwarf stars was successful and to what extent, and which initial conditions seem to be the most favorable. Limitations of the study and potential future work are discussed at the end.

\subsection{Comparison to the known systems} \label{comparison}
\begin{figure}
  \resizebox{\hsize}{!}{\includegraphics{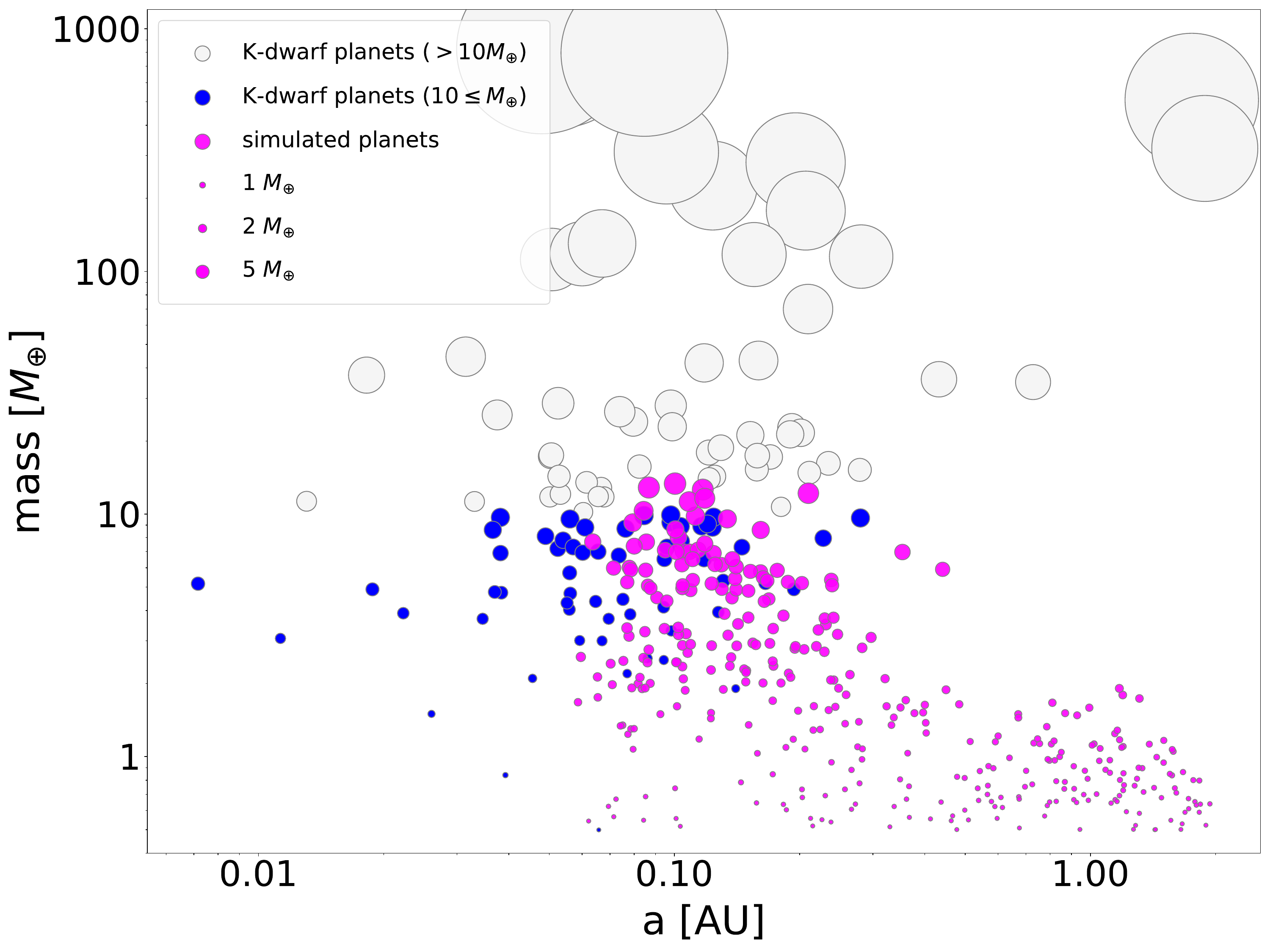}}
  \caption{Masses versus semimajor axes of planets of our simulated planetary systems (in magenta) compared to the currently known systems around K dwarfs (in blue). Only planets with known masses are presented. Known planets with masses above the super-Earths mass range (above $10~M_{\oplus}$) are indicated in light gray. Here, we use the classification of planets by mass proposed by \cite{stevens2013posteriori}. The point size indicates the masses in $M_{\oplus}$ of the planets. The sample of the known systems around K-dwarf stars was retrieved from \url{https://exoplanetarchive.ipac.caltech.edu/} on December 4, 2022.}
  \label{compared}
\end{figure}

The final visual comparison of the simulated versus observed planet population around K dwarfs is shown in Fig. \ref{compared}. Our focus is mainly on the known planets with a super-Earth mass ($\sim10~M_{\oplus}$) or less, although the simulations produced also several planets with slightly higher masses. There is a significant region where the two samples overlap, which shows that we managed to reproduce the observed population at least partially. The fact that simulations are limited by the inner truncation radius can explain why there are no simulated planets with semimajor axes below 0.05~AU. Our simulated systems also contain many small planets mostly at larger distances, which are basically "undetectable" by the current methods (as discussed in Sect. \ref{introduction}). These planets are located in the bottom right corner of the figure, where no observed planets can be, and has not been found.

Our simulated sample clearly differs from the Solar System planets and contains many closely-packed hot super-Earths; this is expected based on the observed sample and from our initial conditions, and it also confirms the hypothesis that the final systems tend to be compact with short orbital periods (as discussed in Sect. \ref{introduction}). According to \cite{migaszewski2016migration}, if a system ends up in a compact configuration, it might be attributed to multiple chains of MMRs between neighboring planets, arising from the planet–disk interactions causing inward migration at early evolutionary stages of planet formation. This is something what we observe in our simulations. The simulated sample does not consist of a large number of systems (42), but still a basic comparison to the observations is possible. In order to ensure that we make meaningful comparisons, we perform our statistics in several ways. We focus only on planets closer than 0.3~AU, as the K-dwarf sample basically does not include planets located farther from the star, except for a few giant planets (Sect. \ref{introduction}). We also focus only on the super-Earths, so that the larger observed planets are not considered. Additionally, we omit all the planets with masses below 2~$M_\oplus$ as the known population around K-dwarf stars contains almost no planets with such low masses. At last, we need to perform a slight modification to our data set. Due to the limitation of our model, our population is farther from the central star than the known population. To be able to assess whether we managed to reproduce the observations in term of semimajor axes, we need to eliminate this issue. As the inner truncation radius of the simulations is located at 0.05~AU from the star, we move all the simulated planets closer to the star by this distance. Finally, we recalculate the values from Tables \ref{table:1000}, \ref{table:1250}, and \ref{table:1500} based on the new cutoffs and present the updated values in Tables \ref{table:1000_2}, \ref{table:1250_2}, and \ref{table:1500_2}. As mentioned earlier, the average planet mass in the observed K-dwarf sample (when considering only the super-Earths, not the giant planets) is $\langle M_p \rangle$ = $5.9 \pm 2.5~M_{\oplus}$. When analyzing our data, specifically $\langle M_p \rangle$ and $M_{\rm max}$, we see that the simulations starting with $<15~M_{\oplus}$ do not reach the masses necessary to reproduce the average mass of the observed planets. It seems that the model needs a higher initial disk mass. On the other hand, as already shown in Fig. \ref{maxMass}, the most massive planets are generated with the disk mass of 30$~M_{\oplus}$ and above, but typically not with the highest disk mass. It appears as if the chosen initial disk mass values cover the whole range: the lowest do not reproduce the expected outcome, whereas the highest disk masses might be a bit too high. The typical number of planets in a system (when not considering disk mass = $5~M_{\oplus}$ systems) is now 3.1 $\pm$ 1.22, which agrees very well with the value for the observed sample around K dwarfs, which is 3.0 $\pm$ 1.3. 

\begin{table}
\caption{Simulations with 1000 g cm\textsuperscript{-2}. We list the average planet mass $\langle M_p \rangle$ with its standard deviation, the mass of the most massive planet, $M_{\rm max}$, and number of planets, $N$, in each system for the different initial disk masses (IDM) and two star masses. Values were calculated after cutoffs (see Sect. \ref{comparison}).}
\label{table:1000_2}      
\centering                                      
\begin{tabular}{c|c c c|c c c}          
\hline\hline                        
 & \multicolumn{3}{|c|}{Central mass} & \multicolumn{3}{c}{Central mass} \\\newline
 & \multicolumn{3}{|c|}{0.6 [$M_{\odot}$]} & \multicolumn{3}{c}{0.8 [$M_{\odot}$]} \\
\hline  
IDM & $\langle M_p \rangle$ & $M_{\rm max}$ & $N$ & $\langle M_p \rangle$ & $M_{\rm max}$ & $N$ \\\newline
[$M_{\oplus}$] & [$M_{\oplus}$] & [$M_{\oplus}$] & & [$M_{\oplus}$] & [$M_{\oplus}$] & \\
\hline 
10 & 2.09 & 2.09 & 1 & & & \\ 
15 & 2.24 $\pm$ 0.21 & 2.45 & 2 & 2.32 $\pm$ 0.03 & 2.35 & 2 \\   
20 & 3.39 $\pm$ 0.80 & 4.53 & 3 & 3.37 $\pm$ 0.89 & 4.38 & 3 \\   
25 & 2.73 $\pm$ 0.49 & 3.19 & 4 & 3.84 $\pm$ 1.91 & 6.52 & 3 \\
30 & 5.68 $\pm$ 1.31 & 6.99 & 2 & 4.17 $\pm$ 1.52 & 6.18 & 4 \\
35 & 3.55 $\pm$ 1.77 & 6.99 & 5 & 5.47 $\pm$ 0.26 & 5.87 & 4 \\
40 & 5.52 $\pm$ 1.57 & 7.53 & 3 & 5.31 $\pm$ 2.57 & 9.55 & 4 \\
\hline                                             
\end{tabular}
\end{table}

\begin{table}
\caption{Simulations with 1250 g cm\textsuperscript{-2}.}              
\label{table:1250_2}      
\centering                                      
\begin{tabular}{c|c c c|c c c}           
\hline\hline                        
 & \multicolumn{3}{|c|}{Central mass} & \multicolumn{3}{c}{Central mass} \\\newline
 & \multicolumn{3}{|c|}{0.6 [$M_{\odot}$]} & \multicolumn{3}{c}{0.8 [$M_{\odot}$]} \\
\hline  
IDM & $\langle M_p \rangle$ & $M_{\rm max}$ & $N$ & $\langle M_p \rangle$ & $M_{\rm max}$ & $N$ \\\newline
[$M_{\oplus}$] & [$M_{\oplus}$] & [$M_{\oplus}$] & & [$M_{\oplus}$] & [$M_{\oplus}$] & \\
\hline 
10 & 2.00 & 2.00 & 1 & & & \\ 
15 & 2.49 $\pm$ 0.36 & 2.88 & 3 & 3.07 $\pm$ 0.14 & 3.21 & 2 \\  
20 & 3.64 $\pm$ 0.86 & 4.85 & 3 & 2.53 $\pm$ 0.25 & 2.87 & 3 \\  
25 & 4.93 $\pm$ 1.58 & 6.20 & 3 & 4.32 $\pm$ 1.99 & 7.73 & 3 \\
30 & 4.34 $\pm$ 0.89 & 5.07 & 4 & 6.31 $\pm$ 3.97 & 11.61 & 3 \\
35 & 7.15 $\pm$ 2.27 & 10.30 & 3 & 5.79 $\pm$ 4.09 & 12.61 & 4 \\
40 & 3.95 $\pm$ 1.51 & 6.52 & 6 & 4.33 $\pm$ 1.38 & 6.19 & 6 \\
\hline                                             
\end{tabular}
\end{table}

\begin{table}
\caption{Simulations with 1500 g cm\textsuperscript{-2} (left) and 1750 g cm\textsuperscript{-2} (right).}              
\label{table:1500_2}      
\centering                                      
\begin{tabular}{c|c c c|c c c}           
\hline\hline                        
 & \multicolumn{3}{|c|}{Central mass} & \multicolumn{3}{c}{Central mass} \\\newline
 & \multicolumn{3}{|c|}{0.8 [$M_{\odot}$]} & \multicolumn{3}{c}{0.8 [$M_{\odot}$]} \\
\hline  
IDM & $\langle M_p \rangle$ & $M_{\rm max}$ & $N$ & $\langle M_p \rangle$ & $M_{\rm max}$ & $N$ \\\newline
[$M_{\oplus}$] & [$M_{\oplus}$] & [$M_{\oplus}$] & & [$M_{\oplus}$] & [$M_{\oplus}$] & \\
\hline 
10 & 2.90 & 2.90 & 1 & 2.58 $\pm$ 0.10 & 2.68 & 2 \\ 
15 & 2.87 $\pm$ 0.50 & 3.37 & 2 & 3.90 $\pm$ 1.43 & 5.34 & 2 \\   
20 & 5.42 $\pm$ 1.68 & 7.10 & 2 & 6.59 $\pm$ 0.78 & 7.37 & 2 \\   
25 & 4.27 $\pm$ 1.84 & 7.02 & 4 & 5.98 $\pm$ 2.43 & 9.21 & 3 \\
30 & 10.73 $\pm$ 2.13 & 12.86 & 2 & 5.22 $\pm$ 2.36 & 8.64 & 4\\
35 & 6.01 $\pm$ 2.94 & 9.79 & 4 & 8.26 $\pm$ 2.24 & 11.25 & 3 \\
40 & 5.57 $\pm$ 2.04 & 8.03 & 5 & 12.76 $\pm$ 0.58 & 13.34 & 2 \\
\hline
\end{tabular}
\end{table}

\begin{figure*}
\centering
   \includegraphics[width=17cm]{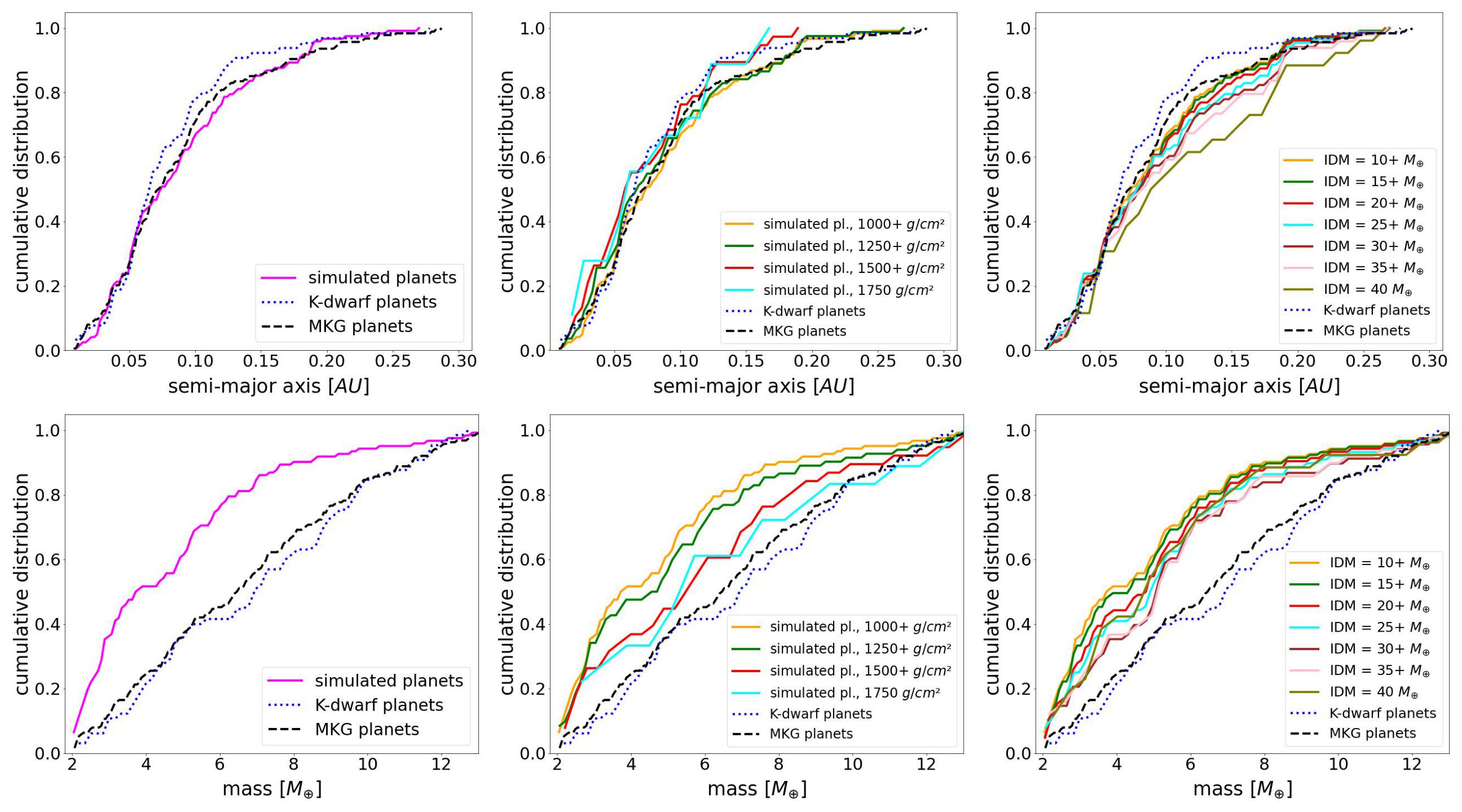}
  \caption{Cumulative distributions of the simulated semimajor axes (the top row) and planetary masses (the bottom row) for the whole population (the left column), for different gas surface densities (the middle column), and for different initial disk masses (IDM, the right column). IDM = 10+ $M_{\oplus}$ means that only systems starting with a disk mass of $10~M_{\oplus}$ or more are considered for that distribution. The same is true for the other IDMs as well as for the gas surface density values, e.g., 1000+ g cm\textsuperscript{-2}. IDM = 5+ $M_{\oplus}$ contains only a few very small planets compared to IDM = 10+ $M_{\oplus}$; therefore, we omitted that distribution. Cumulative distributions for the K-dwarf (blue dotted line) and MKG (dashed black line) samples are displayed for comparison.}
  \label{cumulative}
\end{figure*}

The chaotic nature of the formation process presents a challenge in comparing the outcomes of the simulations to the known population. It also makes it difficult to determine what initial conditions best correspond to reality as similar conditions can produce quite different final architectures of the systems (already shown in Sect. \ref{reproducibility}). Now we describe the model results in the context of our target expectations and variability of the parameter space we explored, and explain why not all models "match" the observed systems and examine how well they reproduce general trends in the currently known population. Investigating the cumulative distribution functions (see Fig. \ref{cumulative}) of the semimajor axes in the simulated systems (the top row), we see that the cumulative distribution of the whole population (on the left) is very close to the observed samples. We used the two-sample Kolmogorov-Smirnov (K-S) test to determine whether the two sets of samples (simulated + K-dwarf sample and simulated + MKG-dwarf sample) come from the same distribution. The p-values for the respective pairs of samples are presented in Tables \ref{tab:pvalues_sma} and \ref{tab:pvalues_m}. In the case of the whole simulated sample versus K-dwarf sample and MKG-dwarf sample, we find no significant difference between the distributions for either of the two sets. The simulated distribution actually follows the MKG sample better than the K-dwarf sample. This just might be due to the fact that the K-dwarf sample is smaller. Distributions for the different gas density values (the top middle plot) are also very close to the observed samples. The K-S test showing no significant difference between the distributions. The simulated distribution with 1250+ g cm\textsuperscript{-2} (i.e., simulations starting with the gas density 1250 g cm\textsuperscript{-2} and above) is actually closest to the K-dwarf sample. Distributions for the different initial disk masses (the top right plot; IDM = 10+ means that only the systems starting with the disk mass of $10~M_{\oplus}$ or more are considered) are still mostly very close to the observed samples with K-S test showing no significant differences, but we see that the higher disk masses, particularly 40~$M_{\oplus}$, are not that similar anymore based on the commonly used 5\% significance level. This might be partially caused by the small size of this particular sample. The simulated distribution for the disk masses of 10+ $M_{\oplus}$ (the same as for the whole simulated sample) is the most similar to the K-dwarf sample. When we examine the cumulative distributions for the planetary masses in the simulated systems (the bottom row), we again look at the whole simulated sample at first (the bottom left plot). Here, the simulated distribution looks very different compared to the observed ones and K-S test actually rejects that the two data sets are coming from the same distribution. This suggests that for these parameters our model is incapable to reproduce the observed population. The other two plots for the masses look more promising, as with the increasing gas surface density and initial disk mass (or more precisely, with removing their lowest values) the simulated distributions are getting closer to the observed distributions. Specifically, the distributions for the different gas densities (the bottom middle plot) show that the curves for the higher values actually resemble the observed distributions much more. Both samples 1500+ and 1750 g cm\textsuperscript{-2} possibly come from the same distribution as the K-dwarf sample. Finally, when we focus on the initial disk masses (the bottom right plot), we see that the higher disk mass values improve the distribution for values up to IDM = 30+ $M_{\oplus}$, above that higher disk mass brings no improvement. Distributions for masses $\leq5~M_{\oplus}$ are different from each other and are mostly getting closer to the observed ones with the increasing disk mass, while above $5~M_{\oplus}$ the cumulative distributions are quite similar for all disk masses, and are different from the observed ones. The distributions are converging, but not to the observations. The cumulative mass distributions for both observed samples increase almost linearly, implying that the probability distribution of the planetary masses is flat; in other words, finding planets with masses between 2 and 12~$M_\oplus$ is all equally probable (uniform distribution). This is not the case for the simulated population. Our cumulative mass distributions show a strong preference for planets with masses $<6$~$M_\oplus$, and fewer planets with larger masses, suggesting that we are missing a mechanism in our simulations for creating more massive planets. The K-S test also shows that our samples do not come from the same distribution as the K-dwarf population. 

For our simulations, we chose two different values of the density for the less massive star with $0.6~M_{\odot}$ and four values for the more massive star with 0.8~$M_{\odot}$. As our test-runs showed no major differences between the outcomes from the two different star masses, we limited the $0.6~M_{\odot}$ star simulations to the two density values to save the computation time. Also, we assumed lower gas density for the lower mass star. However, the unknown mechanism for creating more massive planets might actually be the higher gas density, as we actually manage to reproduce the masses of the observed sample with the values of 1500 g cm\textsuperscript{-2} and above. Both distributions (Fig. \ref{cumulative}, the bottom middle plot) show a trend similar to the uniform distributions of the observed populations. Additionally, extending the simulation running time could further increase the number of more massive planets and reduce the number of less massive planets, and result in a more uniform distribution of the simulated population. Our simulations also produce many low-mass planets with $M<2$~$M_\oplus$, both close to the star and farther away, which are not found in the observed sample. This appears to be due to the observational biases as we show in the following Sect. \ref{synthetic}. Whether these planets are indeed there and not yet discovered, or whether they are not there and our models are insufficient, at the moment we cannot say.

Models presented in this study do not include gas accretion onto the planets or atmosphere formation, which is a current limitation of GENGA (gas accretion is not yet implemented in the code). Even though the simulated planet population had mostly formed before the gas disk dissipated (see Sect. \ref{evolution}), the few planets with the potential of becoming gas giants reached the necessary mass to accrete significant atmospheres only after the gas was mostly gone. So since significant gas accretion is not a process that is expected to take place for the planets formed in our simulations, we assume that the gas then disappeared mainly through the photo-evaporation by the stellar radiation.

\subsection{Synthetic observations and their comparison to the Kepler population} \label{synthetic}
\begin{figure*}
\centering
   \includegraphics[width=17cm]{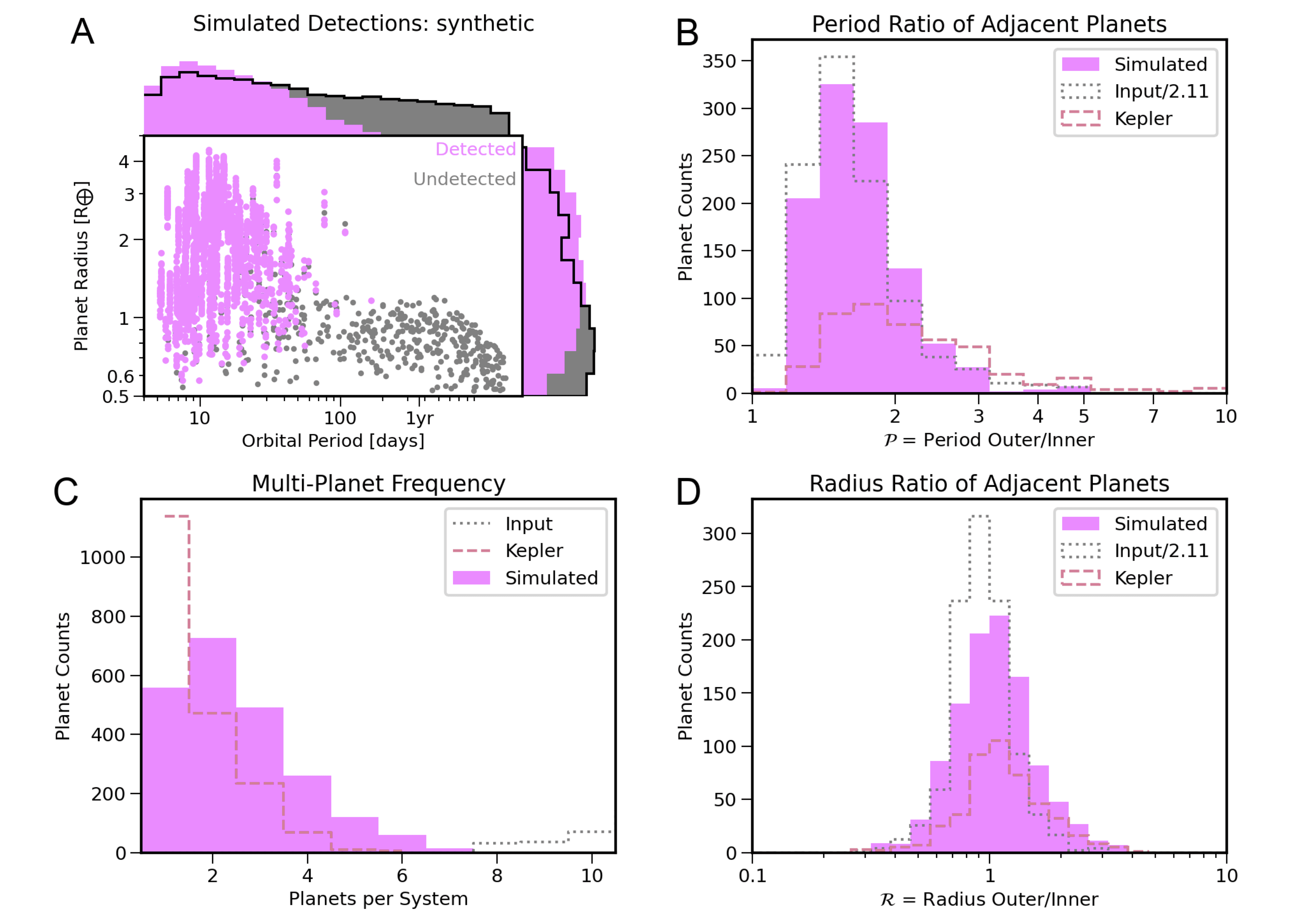}
     \caption{Synthetic EPOS \citep{mulders2018exoplanet,mulders2019exoplanet}  observations of our simulated population (A) and the associated statistics of the systems (B, C, and D). Panel A shows the orbital periods and planet radii (calculated using a mass-radius relation from \citealt{chen2016probabilistic}) of detected planets (in magenta) and undetected planets (in gray). The statistics in panels B, C, and D show the intrinsic distribution of our systems' properties with dotted gray lines, the synthetic observable distribution in magenta, and the distribution observed with Kepler in dashed red lines. Plot B displays the distribution of period ratios between adjacent planet pairs. Plot C shows the frequency of multi-planet systems, and plot D the distribution of radius ratios of the adjacent planet pairs. All statistics are calculated for planets with orbital periods between 1 and 400 days and radii between 0.5 and $5~R_{\oplus}$.}
     \label{fig:synthetic}
\end{figure*}

\begin{figure}
  \resizebox{\hsize}{!}{\includegraphics{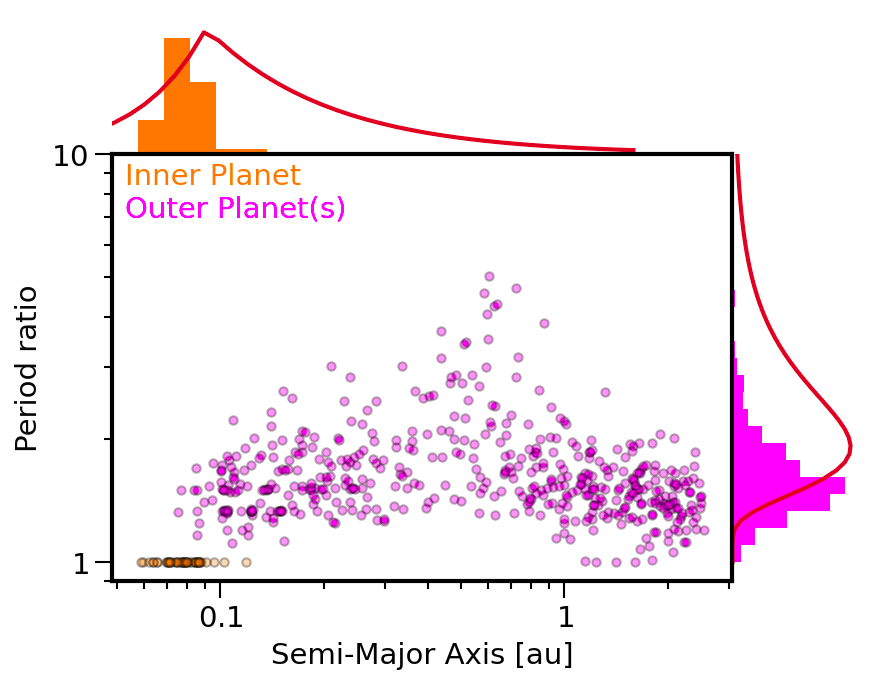}}
  \caption{Orbital period ratios (magenta) and innermost planet locations (orange). This plot was generated using EPOS \citep{mulders2018exoplanet,mulders2019exoplanet}. The period ratios of adjacent planet pairs are plotted at the semimajor axis of the outer planet. The histogram on the right shows the marginalized period ratio distribution compared to that of Kepler. The semimajor axis of the innermost planet in each system is displayed in orange. The histogram on top shows the marginalized distribution of the innermost planets compared to the distribution derived from Kepler (red line). Both Kepler distributions are derived by \cite{mulders2018exoplanet}.}
  \label{Pratio-sma}
\end{figure}

To account for Kepler observational biases, we performed synthetic observations of our systems using the Exoplanet Population Observation Simulator \citep[EPOS;][]{mulders2018exoplanet,mulders2019exoplanet}, designed to simulate survey observations of synthetic exoplanet populations. The survey detection efficiency is the average detection efficiency of all main-sequence stars of spectral types F, G, and K. Figure \ref{fig:synthetic} presents the synthetic observations of our simulated population and the associated statistics of the systems. The code uses planetary radii for the statistics. As our model does not simulate radii, we used a mass-radius relation based on \cite{chen2016probabilistic} to calculate them. All statistics are computed for planets with orbital periods between 1 and 400 days and radii between 0.5 and $5~R_{\oplus}$. Panel A shows the orbital periods and planet radii of detected (magenta) and undetected (gray) planets. We see that the undetected ones are mostly smaller planets of radii less than $\sim 1~R_{\oplus}$ (or a bit above that) and therefore masses less than $\sim 1~M_{\oplus}$, and with orbital periods longer than $\sim$ 100 days or semimajor axes more than $\sim$ 0.4~AU. These cutoffs are similar to the cutoffs used in Sect. \ref{comparison}, and actually exclude the majority of the planets with masses below 2~$M_\oplus$ from our simulated population as most of these low-mass planets are located farther than 0.4~AU from the central star (see again Fig. \ref{mass_distance}). The statistics in panels B, C, and D present the intrinsic distribution of our systems' properties, the synthetic observable distribution, and the distribution observed with Kepler. Plot B displays the distribution of period ratios between adjacent planet pairs. The observable distribution (in magenta) is slightly moved to the higher values compared to the intrinsic distribution (gray dotted line). Debiasing typically shifts the intrinsic period ratio distribution to larger values because planet pairs with one of the planets at larger orbital periods are less likely to be detected \citep[e.g.,][]{brakensiek2016efficient}. The opposite is happening here, since the intrinsic distribution is the "debiased" one. The reason why the data exhibit the opposite trend is the fact that many simulated planets at larger distances from the star have small period ratios (see Fig. \ref{Pratio-sma}), and since they also have lower detection probabilities, they are removed from the observable population, which causes the shift toward higher ratios in the observable distribution. This was observed and explained in \cite{mulders2019exoplanet}. The observable distribution follows the Kepler data (red dashed line) a bit better than the intrinsic distribution. Plot C shows the frequency of multiplanet systems, or the number of detectable planets per star. This statistic traces mainly the mutual inclination distribution together with the spacing between planets, their sizes, and orbital periods. Since usually only a few planets in each system are transiting, the frequency presented here (in magenta) does not really reflect the intrinsic multi-planet frequency (gray dotted line) of our population. However, in Sect. \ref{comparison} after applying the cutoffs to the population, we determined that the typical number of planets in our system is 3.1 $\pm$ 1.22, which seems to be in good agreement with the data in plot C. It is clear that the frequency does not show Kepler dichotomy \citep{johansen2012can}, the large number of systems with single transiting planets versus multiple transiting planets, which is clearly visible in the Kepler distribution (red dashed line). Plot D displays the distribution of radius ratios of the adjacent planet pairs. The intrinsic distribution (gray dotted line) peaks at values slightly lower than 1, whereas the observable distribution (in magenta) peaks at higher values, slightly above 1, corresponding to somewhat larger outer planets compared to the inner ones in the planet pairs. The observable distribution follows the Kepler data (red dashed line) much better than the intrinsic distribution. We will explore planetary radii more in future studies, when we can actually simulate them in our model. Figure \ref{Pratio-sma} shows, apart from orbital period ratios versus semimajor axis, also the innermost planet locations. The histogram on top shows the marginalized distribution of the innermost planets compared to the distribution derived from Kepler, and the histogram on the right shows the marginalized period ratio distribution compared again to that of Kepler. This plot is generated by EPOS, but it displays unprocessed data produced by our model.

\subsection{Study limitations} \label{limitations}
\begin{figure}
  \resizebox{\hsize}{!}{\includegraphics{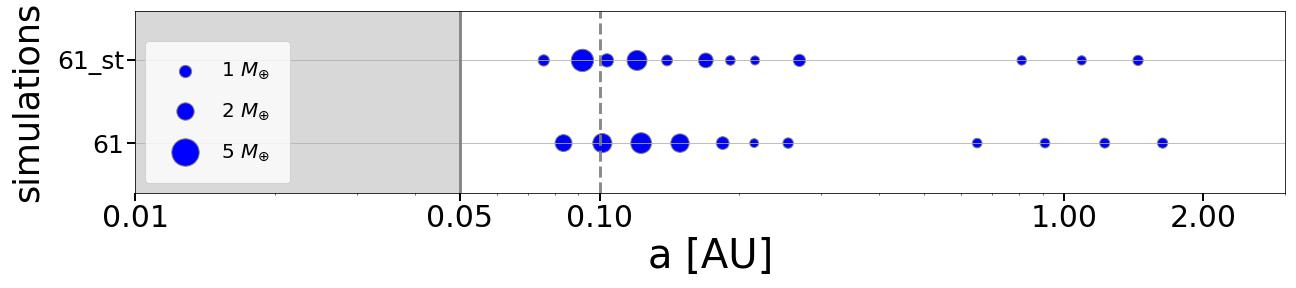}}
  \caption{Planetary systems formed after 10~Myr of simulation with the parameters central mass = $0.8~M_{\odot}$, initial disk mass = $20~M_{\oplus}$, and gas surface density = 1250 g cm\textsuperscript{-2}, with a time step of 0.7305 days (simulation 61) and 0.182625 days (simulation 61\_st).}
  \label{timestep}
\end{figure}

One of the main limitations of this study are the planetary radii not being realistic. As discussed in Sect. \ref{initial}, all the simulated planets have the same density as they formed by accreting planetesimals with the same density, and gravitational compression effects are not taken into account. Therefore, in this paper we do not draw any conclusions from the radii of our population, but we are planning to address them in the future studies. In Sect. \ref{initial}, we also mentioned using a slightly longer time step than is recommended due to extremely long computation time (several months) necessary for our simulations. Here, we compare the results of a run with the recommended time step with a run using the longer time step. Our simulations run with a time step of 0.7305 days. Additionally, we ran one of the simulations with a four-times-shorter time step of 0.182625 days (the numbers are chosen so that we get an integer number for the total simulation time), to show that the outcomes of the simulations do not seem affected by the length of the time step. Figure \ref{timestep} shows the planetary systems formed after 10~Myr of simulation with the parameters: central mass = $0.8~M_{\odot}$, initial disk mass = $20~M_{\oplus}$, gas surface density = 1250 g cm\textsuperscript{-2}, and a time step of 0.7305 days (simulation 61) and 0.182625 days (simulation 61\_st, shorter time step). We see that the architectures of the simulated systems and planet characteristics are similar. Simulations contain almost the same number of planets, 11 and 12, and the total mass of the planets is in both cases very similar, 14.4 and 14.3~$M_{\oplus}$. We see that the mass ratio of adjacent planets is somewhat different, with simulation 61 containing close-in planets with mostly similar masses, while the simulation 61\_st planets are quite different in mass. However, this is a common feature, which we can see in other simulations as well, either in Sect. \ref{effects} or in the various outcomes of 73 Sim simulations in Sect. \ref{reproducibility}. We used the K-S test to compare these two systems and get a p-value of 0.94 and 0.99 for masses and semimajor axes, respectively, which confirms that they are sampled from the same distribution. The simulation with the shorter time step needs a very long computation time (4 times shorter time step results in approximately 4 times longer execution time); therefore, we compared results after 10~Myr, even though in the rest of the paper we use the results after 20~Myr of simulation for simulation 61 (with the longer time step).

Another limitation is connected to the length of the time step, and it is the inner truncation radius set at 0.05~AU. Many of the known exoplanets are located closer to the star than the inner truncation radius of our runs; therefore, in order to reproduce the observed population, the simulations must be able to generate planets even closer to the central mass. However, setting the radius closer to the star will require an even shorter time step, and this will in turn increase the necessary computation time. Our simulations were carried out for 20~Myr due to the project time limit. It would be interesting to explore whether and how the planetary systems would change over even longer timescales and subjected to stellar and planetary tides, since in Sect. \ref{evolution} we see that some of the systems are still evolving after 20~Myr. We ran a third of our simulations (the most AMD-unstable ones) for another 20~Myr, and for the majority of the runs the extended time resulted in no significant changes in the architecture of the systems, their long-term stability, etc. However, some of the simulations show substantial changes in their architecture; therefore, in future studies, we should simulate at least 40-50~Myr of evolution. So this limitation can also be potentially addressed and maybe eliminated in the future work, if we have sufficient computational resources.

\section{Conclusions}
We have explored the effects that various initial model conditions and configurations have on the final outcomes of 20~Myr of planetary accretion around two K-dwarf stars with different masses. We find that: 
\begin{itemize}
  \item The more massive the solid disk we start the simulation with, the more mass will be projected into the total mass of the planets in each simulated system. Generally, our simulations have a high planet formation efficiency as 60\% to 84\% of the initial mass will end up in the planets. Additionally, the scaling between the disk mass and the planet mass (of the largest and second-largest planet) in our systems shows a mostly logarithmic increase with empirical fits of $M_{\rm pl}\propto M_{\rm disk}^{0.6}$ for a $0.6~M_{\odot}$ star and $M_{\rm pl}\propto M_{\rm disk}^{0.1-1.0}$ for a $0.8~M_{\odot}$ star. The comparison with non-migrating simulations shows that migration results in larger variations in planet masses (of the largest planets) and generally a slower increase in planet mass with the initial disk mass.  
  \item We manage to reproduce the main characteristics and architectures of the known systems, and produce mostly long-term stable systems after 20~Myr of evolution with an initial disk mass of $10~M_\oplus$ and above and a gas surface density value of 1500 g cm\textsuperscript{-2} and above. Our simulations also produce many low-mass planets with $M_p<2$~$M_\oplus$, both close to the star and farther away, which are not found in the observed sample. This appears to be due to the observational biases as shown by the performed synthetic observations of our systems.  
  \item The average planet mass in the observed K-dwarf sample (when not considering the giant planets) is $\langle M_p \rangle$ = $5.9 \pm 2.5~M_{\oplus}$. Our data, specifically the average planet mass $\langle M_p \rangle$ and the mass of the most massive planet $M_{\rm max}$, show that simulations that start with a disk mass of $<15~M_{\oplus}$ do not reach the masses necessary to reproduce the average mass of the observed planets. It seems that the model needs a higher initial disk mass.
  \item Our cumulative mass distributions for planets with masses between 2 and 12~$M_\oplus$ show a strong preference for planets with masses $M_p<6$~$M_\oplus$ and a lesser preferences for planets with larger masses. The cumulative mass distributions for the observed samples increase almost linearly. This suggests that we are missing a mechanism for creating more massive planets.
  \item Around 75\% of our systems do not contain multiple-planet resonant chains, and approximately 85\% of the systems underwent at least one late instability. These numbers are consistent with the lower estimate determined based on the observations and would probably increase if the simulation times were extended. 
  \item Earth-mass planets form quickly around 0.6 and 0.8 $M_\odot$ stars, mostly before the gas disk dissipates. The final systems after 20~Myr of evolution contain only a small number of planets with masses $M_p>10$~$M_\oplus$, and these formed after the gas was mostly gone. 
  
\end{itemize}

Future models that span a larger range of stellar and planetary masses and characteristics as well as properties of protoplanetary disks, and which use longer simulation times combined with new observations, will help us further improve our understanding of planet formation around K-dwarf stars.

\section*{Acknowledgments} 
The authors would like to thank the reviewer Sean Raymond for taking the time and effort necessary to review the manuscript. We sincerely appreciate all valuable comments and suggestions, which helped us to improve the quality of the manuscript. This study is supported by the Research Council of Norway through its Centres of Excellence funding scheme, project No. 223272 CEED (P.H., E.M., S.C.W.). It has been done at the Centre for Earth Evolution and Dynamics (CEED) at the University of Oslo. We also acknowledge financial support from the Research Council of Norway through its Centres of Excellence scheme, project number 332523 (PHAB). Our numerical simulations have been performed on Norwegian supercomputers Betzy and Saga operated by Sigma2 as a part of Stephanie C. Werner’s project NN9010K. The authors would like the thank Simon Grimm and Joachim Stadel for allowing them to utilize their software GENGA and for helping with the use of the software. The study has made use of the NASA Exoplanet Archive operated by the California Institute of Technology, and The Extrasolar Planets Encyclopaedia (exoplanet.eu). Additionally, the authors acknowledge the help from the members of the Earth and Beyond group at CEED, particularly Yutong Shan for her helpful comments and discussions.

\bibliographystyle{aa}
\bibliography{References}

\begin{appendix}
\section{Cumulative distributions of the simulated semimajor axes and planetary masses: Statistics}
The two-sample K-S test was used to test whether the two pairs of samples come from the same distribution. The p-values are presented in Tables \ref{tab:pvalues_sma} and \ref{tab:pvalues_m}; IDM = 10+ means that only the systems starting with the disk mass of $10~M_{\oplus}$ or more are considered. Based on the commonly used 5\% significance level, the values above 0.5 confirm that the samples are sampled from the same distribution. In some cases the simulated samples are quite small (the smallest containing only 20-30 planets), but K-S test can be reliably used for the samples of these sizes as well.  

\begin{table}[h]
\caption{Two-sample K-S test statistics (p-values): semimajor axis.}              
\label{tab:pvalues_sma}      
\centering       
\begin{tabular}{c|c|c}    
\hline\hline
 & K-dwarf sample & MKG-dwarf sample \\
\hline 
whole & & \\
simulated sample & 0.37 & 0.78 \\
\hline 
1250+ g cm\textsuperscript{-2} & 0.63 & 0.82 \\
1500+ g cm\textsuperscript{-2} & 0.49 & 0.44 \\
1750 g cm\textsuperscript{-2} & 0.56 & 0.71 \\
\hline 
15+ $M_{\oplus}$ & 0.25 & 0.61 \\
20+ $M_{\oplus}$ & 0.21 & 0.51 \\
25+ $M_{\oplus}$ & 0.18 & 0.21 \\
30+ $M_{\oplus}$ & 0.13 & 0.10 \\
35+ $M_{\oplus}$ & 0.12 & 0.14 \\
40 $M_{\oplus}$ & 0.06 & 0.14 \\
\hline
\end{tabular}
\end{table}

\begin{table}[h]
\caption{Two-sample K-S test statistics (p-values): mass.}              
\label{tab:pvalues_m}      
\centering       
\begin{tabular}{c|c|c}    
\hline\hline
 & K-dwarf sample & MKG-dwarf sample \\
\hline 
whole & & \\
simulated sample & $5.95\times10\textsuperscript{-6}$ & $8.73\times10\textsuperscript{-8}$ \\
\hline 
1250+ g cm\textsuperscript{-2} & $3.11\times10\textsuperscript{-4}$ & $7.69\times10\textsuperscript{-5}$ \\
1500+ g cm\textsuperscript{-2} & 0.25 & 0.37 \\
1750 g cm\textsuperscript{-2} & 0.59 & 0.76 \\
\hline 
15+ $M_{\oplus}$ & $1.08\times10\textsuperscript{-5}$ & $3.30\times10\textsuperscript{-7}$ \\
20+ $M_{\oplus}$ & $9.76\times10\textsuperscript{-5}$ & $9.93\times10\textsuperscript{-6}$ \\
25+ $M_{\oplus}$ & $6.10\times10\textsuperscript{-4}$ & $1.58\times10\textsuperscript{-4}$ \\
30+ $M_{\oplus}$ & $3.17\times10\textsuperscript{-3}$ & $2.03\times10\textsuperscript{-3}$ \\
35+ $M_{\oplus}$ & 0.01 & 0.01 \\
40 $M_{\oplus}$ & 0.02 & 0.06 \\
\hline
\end{tabular}
\end{table}

\end{appendix}
\end{document}